\documentclass[a4paper]{article}
\usepackage{a4wide}
\usepackage{amssymb, amsmath, graphicx,epstopdf, amsfonts}
\usepackage[T1]{fontenc}
\usepackage{lmodern} 
\usepackage{enumitem}
\usepackage{mathpartir}
\usepackage{bm}
\usepackage{url}
\usepackage{cite}
\newcounter{qcounter}
\newcommand\tab[1][1cm]{\hspace*{#1}}

\newtheorem{ttd}{Definition}

\newtheorem{ttp}{Property}
\newtheorem{ttpp}{Proposition}
\newenvironment{proof}{\paragraph{Proof:}}{\hfill$\square$}

\begin{document}

\title{Privacy by Design: On the Formal Design and Conformance Check of  Personal Data Protection Policies and Architectures}

\author{Vinh Thong Ta\\
University of Central Lancashire (UCLan)\\
School of Physical Sciences and Computing (PSC)\\
 Preston, UK\\
vtta@uclan.ac.uk}

\maketitle

\begin{abstract}
The new General Data Protection Regulation (GDPR) will take effect in May 2018, and hence, designing compliant data protection policies and system architectures became crucial for organizations to avoid penalties. Unfortunately, the regulations given in a textual format can be easily misinterpreted by the policy and system designers, which also making the conformance check error-prone for auditors.    
In this paper, we apply formal approach to facilitate systematic design of policies and architectures in an unambiguous way, and provide a framework for mathematically sound conformance checks against the current data protection regulations.   
We propose a (semi-)formal approach for specifying and reasoning about data protection policies and architectures as well as defining conformance between architectures and policies. The usability of our proposed approach is demonstrated on a smart metering service. 
\end{abstract}

\section{Introduction}
\label{sec:int} 

Under the Data Protection Directive 95/46/EC \cite{d95}, the Data Protection Act 1998 \cite{data98}, and the new General Data Protection Regulation (GDPR) \cite{icoper}, personal data is defined as ``any information relating to an identified or identifiable natural person''.  The data protection regulations contain rights for living  individuals who have their data processed (i.e., rights for the data subjects), and enforce responsibilities for the data controllers and the data processors who store, process or transmit such data \cite{d95}. In the USA, personally identifiable information (PII) is used  with a similar interpretation to personal data in Europe \cite{NIST}. 

Unfortunately, despite the mandatory data protection regulations,  there were a huge number of data breach incidents in the past \cite{CNET-012014, Bloomberg-012014, Guardian-032015, Guardian-062015, Guardian-122014, Engaget-072015} and nowadays, such as the recent Cambridge Analytica scandal of Facebook \cite{Faceanalytica17}, where personal data of more than 87 millions Facebook users has been collected and used for advertising and election campaign purposes without a clear data usage consent. One of the main problems was the insufficient check by Facebook on the third party applications. The UK based telecommunication company, TalkTalk, has been reported being penalised several times for failing to protect appropriately customers' data, and put around 21,000 customers' details at risk in the past \cite{Talktalk}. Google also faced lawsuit over collecting personal data without permission several times, and has been reported to illegally gathered the personal data of millions of iPhone users in the UK in one of the recent news \cite{Google}. 

Strengthening the protection of personal data as well as increasing transparency are among some of the main goals of the new General Data Protection regulations that  replaces the Data Protection Directive 95/46/EC. Privacy by design is becoming a legal obligation for service providers  under the Article 25 of the GDPR \cite{Gdpr25}, which requires the design of data protection and control  into the development of business processes for service providers.  According to the BBC news \cite{BBC-012014} in 2014, the new data protection proposals include a clause to prevent European data being shared with another country, as well as providing rights and means for citizens to erase their personal data. 
The regulation also attempts to limit businesses from performing user profiling and demanding appropriate consents before personal data collection (see the Article 7 of the GDPR) \cite{Gdpr7}. 

As the GDPR comes into play in May 2018, numerous summits, conferences and workshops have been organised by data protection bodies to help business and organisations with preparing for the compliant policy and system design. One of the main problems system designers have to face is that regulations given in the textual format is easy to be misinterpreted and, hence, increasing the chance of error in both the designing and auditing phases.    

Some applications based on questionnaires with a huge number of informal questions  (e.g., Simplexxy\footnote{\url{bbj.hu/business/software-offers-help-as-gdpr-compliance-deadline-looms_148061}}) help the policy and system designers to check if their system is compliant with the GDPR, but this still provides only informal hints and directions, instead of helping with system design in an unambiguous way. They do not facilitate any automated reasoning about the data protection properties (e.g., whether a given entity can have a given data based on a policy). 

From the technical perspective, to the best of our knowledge,  only a very limited number of works can be found in the literature that investigate the  formal model or systematic method to design privacy policies and architectures, as well as facilitate compliance check of existing services with the current data protection rules. The main advantage of using formal approaches during the system design is that the compliance with the regulations can be confirmed based on mathematical proofs, unlike the error-prone informal reasoning. 

We propose a variant of semi-formal policy language for specifying data protection policies in a compact way, and semantics with compliance rules for reasoning about different data protection properties of the policy. Furthermore, we also propose a variant of  architecture language, with which system architectures can specified and the data protection properties can be analysed. Finally, we define different mappings and conformance relations between the system architectures and policies. Based on a smart metering service case study, we also demonstrate the usability of our proposed languages.
This work is a modified and improved version of our preprint report first published on Arxiv in 2015~\cite{TaArxiv15}. The main difference between this paper and \cite{TaArxiv15} is that here we mainly concentrate on the data protection properties, while in \cite{TaArxiv15}, we investigated the dynamic nature of policies to model data control rules changing over time (e.g., when the Facebook users change their friend lists, the data control policy will change). 

\section{Contributions and Challenges}
\label{sec:motcont} 

In the following, we provide the main contributions of this paper in Section~\ref{sec:cont}, then we discuss the main challenges in Section~\ref{sec:challenge}.

\subsection{Contributions}
\label{sec:cont}

\begin{figure}[htb!]
    \begin{center}
        \includegraphics[width=0.85\textwidth]{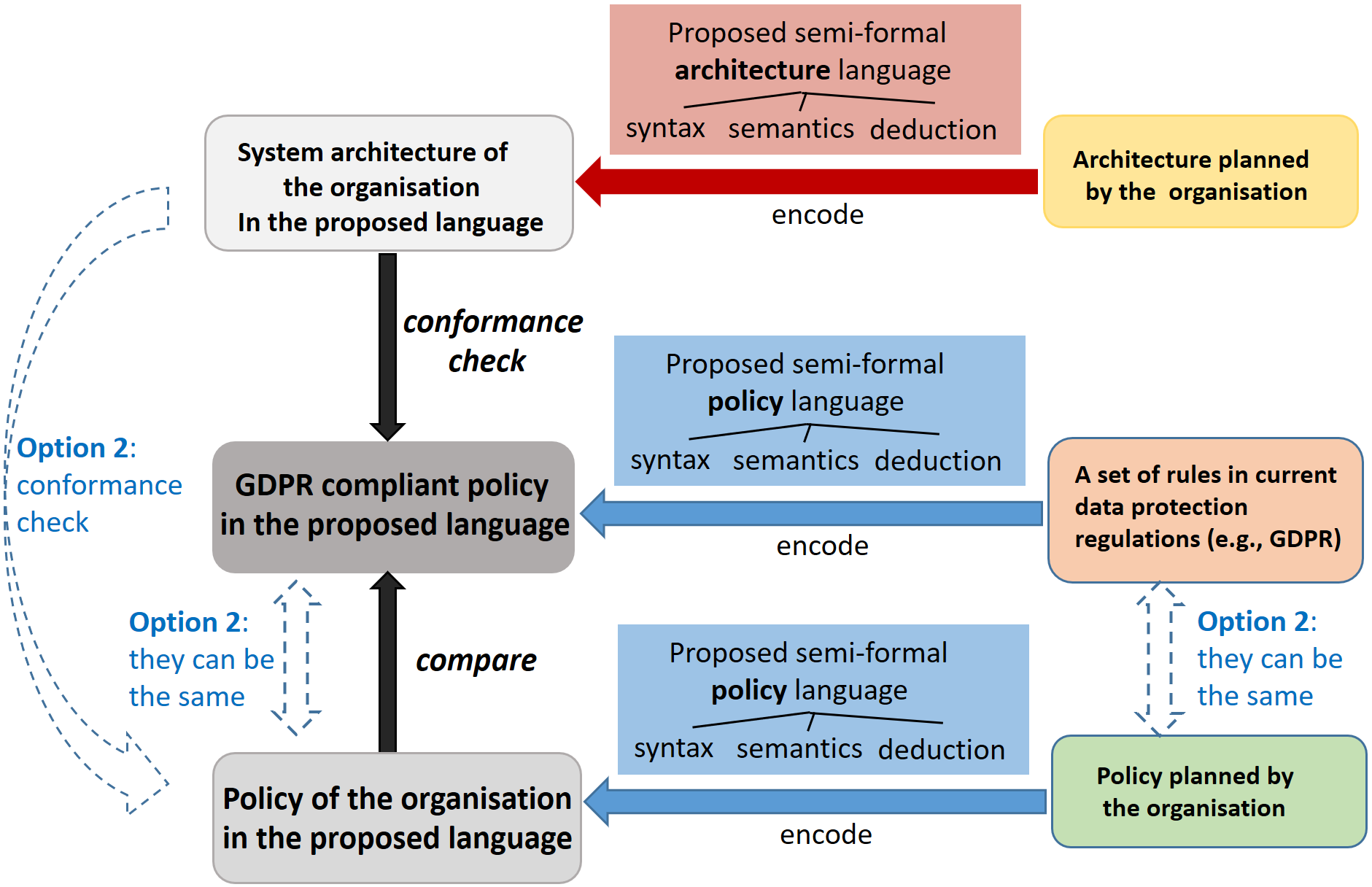}
    \end{center}
    \caption{An overview and the intuition behind the contributions of this paper.}
    \label{fig:contribution}
\end{figure}

Addressing the problems discussed in Section~\ref{sec:int}, this paper proposes a policy and an architecture language for specifying and reasoning about data protection requirements. Our idea and contributions are depicted in the Fig.~\ref{fig:contribution}. 

Our first contribution (Section~\ref{sec:polgen}) is the proposed policy language that enables us to encode the data protection regulations in a more compact and unambigous form.  Our policy languague is a variant of the language proposed by Ta and Butin et al. \cite{TaButin15, ButinFM14}, however, it is more generic than that, because it is crafted for capturing the whole data life-cycle with different syntax, semantics and new definitions and compliance rules. This can help the policy designers to verify if their policy is in compliance with the current regulations. We can also verify which entities can have a given personal data based on the policy.  
 
Our second contribution (Section~\ref{sec:arch0}) is a lower level architecture language for specifying and reasoning about system architectures.  Again, our language is inspired by the works by Ta and Antignac et al. \cite{TaAntignac14, Antignac14}, but specifically designed for data protection regulations with different syntax, semantics, definitions and deduction rules to check certain data protection properties. We will discuss the differences in more details in the Section~\ref{sec:related}. 

Our third main contribution (Section~\ref{conformance}) is a mapping procedure between the policy and architecture levels. In particular, with new definitions, properties and propositions we also provide different conformance checks between the specified architectures and policies. To the best of our knowledge this has not been investigated and proposed before. In one of our previous works  \cite{TaAntignac14}, we have investigated the mapping and conformance check between system architectures and concrete system implementations.    

Finally, in a case study (Section~\ref{pol:smartmeter}) we demonstrate the usage of our languages for defining and analysing a data protection policy and architecture of a smart metering service, as well as the conformance checks between them. 

\subsection{Challenges}
\label{sec:challenge}

The biggest challenge we have to have to face in this paper beside defining an correct syntax and semantics for the languages, is how to find a mapping and conformance relations between a higher level (i.e. more abstract) policy and a lower level system architecture. The main novelty in this work compared to the previous work is that here we specifically crafted the policy and architecture languages syntax and semantics in such a way that appropriate mappings can be defined between a data protection (DPR) policy and an architecture, as well as facilitating conformance checks between the two levels. 

\section{Related Works vs. Our Work}
\label{sec:related}

There are numerous policy languages for modelling privacy, security and access control properties of different systems in the past. For instance, the Platform for Privacy Preferences (P3P) \cite{P3P} was an attempt to enable web users 
to gain control over their private information on online services. In particular, it enables websites 
to express their privacy practices in a standard format that can be retrieved automatically and interpreted 
easily by web client applications. Users will be notified about certain website's privacy policies and have a 
chance to make decision on that. To match and check the privacy preferences of users and web services, the Working 
Group also proposed a P3P Preference Exchange Language (APPEL) \cite{APPEL} integrated into the web clients, with 
which the user can express their privacy preferences that can be matched against the practices set by the online 
services. While the idea was nice, there were several weaknesses, such as the ineffective interactions between P3P 
and APPEL. According to \cite{XPref}, in APPEL, users can only specify what is unacceptable in a policy, and 
simple preferences are hard to express and error prone. Idenfying this, the authors in \cite{XPref} proposed a 
better preference language called XPref giving more freedom for the users, such as allowing acceptable preferences. 

The main differences between these works and our work is that P3P, APPEL, and XPref are mainly designed for web 
applications/services, and the policies are defined in a XML-based language, with resticted options for the 
users. Our work is based on formal languages with more generic syntax and semantics, mathematical definitions and 
proofs, making it applicable for any type of services.  

Another XML-based policy language is the Customer Profile Exchange (CPEcxchange) \cite{CPexchange}, which was designed 
to facilitate business-to-business communication privacy policies (i.e., privacy-enabled global exchange of customer profile 
information). The eXtensible Access Control Markup Language (XACML) \cite{XACML} is a de-facto, XML-based
policy language, specifically designed for access control management in distributed systems. The latest version 3.0 was approved by the 
OASIS standards organization as an international standard on July 2017. Finally, the 
Enterprise Privacy Authorisation Language (EPAL) of IBM \cite{EPAL} was designed to regulate an organisation's 
internal privacy policies. EPAL is partly similar to XACML, however, it mainly focuses on privacy policies instead of 
access control policies in XACML.

A-PPL \cite{APPL} is an accountability policy language specifically designed for modelling data accountability (such as data 
retention, data location, logging and notification) in the cloud. A-PPL is an extension of the the PrimeLife Privacy Policy 
Language (PPL) \cite{PPL}, which enables specification of access and usage control rules for the data subjects and the data 
controller. PPL is built upon XACML, and allows us to define the so-called sticky policies on personal data based on obligations. 
Obligation defines whether the policy language can trigger tasks that must be performed, once some event occurs and the related 
condition is fulfilled. This is also referred to as the Event-Action-Condition paradigm. The Policy Description Language (PDL) \cite{PDL}, 
proposed by Bell Labs, is one of the first policy-based management languages, specifically for network administration. It is declarative and 
is based on the Event-Action-Condition paradigm similar to the PPL language. 

RBAC (Role-Based Access Control) \cite{RBAC} is onw of the most broadly-known role-based access control policy languages.
It uses roles and permissions in the enforced policies, namely, a subject can be assigned roles, and roles can be assigned certain access control permissions. 
ASL (Authorization Specification Language) \cite{ASL} is an another Role-based access control language based on first order logic, and RBAC components. 
Ponder \cite{Ponder} is a declarative and object-oriented policy language, and designed for defining and modelling security policies
using Role-Based Access Control, and security management policies for distributed systems. The policies are defined on roles or group 
of roles. Rei \cite{Rei} is a policy language based on deontic logic, designed mainly for modelling security and privacy properties of 
pervasive computing environments. It's syntax involves obligation and permission, where policies are defined as constraints over pemitted and obligated actions on resources.

To date, formal approaches that support modelling and analysing  
data protection requirements (e.g., GDPR) are less investigated. Some attempts on formalizing privacy and accountability policies and privacy architectures have been made, such as in \cite{TaButin15, ButinFM14} and \cite{TaAntignac14, Antignac14}. In \cite{ButinFM14}, a formal privacy policy language was proposed that defines, for each  type of personal data, authorised purposes, deletion delays, request 
completion delays, admissible contexts and data forwarding policies. Trace compliance properties were defined with respect to data handling events 
and elements of these privacy policies. The correctness properties relating to personal data handling events and system events was also formalized. In \cite{TaButin15}, we modified to the policy language in  \cite{ButinFM14}, and provided its first application on a real-world biometrics surveillance system as a case study.   

This paper is inspired by these previous works, however, the proposed  policy language is more generic, fine-grained and address the data protection requirements on the whole data life-cycle, namely, (i) data collection, (ii) data storage, (iii) data usage, (iv) data forwarding/transfer, and (v) data deletion/retention \cite{ButinTN15}. To the best of our knowledge this has not been proposed yet in the literature. We changed the syntax and semantics elements, along with the corresponding compliance rules and introduce new definitions.  As a result, the policy language proposed in this paper is capable of capturing the policy of more complex systems, compared to the previous approaches. For instance, in \cite{ButinFM14} there is no possibility to specify a sub-policy for data storage. 

Our proposed architecture language is a modified and extended version of the one in \cite{Antignac14}, where the authors addressed the problem of privacy-by-design at the architecture level and proposed a formal approach that facilitates architecture design. In particular, they provided the idea of the architecture language and logic, a dedicated variant of epistemic logics [12], to deal with different aspects of privacy. Basically, an architecture is defined as a set
of architecture relations, which capture the computation and communication 
abilities of each component. For instance, a relation compute$_i$($x$ $=$ $t$) specifies
that a component $C_i$ can compute a value $t$ for the variable $x$. The proposed language mainly focuses on the computation and integrity verification of data based on trust relations. Unlike \cite{Antignac14}, our proposed language focuses primary on data protection regulations, rather than the data integrity perspective. Our language variant proposed in this paper has a more expressive and different syntax and semantics, as well as novel data protection (DPR) compliance rules,  mapping and conformance checks between the architecture and policy levels, which are not given in \cite{Antignac14}.  


\section{Data Protection Regulation on the End-to-End Data Life-cycle}
\label{sec:e2e}

End-to-end personal data protection should cover the whole data life-cycle \cite{ButinTN15}.  Following this, our proposed privacy language enables a fine-grained policy specification on the entire data life-cycle, namely, data collection, storage, usage, deletion, and forwarding.


\begin{itemize}
\item \textit{Data collection}: An important aspect data controllers should follow during personal data collection is \textit{data minimisation}. The 3rd principle of the current data protection regulation says that data controllers should identify the minimum amount of personal data they need to fulfill their purposes \cite{ico1}. Personal data that are not required for the purpose of the services, should not be collected \cite{ButinTN15}. In addition, data controllers should define and declare (i) the purpose of the data collection during, e.g., the registration phase, (ii) what kind of data is collected from users, as well as (iii) ensure that consent is collected from the users whenever required (e.g., for certain type of personal data). Finally, (iv) records (logs) about the data collection procedure should be kept by data controllers.


\item \textit{Data usage}: Data controllers should specify and maintain an unambiguous data usage/processing policy \cite{ButinTN15, EUuse}. The policy should contain details about (i) who can use the given data and (ii) for what purpose. The purpose of usage should be aligned with the purpose of data collection. Further, the (iii) time period for data usage should be specified.  


\item \textit{Data storage}: A clear data storage policy should be maintain and declared. The policy should include \cite{ButinTN15, d95, ico2} (i) where a certain type of particular data is stored, (ii) how a certain type of data is stored, and (iii) how long a particular type of data is stored, and (iv) mechanisms are required for the periodic review of the need of storing the particular personal data.    
   

\item \textit{Data deletion/retention}: Under the Article 17 of the GDPR \cite{ico3} individuals have the right to have personal data erased. This is known as the 'right to be forgotten'. Data controllers should define and declare a precise data deletion policy which specifies (i) who can delete certain type of data, (ii) how the data is deleted (deletion mode, such as deleted from all storage or only partly deleted, as well as has a deletion request has been sent to any third party company to whom the data have been forward), (iii) what is the retention delay, and (iv) the worst-case (global) deletion delay.


\item \textit{Data forwarding}: Data forwarding (transfer) policy should be defined \cite{ButinTN15, d95}, which include (i) a list of third party companies or authorities to whom the data will be forwarded, (ii) the purpose of data  forwarding, and (iii) the logs/records of the data forwarding procedure.      

\item Transparency is an another fundamental requirement in the EU Personal Data Protection Regulation (e.g., Articles 10 and 11 of Directive 95/46/EC)\cite{EUtrans, d95}, which basically says that privacy notices and information about the data collection should be made available to the customers. 


\end{itemize}
\section{$\mathcal{P}_{DC}$ : A Policy Language for Data Protection}
\label{sec:polgen} 
We propose a high-level privacy policy language (called $\mathcal{P}_{DC}$) to specify and reason about the data protection requirements. The data subjects in our model are system users whose personal data has been collected by the data controller or input to the system by either the data subject or someone else (e.g., meter reading). The data controllers and data processors are service providers/organisations who collect, calculate/process and use the personal data about the data subjects.  For simplicity, in this paper, we merge and use interchange the data controller and data processor, while in real life they can be two different entities\footnote{Note that in the new GDPR the role and responsibility of the data processor will be more relevant than the data controller was in the old DPR}.  

The proposed policy language captures the sub-policies defined on the entire data life-cycle. Policy designers can use our language to define their data collection, usage, storage, deletion/retention, and forwarding sub-policies.

\subsection{Policy Syntax}
\label{sec:syntaxpp}

The policy  specification concept is based on our previous work \cite{TaButin15}, but with the modified and extended (more expressive) syntax and semantics elements, crafted for handling personal data protection properties. 
Our language can be adapted by any type of data controller, however, for simplicity, in the following, let us refer to a data controller as service provider. A service provider $SP$ has a finite set of $n$ different services, \textit{PServices}$^{SP}_{\mathcal{P}\mathcal{L}}$ = $\{$\textit{serv}$_1$, \dots, \textit{serv}$_n$$\}$, the corresponding finite set of all data types required by the $n$ services, \textit{TYPE}$^{SP}_{\mathcal{P}\mathcal{L}}$ = $\{$$\theta_1$, \dots, $\theta_m$$\}$, as well as the finite set of different entities, \textit{Entity}$^{SP}_{\mathcal{P}\mathcal{L}}$ = $\{$\textit{sp}, \dots, \textit{ow}, \textit{or}, \textit{tar}, \dots, $i$, $j$$\}$ defined in the system of $SP$. 

Let $\mathcal{C}$$\mathcal{D}$ be a set of all the possible finite sets of collected data, and $cd$ $\in$ $\mathcal{C}$$\mathcal{D}$. A data $dt$ $\in$ $cd$ is defined as the tuple ($ow$, $ds$, $\theta$), where $ow$ $\in$ \textit{Entity}$^{SP}_{\mathcal{P}\mathcal{L}}$ is the owner of the data who inputs this data into the system\footnote{In our model, for each data there is only one $ow$ who inputs this data into the system (e.g.,  uploads a photo or submit a name), the user who obtains the data during the system run is not considered as $ow$.}, $ds$ $\in$ \textit{Entity}$^{SP}_{\mathcal{P}\mathcal{L}}$ specifies a data subject included in this data, $\theta$ $in$ \textit{TYPE}$^{SP}_{\mathcal{P}\mathcal{L}}$ is the type. We note that in our model the tuple ($ow$, $ds$, $\theta$) remains unchanged during the system run.  

In the following, we define a set of \textit{data protection policies} on the entire data life-cycle (data collection, usage, storage, deletion, and forwarding).    

\begin{ttd}
(Data Protection Policy). The syntax of the data protection policies are defined as the following tuples:   

\begin{figure}[htbp]
\centering
\fbox{\begin{minipage}{11.87 cm}
\begin{tabbing}    
    \=1\=1\=1\=1\= \kill
    \> 1. \textit{POL} $=$ Pol$_{Col}$ $\times$ Pol$_{Use}$ $\times$ Pol$_{Str}$ $\times$ Pol$_{Del}$ $\times$ Pol$_{Fw}$.\\\\ 
 \=1\=1\=1\=1\= \kill
    \> 2. Pol$_{Col}$ $=$ Cons$_{col}$ $\times$ CPurp.\\\\ 
 \=1\=1\=1\=1\= \kill
    \> 3. Pol$_{Use}$ $=$ Cons$_{use}$ $\times$ UPurp $\times$ WhoUse.\\\\ 
 \=1\=1\=1\=1\= \kill
    \> 4. Pol$_{Str}$ $=$  W$_{st}$ $\times$ H$_{st}$ $\times$ RDate$_{st}$.\\\\ 
 \=1\=1\=1\=1\= \kill
    \> 5. Pol$_{Del}$ $=$  Who$_{del}$ $\times$ How$_{del}$ $\times$ Ret$_{del}$ $\times$ Global$_{del}$.\\\\ 
 \=1\=1\=1\=1\= \kill
    \> 6. Pol$_{Fw}$ $=$   Purp$_{fw}$ $\times$ List$_{3rd}$.
\end{tabbing}
\end{minipage}
}
\label{fig:dcpol}
\end{figure}
\end{ttd}

We define the set of high-level data protection policies (\textit{POL}), which are composed of the five sub-policies based on the end-to-end data life-cycle. Namely, the sets of data collection (\textit{Pol}$_{Col}$), data usage  (\textit{Pol}$_{Use}$), data storage (\textit{Pol}$_{Str}$), data deletion (\textit{Pol}$_{Del}$), data forwarding policies (\textit{Pol}$_{Fw}$). 

Data collection policy includes consent required (Cons$_{col}$) for certain data types (e.g., personal data) and collection purposes (CPurp). Data usage policy specifies the  consent (Cons$_{use}$) of data usage for some data types (e.g., personal data), the purpose of data usage (UPurp), and the set of entities who are allowed to use the data (WhoUse). Data storage policy specifies where the data is stored (W$_{st}$), how the data is stored (H$_{st}$), and the date for reviewing the necessity of storing the data further (RDate$_{st}$). The data deletion policy specifies who can delete the data (Who$_{del}$), how the data is deleted (How$_{del}$), the retention period (Ret$_{del}$), and the global (worst-case) deletion delay (Global$_{del}$). Finally, the data forwarding policy involves the purpose of the data forwarding (Purp$_{fw}$), and all the third party entities to which the data will be forwarded (List$_{3rd}$).   

Each data protection policy is defined on a data type ($\theta$), specifically, let $\pi_{\theta}$, $\pi_{\theta}$ $\in$ \textit{POL}, be a policy defined on a data type $\theta$, and the five sub-policies $\pi_{col}$ $\in$ \textit{Pol}$_{Col}$, $\pi_{use}$ $\in$ \textit{Pol}$_{Use}$, $\pi_{str}$ $\in$ \textit{Pol}$_{Str}$, $\pi_{del}$ $\in$ \textit{Pol}$_{Del}$, $\pi_{fw}$ $\in$ \textit{Pol}$_{Fw}$, where  

\begin{center}
$\pi_{\theta}$ $=$ ($\pi_{col}$, $\pi_{use}$, $\pi_{str}$, $\pi_{del}$, $\pi_{fw}$),   
\end{center} 
 
and each sub-policy of $\theta$ is defined as follows: 
 
\begin{enumerate}
\item $\pi_{col}$ = (\textit{cons}, \textit{cpurp}), with \textit{cons} $\in$ \{$Y$, $N$\}. This specifies that if a consent is required to be collected from users (Y for Yes) or not (N for No) for this type of data, and \textit{cpurp} = (\textit{purpset}$_c$, \textit{decl}), where \textit{purpset}$_c$ is a set of collection purposes, as well as \textit{decl} $\in$ \{$Y$, $N$\}. \textit{decl} specifies if the collection purposes is declared for the users (visibility), which again, can take the value of Y (Yes) or N (No). Finally, purposes can be strings that uniquely define the purpose.  

In the sequel, we discuss some further requirements for data collection.     
 
\begin{itemize}
\item \textit{Data Minimisation}: Given a service provider, the data collection purposes are usually determined by the services available tocustomers after the registration phase.   
Data minimisation depends strongly on the purposes of the data collection. For a given service provider, let the set of purposes of data collection be \textit{cpurp} $\in$ CPurp, and \textit{pservices}$_e$ $\in$ \textit{Services}$^{SP}_{\mathcal{P}\mathcal{L}}$ be the set of all functions/procedures required by the services chosen by an entity $e$. Ideally, in strict term, we have the constraint \textit{cpurp} $\subseteq$ \textit{pservices}$_e$, which basically says that the data collection purposes are to enable the entity $e$ to use its chosen services.       

\item \textit{Data types to be collected}: Let \textit{typecol}$_e$ be a set of collected data types by a given service provider from user $e$. Also, let \textit{typeserv}$_{U}$ be the set of data types required by the services chosen by $U$. In an ideal case, the service provider should collect only the data of types required by the services chosen by $U$, namely, we have the constraint \textit{typecol}$_U$ $\subseteq$ \textit{typeserv}$_{U}$.

\item \textit{Consent}:  For a given service provider and user $U$ let \textit{constype}$_U$ be a set of data types that requires consent for collection (e.g., sensitive data in case of health-care services), we have \textit{constype}$_U$ $\subseteq$ \textit{typecol}$_U$. 

\item \textit{Transparency}: Information about the data collection and usage purposes, as well as information about the data storage, deletion and forwarding should be available tocustomers/users. Transparency is captured by the variable \textit{decl} (declaration) in the policy language syntax. 

\end{itemize}  

\item $\pi_{use}$ = (\textit{cons}, \textit{upurp}, \textit{whouse}), with  policy for consent collection, \textit{cons} $\in$ \{$Y$, $N$\}, and policy for usage purposes,  \textit{upurp} = (\textit{purpset}$_u$, \textit{decl}), where \textit{purpset}$_u$ is a set of usage purposes, and \textit{decl} $\in$ \{$Y$, $N$\} is similar to above. Finally, the policy on who can use the data,  \textit{whouse} = (\textit{who}, \textit{decl}), where \textit{who} is a set of entities who are allowed to use the data.  

\item $\pi_{str}$ = (\textit{wh},  \textit{ho},  \textit{revdate}), with 
\begin{itemize}
\item \textit{wh} is a set of places where the data $dt$ is stored, for instance, in a client computer (\textit{clientloc}) or in the service provider's (\textit{sploc}) 
servers, namely, \textit{wh} $\in$ \{(\textit{clientloc}, \textit{places}, \textit{decl}), (\textit{sploc}, \textit{places}, \textit{decl})\}, where \textit{places} is a set of places where the data will be stored.  Again, \textit{decl} specifies the declaration of this information for users.     

\item \textit{ho} is a set of methods in which the data $dt$ is stored, 
for example, in hidden (\textit{hidden}) or non-hidden (\textit{nohidden}) forms. Hidden form means that the data should not be read by the service provider, while non-hidden form means that it can be available in plaintext for the service provider.  For example, in the first case, the data can be encrypted using the key not available to the service provider, while in the latter case, the data can be stored encrypted with the service provider's key, or in plaintext. Namely, we have \textit{ho} $\in$ \{(\textit{hidden}, \textit{decl}), (\textit{nohidden}, \textit{decl})\}.

\item \textit{revdate} is the date for periodic review of the need of storage of the data $dt$ in certain \textit{places}, namely, \textit{revdate} = (\textit{date}, \textit{places}, \textit{decl}), where \textit{date} is a time interval within which the review has to be done.  

\item  \textit{decl} $\in$ \{$Y$, $N$\}.
\end{itemize}

\item $\pi_{del}$ = (\textit{how}, \textit{deld},   \textit{gdeld}), with 
\begin{itemize}
\item \textit{how} defines the deletion approaches for this type of data. and \textit{how} $=$ (\{$p1$, $p2$\}, \textit{decl}), where $p1$ $\in$ \{\textit{aut}, \textit{man}\},  $p2$ $\in$ \{\textit{full}, (\textit{partly}, \textit{purposes})\}. Parameter $p1$ can be \textit{aut} which refers to the automated deletion mode, initiated automatically by the service provider, for instance, the deletion delay after some certain pre-defined events. \textit{man} refers to a manual deletion initiated by a user.  Parameter $p2$ captures whether the data is fully deleted, namely, deleted from all the main or backup servers.   

\item \textit{deld} = ($dd$, \textit{decl}) represents the delay for the deletion after the initiation (e.g., when a user 
has clicked on a delete button, or a user has unregistered from the service). The value of this delay can be \textit{ND}, refers to ``\textit{Not Defined}'', or a certain time period (e.g., 1 day, 1 min, 1 sec, etc.), or \textit{DF} which refers to ``\textit{Defined}-without a specific numerical time frame'' (e.g., \textit{DF} can be ``until required for the services").        

\item \textit{gdeld} = ($gd$, \textit{decl}), where $gd$ is a global (worst-case) deletion delay within which the data should be deleted after the customer/client unregister from the service. $gd$ can be a numerical value (e.g., 5 years, 1 hour, etc), or \textit{ND} and \textit{DF} for ``Not Defined" and ``Defined without a specific numerical time frame", respectively.       

\item  \textit{decl} $\in$ \{$Y$, $N$\}. 
\end{itemize}

\item $\pi_{fw}$ = (\textit{cons}, \textit{fwpurp}, \textit{3rdparty}), where \textit{cons} captures if consent is required or not, \textit{fwpurp} is a set of purposes of data forwarding,  \textit{3rdparty} is a set of third party authorities and companies to whom the data has been forwarded.   

\end{enumerate}

Finally, let us assume a finite set \{$\theta_1$, \dots, $\theta_m$\} of all data types supported by the service of a given provider $SP$. 

\begin{figure}[htbp]
\centering
\fbox{\begin{minipage}{10.7 cm}
Then, the default data protection (DPR) policy for $SP$ is defined by the set
\begin{center}
$\mathcal{P}\mathcal{L}$ $=$ $\{$$\pi_{\theta_1}$, \dots, $\pi_{\theta_m}$$\}$.
\end{center}
\end{minipage}
}
\label{fig:polsp}
\end{figure}

\subsection{Policy Semantics}
\label{sec:semantic0} 

\subsubsection{Abstract Events}
\label{sec:aevents} 

To reason about the system operation and its compliance with a given DPR policy we introduce abstract events \cite{TaButin15}. Each abstract event captures an  activity that happens during the system run. These events are abstract, because they specify high-level actions happening during the system  operation, ignoring the low-level system internals such as writing to a memory space (e.g., at the system log level) or the protocol level (e.g., a concrete implementation of the system).
   
Events are defined by the tuples starting with an event name capturing the activity carried out by different entities, followed by the time of the event, and further parameters required by that activity.     

Our language includes the pre-defined activities \textit{own}, \textit{register}, \textit{store}, \textit{storerev}, \textit{collect},  \textit{cconsent}, \textit{uconsent}, \textit{fwconsent}, \textit{declare}, \textit{use}, \textit{deletereq}, \textit{mandelete}, \textit{autdelete}, \textit{forward}, and \textit{register} which are related to the data protection regulation. For each data $dt$ defined by the tuple of ($ow$, $ds$, $\theta$) and $\pi_{\theta}$, we have the following events:

\begin{list}
{\bfseries{}\textit{Ev}\textit{\arabic{qcounter}}:~}
{
\usecounter{qcounter}
}
\item (\textit{own}, $t$, \textit{or}, $\theta$, $v$) 
\item (\textit{register}, $t$, \textit{or}, \textit{pservices}, \textit{typeserv}) 
\item (\textit{store}, $t$, \textit{or}, $\theta$, $v$, \textit{places})
\item (\textit{storerev}, $t$, $sp$, $\theta$, $v$, \textit{places})
\item (\textit{collect}, $t$, \textit{sp}, \textit{or}, $\theta$,  $v$,   \textit{purposes})
\item (\textit{cconsent}, $t$, $sp$, \textit{tar}, $\theta$, $v$,  \textit{purposes}, $\pi_{\theta}$)
\item (\textit{uconsent}, $t$, $sp$, \textit{tar}, $\theta$, $v$, \textit{purposes}, $\pi_{\theta}$)
\item (\textit{fwconsent}, $t$, $sp$, \textit{tar}, $\theta$, $v$, \textit{purposes}, $\pi_{\theta}$)
\item (\textit{declare}, $t$, $sp$, \textit{tar}, $\theta$, $v$, \textit{params}, $\pi_{\theta}$) 
\item (\textit{use}, $t$, \textit{or}, $\theta$, $v$, \textit{purposes}, \textit{bywhom}) 
\item (\textit{deletereq}, $t$, \textit{or}, $\theta$, $v$, $\pi_{\theta}$) 
\item (\textit{mandelete}, $t$, \textit{or}, $\theta$, $v$, \textit{places}, $\pi_{\theta}$) 
\item (\textit{autdelete}, $t$, \textit{sp}, $\theta$, $v$, \textit{places}, $\pi_{\theta}$) 
\item (\textit{forward}, $t$, $sp$, \textit{or}, $\theta$, $v$, \textit{purposes}, \textit{towhom}, $\pi_{\theta}$)
\item (\textit{unregister}, $t$, \textit{or}, \textit{pservices}, \textit{typeserv})
\end{list}

In the sequel, we refer to each element $e$ of a tuple \textit{tup} as \textit{tup}.$e$, for example, we can refer to $\pi_{str}$ in $\pi_{\theta}$ as $\pi_{\theta}$.$\pi_{str}$. Event \textit{own} captures the fact that the data of type $\theta$ and value $v$ is initially owned (possessed) by $or$. Event \textit{store} specifies that the data of type $\theta$ and value $v$ is 
stored at the places \textit{places}. Event \textit{storerev} captures when a periodic review has been done at time $t$ about whether data $dt$ should be stored further at the places \textit{places}.   

Event \textit{collect} specifies the event when the data of type $\theta$ and value $v$ has been collected by the service provider $sp$ from entity \textit{or} at time $t$,  for the purposes \textit{purposes}. Event \textit{cconsent} specifies when consent required for the data collection purposes \textit{purposes} for data the data of type $\theta$ and value $v$ has been collected from the entity \textit{tar} at time $t$ (i.e., the customer/user gived consent to the service provider) according to the policy $\pi_{\theta}$. Event \textit{uconsent} captures when consent has been collected by the service provider at time $t$ for the data usage purposes \textit{purposes}. Event \textit{fwconsent} is similar to the previous two events but related to data forwarding instead.  

Event \textit{declare} captures the event when certain set of parameters \textit{params} of policy a $\pi_{\theta}$,  have been declared at time $t$ according to the policy $\pi_{\theta}$. It specifies the declaration of  every parameter $p$, $p$ $\in$ \textit{params}, e.g., when $p$ = \textit{fwpurp} in $\pi_{fw}$ (i.e., $p$ = $\pi_{\theta}$.$\pi_{fw}$.\textit{fwpurp}).  We note that consent collection and declaration of the two sets \textit{purposes} and \textit{params}, normally happen only once during the data collection phase. The event \textit{use} captures that the data of type $\theta$ and value $v$ collected from   \textit{or} has been used  by entities in the set \textit{bywhom} at the time $t$ for certain purposes \textit{purposes}.    

In \textit{deletereq} the customer/client \textit{or} (\textit{or} $\neq$ \textit{sp}) initiates the request towards the service provider (e.g., when a user clicks on the deletion button or the unregister/deactivate account button) from user \textit{or} at time $t$ to delete the data of type $\theta$ and value $v$.  Note that in these cases,  \textit{deletereq} only takes place when manual deletion mode is active (i.e., $\pi_{\theta}$.$\pi_{del}$.\textit{how}.\textit{p1} $=$ \textit{man}) and when the data is stored at the service provider's location, or has been forwarded to the third parties.

The event 
\textit{mandelete} captures the fact that the data of type $\theta$ and value $v$ is deleted by \textit{or} at time $t$. This can happen either by a client (e.g., when the data is stored locally at the client side), or it can be done by the service provider after the delete request has been sent by the user (i.e., this time \textit{or} can be either client or service provider).  Event \textit{autdelete} refers to the automated deletion by the service provider (\textit{sp}). Event 
\textit{forward} captures that at time $t$ the service provider forwards the data of type $\theta$ and value $v$ from \textit{or} to the third parties specified in the set \textit{towhom}.

\begin{ttd} $($Trace$)$. 
A trace $\tau$ is a sequence of abstract events.  
\end{ttd}

Traces capture the current operation of a system, which can be either compliant or non-compliant with the data protection requirements.     
In order to define the notion of compliant traces, we need to introduce first the notion of abstract states for capturing the states of data.        

\begin{ttd} $($Abstract state$)$.
The abstract state of a system is a function 

\noindent $\sum$ $:$ $($\textit{User} $\times$ \textit{User} $\times$ \textit{Type}$)$ 
$\rightarrow$ $($\textit{Time} $\times$ \textit{Value} $\times$ \textit{Places} $\times$ \textit{Policy} $\times$ \textit{Has}$)$. 
\end{ttd}

The abstract state associated with the data $dt$ whose owner is \textit{ow}, of type $\theta$, and containning information about the data subject \textit{ds} (i.e., $dt$ = ($ow$, $ds$, $\theta$)), is composed of a timestamp, the current value of the data, the places where the data is currently stored (\textit{Places}), the current policy attached to this data, and a set of users who have this data at this time. Specifically:  

\begin{center} 
($ow$, $ds$, $\theta$) $\rightarrow$ ($t$, $v$, \textit{plc}, $\pi_{\theta}$, $\mathcal{H}_{\textit{has}}$), 
\end{center} 

where $\pi_{\theta}$ is an instance of the corresponding default policy $\pi_{\theta}$. The notation $\sum$[($ow$, $ds$, $\theta$) $\rightarrow$ ($t$, $v$, \textit{plc}, $\pi_{\theta}$, $\mathcal{H}_{\textit{has}}$)] captures a state update of the data of ($ow$, $ds$, $\theta$) to the state on the rightside, and is 
used to denote a state $\sum'$ similar to $\sum$ except that its value is  
($t$, $v$, \textit{plc}, $\pi_{\theta}$, $\mathcal{H}_{\textit{has}}$).  

The semantics of an event at a given 
position $j$ in a trace is specified by the function $S_A$$:$ (\textit{Event} $\times$ $N$) $\rightarrow$ 
\textit{AbstractState} $\rightarrow$ \textit{AbstractState}. Specifically, we have:

\begin{tabbing}  
    \=1\=1\=1\=1\= \kill
    1. $S_A$(\textit{own}, $t$, \textit{or}, $\theta$), $j$)$\sum$\\
    \=123\=1\=1\=1\= \kill
		\>\> $=$ $\sum$[($ow$, $ds$, $\theta$) $\rightarrow$ ($t$, $v$, \textit{plc}, $\pi_{\theta}$, $\mathcal{H}_{\textit{has}}$ $\cup$ \{\textit{or}\})]\\\\ 
 2. $S_A$((\textit{register}, $t$, \textit{or}, \textit{pservices}, \textit{typeserv}), $j$)$\sum$ $=$ $\sum$\\\\
\=1\=1\=1\=1\= \kill
   3. $S_A$((\textit{store}, $t$, $dt$, \textit{places}, $v$), $j$)$\sum$\\ 
		\=123\=1\=1\=1\= \kill
		\>\> $=$ $\sum$[($ow$, $ds$, $\theta$) $\rightarrow$ ($t$, $v$,  \textit{plc} $\cup$ \textit{places}, $\pi_{\theta}$, $\mathcal{H}_{\textit{has}}$ $\cup$ \{\textit{sp}\})]\\
		\=123\=1\=1\=1\= \kill
		\>\>\>\> if $\pi_{\theta}$.$\pi_{str}$.\textit{ho} $=$ (\textit{nohidden}, \textit{decl}) $\wedge$ $\pi_{\theta}$.$\pi_{str}$.\textit{wh} $=$ (\textit{sploc}, \textit{places}, \textit{decl})\\
		\=123\=1\=1\=1\= \kill	 
		\>\> $=$ $\sum$[($ow$, $ds$, $\theta$) $\rightarrow$ ($t$, $v$,  \textit{plc} $\cup$ \textit{places}, $\pi_{\theta}$, $\mathcal{H}_{\textit{has}}$)]\\
		\=123\=1\=1\=1\= \kill
		\>\>\>\> if $\pi_{\theta}$.$\pi_{str}$.\textit{ho} $=$ (\textit{hidden}, \textit{decl}) $\wedge$ $\pi_{\theta}$.$\pi_{str}$.\textit{wh} $=$ (-, \textit{places}, \textit{decl})\\
		\=123\=1\=1\=1\= \kill	 
		\>\> $=$ $\sum$, else.\\\\
\=1\=1\=1\=1\= \kill
	4.	$S_A$((\textit{storerev}, $t$, $dt$, $v$, \textit{places}), $j$)$\sum$ $=$ $\sum$.\\\\			 
\=1\=1\=1\=1\= \kill
	5.$S_A$((\textit{collect}, $t$, $sp$, \textit{or}, $dt$, $v$,  \textit{purposes}), $j$)$\sum$\\ 
		\=123\=1\=1\=1\= \kill
		\>\> $=$ $\sum$[($ow$, $ds$, $\theta$) $\rightarrow$ ($t$, $v$,  \textit{plc}, $\pi_{\theta}$, $\mathcal{H}_{\textit{has}}$ $\cup$ \{sp\})].\\\\		
\=1\=1\=1\=1\= \kill
	6.	$S_A$((\textit{cconsent}, $t$,  $sp$, \textit{tar}, $\theta$, $v$, \textit{purposes}, $\pi_{\theta}$), $j$)$\sum$ $=$ $\sum$.\\\\
\=1\=1\=1\=1\= \kill
	7.	$S_A$((\textit{uconsent}, $t$, $sp$, \textit{tar}, $\theta$, $v$, \textit{purposes}, $\pi_{\theta}$), $j$)$\sum$ $=$ $\sum$.\\\\
8.	$S_A$((\textit{fwconsent}, $t$, $sp$, \textit{tar}, $\theta$, $v$, \textit{purposes}, $\pi_{\theta}$), $j$)$\sum$ $=$ $\sum$.\\\\
\=1\=1\=1\=1\= \kill
	9.	$S_A$((\textit{declare}, $t$, $sp$, \textit{tar},  $v$, \textit{params}, $\pi_{\theta}$), $j$)$\sum$ $=$ $\sum$.\\\\
\=1\=1\=1\=1\= \kill
	10.	$S_A$((\textit{use}, $t$, \textit{or}, $dt$, $v$, \textit{purposes}, \textit{bywhom}, $j$)$\sum$ $=$ $\sum$.\\\\	
\=1\=1\=1\=1\= \kill
	11.	$S_A$((\textit{deletereq}, $t$, \textit{or}, $dt$, $v$, $\pi_{\theta}$), $j$)$\sum$ $=$ $\sum$, if  $\pi_{\theta}$.$\pi_{del}$.\textit{how}.$p1$ =  \textit{man}, and\\
	\=12345678901234567890123456789012345678\=1\=1\= \kill
		\>\> 
	 $\pi_{\theta}$.$\pi_{str}$.\textit{wh} =  (\textit{sploc}, \textit{places}, \textit{decl}).\\\\
				\=1\=1\=1\=1\= \kill
	12.	$S_A$((\textit{mandelete}, $t$, \textit{or}, $dt$, $v$,  \textit{places}, $\pi_{\theta}$) , $j$)$\sum$  
		\\ 
		\=123\=1\=1\=1\= \kill
		\>\> $=$ $\sum$[($ow$, $ds$, $\theta$) $\rightarrow$ $\bot$], if  $\pi_{\theta}$.$\pi_{del}$.\textit{how}.$p2$ =  \textit{full}.  \\\\	
\=1\=1\=1\=1\= \kill
	13.	$S_A$((\textit{autdelete}, $t$, \textit{sp}, $dt$, $v$,  \textit{places}, $\pi_{\theta}$) , $j$)$\sum$ \\ 
		\=123\=1\=1\=1\= \kill
		\>\>$=$ 
		$\sum$[($ow$, $ds$, $\theta$) $\rightarrow$ $\bot$], if  $\pi_{\theta}$.$\pi_{del}$.\textit{how}.$p2$ =  \textit{full}. \\\\
				\=1\=1\=1\=1\= \kill
	14.	$S_A$((\textit{mandelete}, $t$, \textit{or}, $dt$, $v$,  \textit{places}, $\pi_{\theta}$) , $j$)$\sum$  
		\\ 
		\=123\=1\=1\=1\= \kill
		\>\> $=$ $\sum$[($ow$, $ds$, $\theta$) $\rightarrow$ ($t$, $v$, \textit{plc} $\backslash$ \textit{places}, $\pi_{\theta}$, \textit{Owner}(\textit{plc} $\backslash$ \textit{places}))], \\ 
		\=123\=1\=1\=1\= \kill
		\>\> if  $\pi_{\theta}$.$\pi_{del}$.\textit{how}.$p2$ =  \textit{partly} and \textit{or} = \textit{sp}.  \\\\	
\=1\=1\=1\=1\= \kill
	15.	$S_A$((\textit{autdelete}, $t$, \textit{sp},  $dt$, $v$,  \textit{places}, $\pi_{\theta}$) , $j$)$\sum$ \\ 
		\=123\=1\=1\=1\= \kill
		\>\> $=$ $\sum$[($ow$, $ds$, $\theta$) $\rightarrow$ ($t$, $v$, \textit{plc} $\backslash$ \textit{places}, $\pi_{\theta}$,  \textit{Owner}(\textit{plc} $\backslash$ \textit{places}))], \\ 
		\=123\=1\=1\=1\= \kill
		\>\> if  $\pi_{\theta}$.$\pi_{del}$.\textit{how}.$p2$ =  \textit{partly}.  \\\\ 
\=1\=1\=1\=1\= \kill
	16. $S_A$((\textit{forward}, $t$, $sp$, \textit{or}, $dt$, $v$,   \textit{purposes}, \textit{towhom}, $\pi_{\theta}$), $j$)$\sum$\\ 
		\=123\=1\=1\=1\= \kill
		\>\> $=$ $\sum$[($ow$, $ds$, $\theta$) $\rightarrow$ ($t$, $v$,  \textit{plc}, $\pi_{\theta}$, $\mathcal{H}_{\textit{has}}$ $\cup$ \textit{towhom})].\\\\ 
		17. $S_A$((\textit{unregister}, $t$, \textit{or}, \textit{pservices}, \textit{typeserv}), $j$)$\sum$ $=$ $\sum$.
\end{tabbing}

The events \textit{storerev}, \textit{cconsent}, \textit{uconsent}, \textit{declare} and \textit{use} do not change the current state of a data. 
As a result of the event \textit{own} the owner of the data \textit{dt} is added to the set $\mathcal{H}_{has}$. The event \textit{collect} just adds the service provider \textit{sp} to $\mathcal{H}_{has}$. The event \textit{store} adds the service provider \textit{sp} to $\mathcal{H}_{has}$ and also adds  the set of storing places, \textit{places}, to \textit{plc}, either in case $dt$ is stored unhidden from the service provider (e.g., as plaintext at the service provider location or encrypted with the service provider's key). Otherwise, if the data is stored hidden, then the service provider does not have it. Otherwise, the state is unchanged. Event \textit{deletereq} does not change the current state, however, it only takes place when the deletion policy allows manual deletion initiated by users. Events \textit{mandelete} and  \textit{autdelete} capture the deletion event itself in case the deletion has been manually triggered by users (as a result of \textit{deletereq}) and automatically set by the service provider (without \textit{deletereq}), respectively.  For \textit{mandelete} and  \textit{autdelete}, we further distinguish the case of full and partly deletion modes (defined by $\pi_{\theta}$.$\pi_{del}$.\textit{how}.$p2$). In the first case, the current state of $dt$ is  replaced with the undefined state $\bot$, representing full deletion from the system. Note that here, we abstract away from some activities such as (1) the deletion request sent by the service provider to the third party (in case they have the data), and (2)  fully deletion of the data from the third party servers. In case of partly deletion, the new state of the $dt$ will be the state in which \textit{places} is removed from the places in which  $dt$ is currently stored. These places can be e.g., main servers, backup servers of the service provider. As a result of the deletion partly event (either \textit{autdelete} or \textit{mandelete}), only the entities who own the places \textit{plc} $\backslash$ \textit{places}, \textit{Owner}(\textit{plc} $\backslash$ \textit{places}), will have the data.

\subsection{Trace Compliance}
\label{sec:compliance0}

Compliance for traces of events is defined to capture the system operation that respects data control and accountability requirements.  
We define the most relevant trace compliance rules $C_1$ - $C_{10}$, which are related directly to the data protection properties. 
The high-level interpretation of each rule is as follows: 

\begin{itemize}
\item $C_1$: A perioric review on data storage (event \textit{storerev}) can only take place in the specified time interval according to the policy $\pi_{\theta}$.$\pi_{str}$.\textit{revdate}.   
\item $C_2$: For a $dt$ = ($ow$, $ds$, $\theta$), if  $\pi_{\theta}.\pi_{\ast}$.\textit{param}.\textit{decl} = \textit{Y} (where $\pi_{\ast}$ can be any from the five sub-policies), then the corresponding value \textit{declval} must be declared before the collection of $dt$. For example, in the collection policy, $\pi_{col}$.\textit{cpurp} = (\textit{purpset}$_c$, \textit{decl}) defined above, if  \textit{decl} = \textit{Y}, then the corresponding \textit{purpset}$_c$ should be declared.   
\item $C_3$: For a $dt$, if \textit{cons} = \textit{Y} in $\pi_{\theta}$.$\pi_{col}$, then consent must be collected before the collection of $dt$ itself.   
\item $C_4$: For a $dt$, if \textit{cons} = \textit{Y} in $\pi_{\theta}$.$\pi_{use}$, then the consent must be collected before the usage of $dt$, for the specified purposes. 
\item $C_5$: A data $dt$ = ($ow$, $ds$, $\theta$) is only used for purposes defined in the corresponding policy. 
\item $C_{6}$: The manual deletion of a data $dt$ $=$ ($ow$, $ds$, $\theta$) must happen within the specified deletion delay, $dd$, and places \textit{places},   
after the deletion request (event \textit{deletereq}) has been initiated.
\item $C_{7}$: The automated deletion of a data $dt$ $=$ ($ow$, $ds$, $\theta$) must happen  within the time interval specified by the global deletion policy since data collection. 
\item $C_{8}$: Whenever the data $dt$ is forwarded by the service provider to a set of third parties, the forwarding purposes and the addressee should conform with the defined policy of $dt$.  
\item $C_{9}$: Whenever the data $dt$ is forwarded by the service provider to a set of third parties, appropriate consent should be collected before.    
\item $C_{10}$: Whenever the data $dt$ is collected, the set of collection purposes is a subset of the user chosen services, and the type of the collected data (i.e., $\theta$ in $dt$) is an element of the set of data types needed for the user chosen services.    
\end{itemize}
  
Following \cite{TaButin15}, the current state after the execution of a trace $\tau$ $=$ [$ev_1$,\dots, $ev_n$] is 
defined as $F_A$($\tau$, $1$)$\sum_0$ with $\forall$ ($ow$, $ds$  $\theta$):
$\sum_0$($ow$, $ds$  $\theta$) $=$ $\bot$, and 

\begin{itemize} 
\item $F_A$($[\ ]$, $n$)$\sum$ $=$ $\sum$.
\item $F_A$($[ev_1, \dots, ev_m]$, $n$)$\sum$ $=$ $F_A$($[ev_2, \dots, ev_m]$, $n+1$)($\mathcal{S}_A$($ev_1$, $n$)$\sum$)  
\end{itemize} 

where, \textit{State}$_A$($\tau$, $i$) $=$ $F_A$($\tau_{|i}$, $1$)$\sum_0$ with 
$\tau_{|i}$ $=$ $\tau_1$ \dots $\tau_i$ representing the prefix of length $i$ of $\tau$.


\begin{figure}[htbp]

\fbox{\begin{minipage}{11.87 cm}

\begin{itemize}
\item $C_1: \tau_i = (\textit{storerev}, t', dt, v, \textit{places})$ $ 
\wedge \ \textit{State}_A(\tau, i-1)(ow, ds, \theta) = (t, v, \textit{plc}, \pi_{\theta}, \mathcal{H}_{\textit{has}}) \Longrightarrow
t' \in \pi_{\theta}.\pi_{str}.\textit{revdate}$\\

\item $C_2: \tau_i = (\textit{collect}, t',  sp, \textit{or}, dt, v,   \textit{purposes})$ $ 
\wedge \ \textit{State}_A(\tau, i-1)(ow, ds, \theta) = (t, v, \textit{plc}, \pi_{\theta}, \mathcal{H}_{\textit{has}})\ \wedge \ (\pi_{\theta}.\pi_{\ast}.\textit{param}.\textit{decl} = \textit{Y})\ \Longrightarrow
\exists \ k \ | \ \exists \ t'' \ | \ \tau_k =  (\textit{declare}, t'', sp,  \textit{tar}, dt, v,  \pi_{\theta}.\pi_{\ast}.\textit{param}.\textit{declval},\pi_{\theta})$ $\wedge$ $(t'' < t')$  $\wedge$ (\textit{tar} $=$ \textit{or}), where $\pi_{\ast}$ $\in$ \{$\pi_{col}$, $\pi_{use}$, $\pi_{str}$, $\pi_{del}$, $\pi_{fw}$\}\\  


\item $C_3: \tau_i = (\textit{collect}, t', sp, \textit{or}, dt, v,  \textit{purposes})$ $ 
\wedge \ \textit{State}_A(\tau, i-1)(ow, ds, \theta) = (t, v, \textit{plc}, \pi_{\theta}, \mathcal{H}_{\textit{has}})\ \wedge \ (\pi_{\theta}.\pi_{col}.\textit{cons} = \textit{Y}) \ \Longrightarrow
\exists \ k \ | \ \exists \ t'' \ | \ \tau_k =  (\textit{cconsent}, t'', sp,  \textit{tar}, dt, v,  \textit{purposes}, \pi_{\theta})$ $\wedge$ $(t'' < t')$ $\wedge$ (\textit{tar} $=$ \textit{or}) $\wedge$ (\textit{purposes} $\subseteq$ $\pi_{\theta}$.$\pi_{col}$.\textit{cpurp}.\textit{purpset}$_c$)\\

\item $C_4: \tau_i = (\textit{use}, t', \textit{or}, dt, v,   \textit{purposes}, \textit{bywhom})$ $ 
\wedge \ \textit{State}_A(\tau, i-1)(ow, ds, \theta) = (t, v, \textit{plc}, \pi_{\theta}, \mathcal{H}_{\textit{has}})\ \wedge \ (\pi_{\theta}.\pi_{use}.\textit{cons} = \textit{Y}) \ \Longrightarrow
\exists \ k \ | \ \exists \ t'' \ | \ \tau_k =  (\textit{uconsent}, t'', sp, \textit{tar}, dt, v,  \textit{purposes}, \pi_{\theta})$ $\wedge$ $(t'' < t')$ $\wedge$ (\textit{tar} $=$ \textit{or}) $\wedge$ (\textit{purposes} $\subseteq$ $\pi_{\theta}$.$\pi_{use}$.\textit{upurp}.\textit{purpset}$_u$)\\

\item $C_5: \tau_i = (\textit{use}, t', \textit{or}, dt, v,   \textit{purposes}, \textit{bywhom})$ $ 
\wedge \ \textit{State}_A(\tau, i-1)(ow, ds, \theta) = (t, v, \textit{plc}, \pi_{\theta}, \mathcal{H}_{\textit{has}}) \Longrightarrow
\textit{purposes} \subseteq \pi_{\theta}.\pi_{use}.upurp.purpset_u$ $\wedge$ \textit{bywhom} $\subseteq \pi_{\theta}.\pi_{use}.\textit{whouse}.\textit{who}$ $\wedge$ \textit{bywhom} $\subseteq \mathcal{H}_{\textit{has}}$\\

\item $C_{6}: \tau_i = (\textit{deletereq}, t', \textit{or}, dt, v, \pi_{\theta}) 
\ \wedge \ \textit{State}_A(\tau, i-1)$
$(ow, ds, \theta)$ $=  (t, v,  \textit{plc}, \pi_{\theta}, \mathcal{H}_{\textit{has}}) \Longrightarrow \exists \ k \ | \ \exists \ t'' \ | \ \tau_k = (\textit{mandelete}, t'', \textit{or}, dt, v, \textit{places}, \pi_{\theta}) \ \wedge \ (t' < t'' \leq t' + \pi_{\theta}.\pi_{del}.deld.dd$)\\

\item $C_{7}: \tau_i = (unregister, t', \textit{or}, , \textit{pservices}, \textit{typeserv})$ $ 
\wedge \ \textit{State}_A(\tau, i-1)$
$(ow, ds, \theta)$ $=  (t, v,  \textit{plc}, \pi_{\theta}, \mathcal{H}_{\textit{has}}) \Longrightarrow \exists \ k \ | \ \exists \ t'' \ | \ \tau_k = (\textit{autdelete}, t'', \textit{or}, dt, v, \textit{places}, \pi_{\theta}) 
 \ \wedge \ (t' < t'' \leq t' + \pi_{\theta}.\pi_{del}.\textit{gdeld}.\textit{dcond}$). \\

\item $C_8: \tau_i = (\textit{forward}, t', sp, \textit{or}, dt, v,  \textit{purposes}, \textit{towhom}, \pi_{\theta})$ $ 
\wedge \ \textit{State}_A(\tau, i-1)(ow, ds, \theta) = (t, v, \textit{plc}, \pi_{\theta}, \mathcal{H}_{\textit{has}}) \Longrightarrow
\textit{purposes} \subseteq \pi_{\theta}.\pi_{fw}.\textit{fwpurp}$ $\wedge$ \textit{towhom} $\subseteq$ $\pi_{\theta}.\pi_{fw}.\textit{3rdparty}$\\

\item $C_{9}: \tau_i = (\textit{forward}, t', sp, \textit{or}, dt, v,  \textit{purposes}, \textit{towhom}, \pi_{\theta})$ $ 
\wedge \ \textit{State}_A(\tau, i-1)(ow, ds, \theta) = (t, v, \textit{plc}, \pi_{\theta}, \mathcal{H}_{\textit{has}})\ \wedge \ (\pi_{\theta}.\pi_{fw}.\textit{cons} = \textit{Y})) \ \Longrightarrow
\exists \ k \ | \ \exists \ t \ | \ \tau_k =  (\textit{fwconsent}, t, \{\textit{tar}_i\}, dt, v,  \textit{purposes}, \pi_{\theta})$ $\wedge$ $(t < t')$   $\wedge$ (\{\textit{tar}$_i$\} $\subseteq$ \textit{or} $\cup$ \{\textit{or}$_j$\}) $\wedge$ (\textit{purposes} $\subseteq$ $\pi_{\theta}$.$\pi_{fw}$.\textit{fwpurp})\\

\item $C_{10}: \tau_i = (\textit{collect}, t', sp, or, dt, v,   \textit{purposes})$ $ 
\wedge \ \textit{State}_A(\tau, i-1)(ow, ds, \theta) = (t, v, \textit{plc}, \pi_{\theta}, \mathcal{H}_{\textit{has}})\ \Longrightarrow$ $\exists \ k \ | \ \exists \ t \ | \ \tau_k$ =  (\textit{register}, $t$, \textit{or}, \textit{pservices}, \textit{typeserv}) $\wedge$ $(t < t')$   $\wedge$ (\textit{purposes} $\subseteq$ \textit{pservices}) $\wedge$ ($\theta$ $\in$ \textit{typeserv})\\
\end{itemize}

\end{minipage}
}
\caption{Trace compliance rules for a policy.}\label{fig:compsn0}
\end{figure}

In Fig.~\ref{fig:compsn0}, the rule $C_1$ says that if at  $\tau_i$, the $i$-th step of the trace (i.e., the system operation at time $t'$), there is an event  \textit{storerev}, then $t'$ should be in the predefined revision time. The rules $C_2$ and $C_3$ say that if at the $i$-th step of the trace there is a collection event, then sometimes before that there should be a corresponding declaration and consent collection, respectively, and the entity from whom the consent has been collected (\textit{tar}) should be the same as the entity from whom the data has been collected (\textit{or}).  $C_4$ is similar to the previous two rules but related to the data usage.  $C_5$ says that if the data $dt$ is used by the service provider \textit{sp} for some \textit{purposes} at the $i$-th event in a trace, and the policy at the $(i-1)$th state is  $\pi_{\theta}$, then \textit{purposes} must be the subset of $\pi_{\theta}.\pi_{use}.upurp.purpset_u$ and the entities in the set \textit{towhom} must have $dt$. The rule $C_6$ says that whenever there is a deletion request for $dt$ this must be done within the specified deletion delay. The rule $C_7$ states that whenever an user unregisters from the service, the wors-case deletion must be done within the specified deletion delay. The rules $C_8$ and $C_{9}$ define policy compliance for data forwarding. $C_8$ says that the forwarding purposes and the addressee in the event must be subsets of the specified purposes and addressee in the policy. $C_9$ says that if the consent option \textit{cons} is set to \textit{cons} = \textit{Y} in the policy, then a corresponding consent must be collected before $dt$ is forwarded. Finally, $C_{10}$ captures a strict aspect of the data minimisation requirement.  

The formalization of these compliance rules can be found in Fig.~\ref{fig:compsn0}. Trace compliance is defined with respect to the rules above. In the following definition and properties, let $dt|_{\theta}$ represents any data of type $\theta$ and $dt$ refers to any tuple of given values of $ow$, $ds$ and $\theta$. 

\begin{ttd}\label{def:sysimp}
$($\textbf{\emph{Low-level system implementation}}$)$ \noindent Let \textit{SysImp} be a system implementation, which is  a composition of  specific protocols and procedures. Further, let  
\textit{SysImpOp}($\theta$) be the system operation with a data $dt$ of type $\theta$, and a set of low-level actions \textit{AC}$_{\textit{sysimp}}(\theta)$ = $\{$$\alpha^{sys}_1$,\dots, $\alpha^{sys}_m$$\}$ for some finite $m$ that can happen during any possible system operation sequences.

\textit{SysImpOp}($\theta$) is a set of all possible  sequences of low-level actions, namely, we have \textit{SysImpOp}($\theta$) = $\{$$\textit{Seq}_{\textit{AC}_{\textit{sysimp}}(\theta)}$$\}$. Each operation sequence is a trace of subsequent low-level actions performed at a given time value: 

\begin{center}
$\textit{Seq}_{\textit{AC}_{\textit{sysimp}}}$ = $\textit{state}_i$ $\stackrel{\alpha^{sys}_i, t_i}{\longrightarrow}$ $\textit{state}_{i+1}$ $\stackrel{\alpha^{sys}_{i+1}, t_{i+1}}{\longrightarrow}$ \dots $\stackrel{\alpha^{sys}_{i+n-1}, t_{i+n-1}}{\longrightarrow}$ $\textit{state}_{i+n}$.    
\end{center}

Low-level actions can be, for example, sending or receiving some message, computing some function or a verification of the received message, etc. For illustration purposes, let us consider the action trace $\textit{Seq}_{\textit{AC}_{\textit{sysimp}}(\theta)}$ above, and as an example, let $\alpha^{sys}_i$ = \textit{send}(\textit{or}, \textit{tar}, $V$:$\theta$, \textit{purposes}) captures that an entity \textit{or} sends a message of value $V$ and type $\theta$ to another entity \textit{tar}, while $\alpha^{sys}_{i+n-1}$ = \textit{recv}(\textit{tar}, \textit{or}, $V$:$\theta$, \textit{purposes}) captures the reception of that message by \textit{tar}. Hence, if ($\theta$ = useconsent) and (\textit{or} = \textit{cust}), (\textit{tar} = \textit{sp}), then state $\textit{state}_{i+n}$ captures  that 
the usage consent of data $V$ has been collected by the service provider at time $t_{i+n-1}$.  

We let \textit{SysImpOp} be the system operation for all the supported types, namely, \textit{SysImpOp} = $\bigcup_{\forall \theta \in \textit{TypeSys}}\{\textit{Seq}_{\textit{AC}_{\textit{sysimp}}(\theta)}\}$. 
\end{ttd}

In the following definitions, we let 

\begin{center}
\noindent\fbox{%
    \parbox{8cm}{%
  \ \ \ \ $\mathcal{C}$ = \{$C_1$, $C_2$, $C_3$, $C_4$, $C_5$, $C_6$, $C_7$, $C_8$, $C_9$, $C_{10}$\}
     }
}
\end{center} 

\noindent \textbf{Trace rule-compliance relation}: Let us denote by $\textit{seq}_{\theta}$ $\prec^{rule}_{pol}$ $C_i$ the fact that a given operation trace $\textit{seq}_{\theta}$$\in$ \textit{SysImpOp}($\theta$) of a system implementation \textit{SysImp}, on a data of type $\theta$, is in compliance with a rule $C_i$ $\in$ $\mathcal{C}$.

\begin{ttd}$($\textbf{trace $\pi_{\theta}$-compliance}, $\textit{seq}_{\theta}$ $\prec_{pol}$ $\mathcal{C}$$)$

\noindent An operation trace $\textit{seq}_{\theta}$$\in$ \textit{SysImpOp}($\theta$) of a system implementation \textit{SysImp}, on a data of type $\theta$, is in compliance with a DPR sub-policy $\pi_{\theta}$ if it satisfies all the properties in $\mathcal{C}$, $\forall$$C_i$ $\in$ $\mathcal{C}$: $\textit{seq}_{\theta}$ $\prec^{rule}_{pol}$ $C_i$.  This is denoted by $($$\textit{seq}_{\theta}$ $\prec_{pol}$ $\mathcal{C}$$)$.  
\end{ttd}

The DPR trace compliancy is relevant for verifying if a specific system implementation (e.g., protocol, system log procedure) complies with a given policy. 

\begin{ttd}$($\textbf{system $\pi_{\theta}$-compliance}, \textit{SysImp} $\prec^{sys}_{pol}$ $\mathcal{C}$$)$

\noindent A system implementation \textit{SysImp} is compliant with a DPR sub-policy $\pi_{\theta}$ if $\forall$$\textit{seq}_{\theta}$ $\in$ \textit{SysImpOp}($\theta$): $\textit{seq}_{\theta}$ $\prec_{pol}$ $\mathcal{C}$.  We denote this by $($\textit{SysImp} $\prec^{sys}_{pol}$ $\mathcal{C}$$)$.   
\end{ttd}

\begin{ttd}$($\textbf{system $\mathcal{P}$$\mathcal{L}$-compliance, \textit{SysImp} $\models_{pol}$ $\mathcal{P}$$\mathcal{L}$}$)$

\noindent A system implementation \textit{SysImp} is compliant with a policy $\mathcal{P}$$\mathcal{L}$ if $\forall$$\textit{seq}_{\theta}$ $\in$ \textit{SysImpOp}($\theta$) and $\forall$$\pi_{\theta}$ $\in$ $\mathcal{P}$$\mathcal{L}$: $\textit{seq}_{\theta}$ $\prec_{pol}$ $\mathcal{C}$. This is denoted by \textit{SysImp} $\models_{pol}$ $\mathcal{P}$$\mathcal{L}$.   
\end{ttd}

\begin{ttp}\label{prop:has1}\textbf{$($Has $dt|_{\theta}$ up to $\pi_{\theta}$$)$}

An entity $e$, $e$ $\in$ \textit{Entity}$_{\mathcal{P}\mathcal{L}}$, has a data of type up to  the policy $\pi_{\theta}$ iff 

$\exists$ $\tau$, $i$, $ow$, $ds$: $\textit{State}_A(\tau, i)(ow, ds, \theta) = (t, v, \textit{plc}, \pi_{\theta}, \mathcal{H}_{\textit{has}})$ $\wedge$ $(e$ $\in$ $\mathcal{H}_{has})$. We denote this by $[e$ has $dt|_{\theta}]$$_{\pi_{\theta}}$. 

If this does not satisfy/hold, then we denote it by $[e$ $\neg$has $dt|_{\theta}]$$_{\pi_{\theta}}$. 
\end{ttp}

\begin{ttp}\label{prop:has2}\textbf{$($Has $dt$ up to $\pi_{\theta}$$)$}

An entity $e$, $e$ $\in$ \textit{Entity}$_{\mathcal{P}\mathcal{L}}$, has the data $dt$ = ($ow$, $ds$, $\theta$) for given $ow$ and $ds$, up to the policy $\pi_{\theta}$ iff  

$\exists$ $\tau$, $i$: $\textit{State}_A(\tau, i)(ow, ds, \theta) = (t, v, \textit{plc}, \pi_{\theta}, \mathcal{H}_{\textit{has}})$ $\wedge$ $(e$ $\in$ $\mathcal{H}_{has})$. We denote this by $[e$ has $dt]$$_{\pi_{\theta}}$.

\end{ttp}    

\begin{ttp}\label{prop:has3}\textbf{$($Has $dt|_{\theta}$ up to $\mathcal{P}\mathcal{L}$$)$}
If $\exists$$e$, $dt|_{\theta}$: $e$ $\in$ \textit{Entity}$_{\mathcal{P}\mathcal{L}}$ and $dt|_{\theta}$ $\in$ \textit{Data}$_{\mathcal{P}\mathcal{L}}$ and $[e$ has $dt|_{\theta}]$$_{\pi_{\theta}}$, then we have $[e$ has $dt|_{\theta}]$$_{\mathcal{P}\mathcal{L}}$.   

If this does not satisfy/hold, then we denote it by $[e$ $\neg$has $dt|_{\theta}]$$_{\mathcal{P}\mathcal{L}}$. 
\end{ttp}

\begin{ttp}\label{prop:has4}\textbf{$($Has $dt$ up to $\mathcal{P}\mathcal{L}$$)$}
If $\exists$$e$, $dt$ = $(ow$, $ds$, $\theta)$: $e$ $\in$ \textit{Entity}$_{\mathcal{P}\mathcal{L}}$ and $dt$ $\in$ \textit{Data}$_{\mathcal{P}\mathcal{L}}$ and $[e$ has $dt]$$_{\pi_{\theta}}$, then we have $[e$ has $dt]$$_{\mathcal{P}\mathcal{L}}$.   
\end{ttp}   

An entity in the system is allowed to have the data $dt$ if and only if this is explicitly specified in the policy $\pi_{\theta}$. Therefore, in case a system allows unathorised access to a data $dt$, it violates the policy  $\pi_{\theta}$. The main difference between the Property~\ref{prop:has1} and \ref{prop:has2} as well as between  Property~\ref{prop:has3} and \ref{prop:has4} is that in the first cases the property is defined on any data of type $\theta$, regardless of the value of $ow$ and $ds$, while the latter cases are defined on the data with given $ow$, $ds$ and $\theta$. 

\section{The Corresponding Architecture Level}
\label{sec:arch0}

To check the conformance of a system architecture with a given data protection (DPR) policy, we propose an architecture language, called  $\mathcal{A}_{DC}$. $\mathcal{A}_{DC}$ follows a similar concept of the language outlined in \cite{TaAntignac14}, however, it has been modified accordingly for our purposes. Unlike the architecture language outlined in \cite{Antignac14, TaAntignac14}, which mainly focuses on data computation and verification of the data integrity and trust relation, our $\mathcal{A}_{DC}$ primary captures the personal data protection properties on the end-to-end data life-cycle to fit with the policy language. Hence, $\mathcal{A}_{DC}$ defines different syntax and semantics elements compared to \cite{TaAntignac14, Antignac14}.  

\subsection{Architectures Syntax}
\label{sec:syntax0}

We assume that an architecture for a service provider \textit{SP} is composed of a finite set of entities, \textit{Entity}$^{SP}_{\mathcal{P}\mathcal{A}}$ = \{$E_{i_1}$, \dots, $E_{i_m}$\}.  Also let \textit{TYPE}$^{SP}_{\mathcal{P}\mathcal{A}}$ be the set of all the supported types in the architecture.    

\begin{figure}[htbp]
\centering
\fbox{\begin{minipage}{11.87 cm}
\begin{tabbing}    
    \=123456\=1\=1\=1\= \kill
		\>\> \underline{\textit{PatOf}}: \\\\
		\>\> \textit{PartOf}: $E_e$ $\in$ \textit{Entity}$^{SP}_{\mathcal{P}\mathcal{A}}$ $\rightarrow \{$$E_e$ $\in$ \textit{Entity}$^{SP}_{\mathcal{P}\mathcal{A}}$\}\\\\
    \=123456\=1\=1\=1\= \kill
		\>\> \underline{\textit{Terms}}: \\\\    
    \=123456\=1\=1\=1\= \kill
    \>\> $T$ ::= $BT$ $|$ $D$ $|$ \textit{Decl} $|$ \textit{Cp}($\tilde{X}$) \\\\
    \=123456\=1\=1\=1\= \kill
		\>\> $id$ ::= $xk$ $|$ $c$, where $k$ and $c$ can be any constant.\\\\ 
 \=123456\=1\=1\=1\= \kill
    \>\> $BT$ ::= $\tilde{X}$ $|$ $F(BT_1, \dots, BT_n)$ $|$ $D$\\\\
    \=123456\=1\=1\=1\= \kill
    \>\> $\tilde{X}$ ::= $X^{xds}_{ow, \theta}$ $|$ $X_{|\theta}$\\\\
	\=123456\=1\=1\=1\= \kill
		\>\> $D$ ::=  $V^{ds}_{ow, \theta}$ \\\\ 
		\=123456\=1\=1\=1\= \kill
		\>\> $C$ ::=  \textit{Vpurp}$_j$ $|$ \textit{dd}$_j$  $|$ $TT_j$\\\\ 
		\=123456\=1\=1\=1\= \kill
		\>\> \underline{\textit{Destructors}}: \\\\
		\=123456\=1\=1\=1\= \kill
		\>\>  $G$($\tilde{X}_1$, \dots, $\tilde{X}_n$) $\rightarrow$ $\tilde{X}$\\\\ 
		\=123456\=1\=1\=1\= \kill
		\>\> \underline{\textit{Type}}: \\\\
\=123456\=1\=1\=1\= \kill
		\>\> TYPE($T$) ::= $\theta$,  where $\theta$ $\in$ \textit{TYPE}$^{SP}_{\mathcal{P}\mathcal{A}}$\\\\
		\=123456\=1\=1\=1\= \kill
		\>\> \underline{\textit{Owner}}: \\\\
\=123456\=1\=1\=1\= \kill
		\>\> OWNER($\tilde{X}$) = $E_{ow}$,  where $E_{ow}$ $\in$ \textit{Entity}$^{SP}_{\mathcal{P}\mathcal{A}}$. 
\end{tabbing}
\end{minipage}
}
\caption{Terms, Destructors and Types.}\label{fig:archterms}
\end{figure}

We assume a finite set of data subjects ($ds$ $\in$ \textit{Entity}$^{SP}_{\mathcal{P}\mathcal{A}}$), and data owners $ow$ $\in$ \textit{Entity}$^{SP}_{\mathcal{P}\mathcal{A}}$), as well as a finite set of types ($\theta$ $\in$ \textit{TYPE}$^{SP}_{\mathcal{P}\mathcal{A}}$).  We also assume a finite set of variables \textit{Var}, $\tilde{X}$ $\in$ \textit{Var}, and values \textit{Val}, $D$ $\in$ \textit{Val}.  

\textbf{PartOf}: \textit{PartOf} is a function that expects an entity as input and returns a set of other entities in the same architecture (which are parts of the input entity). For example, if $E_m$ and $E_{p}$ represent a smart meter, and a digital panel (tablet), respectively, and we want to specify that they are parts of the service provider, $E_{sp}$, then, we let \textit{PartOf}($E_{sp}$) = \{$E_{m}$, $E_{p}$\}. This function is important for conformance check as several entities may be parts of a same entity, hence, for example, if the smart meter or the panel has the customer data, then it also means that the service provider has this data. 

\textbf{Terms:} Terms model any data defined in the architecture. Term $T$ is defined as in the Fig.~\ref{fig:archterms}.  A term can be a variable ($X^{xds}_{ow, \theta}$, $X_{|\theta}$) that represents some data, a term can also be a data constant ($V^{ds}_{ow, \theta}$) and other constant ($D$) that represent the value of a data,  as well as  a copy of any data $\tilde{X}$ (denoted by \textit{Cp}($\tilde{X}$). Finally, $id$ can be an index variable $xk$ or an index constant $c$. During the system operations,  variables will be bounded to values, and index constant will be bounded to index variables. 

For each entity $E$ we define a finite set of variables (i.e., data) \textit{Var} $=$ \{$\tilde{X}$ $|$ TYPE($\tilde{X}$) = $\theta$, $\theta$ $\in$ \textit{TYPE}$^{SP}_{\mathcal{P}\mathcal{A}}$\} of type $\theta$ that it owns and inputs it into the system.    
A variable $X^{xds}_{ow, \theta}$ $\in$ \textit{Var} represents any kind of data supported by a service provider, such as the users' basic information, photos, videos, posts, energy consumption data, etc. We distinguish $X^{xds}_{ow, \theta}$ and $X_{|\theta}$, where in the second case only the type of the data is given. For example, $X_{|\theta_{Skey}}$ and $X_{|\theta_{Ckey}}$ represent a service provider's crypto key and the customer's key, respectively. $xds$ is an index variable that captures the data subject included in a given data. Index constants $ow$ and $\theta$ denotes the owner and the type of a given data, respectively. For instance, $X^{xds}_{\textit{customer1}, \theta_{photo}}$ represents a photo of the  customer1 about a data subject $xds$. During the system operation, specific values will be given to the index variable $xds$ and the variable $X$ itself  (as we may have different photos during different system runs). Similarly, data type can be anything, such as basic personal information, energy consumption data, etc. $F$ is a function that can be, for instance, encryption (symmetric,  asymmetric, homomorphic), hash, digital signature. Function \textit{Cp}($\tilde{X}$) returns a copy of $\tilde{X}$.
 
A variable $X^{xds}_{ow,\theta}$  is given specific data value $V^{ds}_{ow,\theta}$ during a system run, such that $ow$ and  $\theta$ in $X$ are the same to those in $V$, while the index variable $xds$ will be bounded to a constant $ds$. In $V^{ds}_{ow, \theta}$, the index constant $ow$ denotes the owner of the data, $ds$ denotes the data subject included in the data, and $id$ is a unique ID of the data in the system.  Note that $ds$ and $ow$ can be the same, e.g., when $ow$ owns an energy consumption data about itself. Note that the type of $V^{ds}_{ow, \theta}$ should be the same as   the type of the variable to which it is bound. We also define other constants such as a purpose (\textit{Vpurp}$_j$), a deletion delay (\textit{dd}$_j$), or a time value (\textit{TT}$_j$). 

\textbf{Destructors:} A destructor represents an evaluation of a function. For instance, if the function $F$ is an encryption or a digital signature, then the corresponding destructor $G$ is the decryption or signature verification procedure. More precisely, if $\tilde{X}_1$ $=$ \textit{Enc}($\tilde{X}$, $X_{|\theta_{Skey}}$) that represents the encryption of data $\tilde{X}$ with the server key $X_{|\theta_{Skey}}$, and $\tilde{X}_2$ $=$ $X_{|\theta_{Skey}}$, then $G(\tilde{X}_1, \tilde{X}_2)$ $\rightarrow$ $\tilde{X}$ is \textit{Dec}(\textit{Enc}($\tilde{X}$, $X_{|\theta_{Skey}}$), $X_{|\theta_{Skey}}$) $\rightarrow$ $\tilde{X}$.  Note that not all functions have a corresponding destructor, e.g., in case $\tilde{X}_1$ is a one-way cryptographic hash function, $\tilde{X}_1$ $=$ \textit{Hash}($\tilde{X}$), then due to the one-way property there is no destructor (reverse procedure) that returns 
$\tilde{X}$ from the hash $\tilde{X}_1$.


\textbf{Architecture:} An architecture $\mathcal{P}\mathcal{A}$ is defined by a set of \textit{activities} of users or service providers  
(denoted by $\{\mathcal{F}\}$). The formal definition of privacy architectures is given as follows: 

\begin{figure}[htbp]
\centering
\fbox{\begin{minipage}{11.87 cm}
\begin{tabbing}  
    \=1\=1\=1\= \kill\\
    \>\> $\mathcal{P}\mathcal{A}$ ::= $\{\mathcal{F}\}$\\\\
    \=1\=1\=1\=1\= \kill
    \>\> $\mathcal{F}$ ::= \textit{Own}$_{e}$($\tilde{X}$)\\
    \=1\=1\=1\=1\= \kill
		\>\>\ \ \ \ \ \ \ \ \ $|$ 
    \textit{Register}$_{e1,e2}$(\textit{pservices}, \textit{typeserv})\\
    \=1\=1\=1\=1\= \kill
		\>\>\ \ \ \ \ \ \ \ \ $|$ 
    \textit{Compute}$_{e}$($\tilde{X}$ $=$ $BT$)\\
    \=1\=1\=1\=1\= \kill
		\>\>\ \ \ \ \ \ \ \ \ $|$ 
    \textit{Receive}$_{e1,e2}$($\tilde{X}$)\\
		\=1\=1\=1\=1\= \kill
		\>\>\ \ \ \ \ \ \ \ \ $|$ 
    \textit{Store}$_{e}$($\tilde{X}$, \textit{Places})\\
		\=1\=1\=1\=1\= \kill
		\>\>\ \ \ \ \ \ \ \ \ $|$ \textit{Storerev}$_{e}$($\tilde{X}$, \textit{Places}, \textit{TT})\\
		\=1\=1\=1\=1\= \kill
		\>\>\ \ \ \ \ \ \ \ \ $|$  \textit{Collect}$_{e1,e2}$($\tilde{X}$, \textit{Purposes}) \\
	\=1\=1\=1\=1\= \kill
		\>\>\ \ \ \ \ \ \ \ \
		$|$ \textit{Use}$_{\textit{bywhom}}$($\tilde{X}$, \textit{Purposes}) \\
\=1\=1\=1\=1\= \kill
		\>\>\ \ \ \ \ \ \ \ \
		$|$ \textit{Forward}$_{e,\textit{towhom}}$($\tilde{X}$, \textit{Purposes}) \\
		\=1\=1\=1\=1\= \kill
		\>\>\ \ \ \ \ \ \ \ \ $|$  \textit{CConsent}$_{e1,e2}$($\tilde{X}$, \textit{Purposes})\\
		 \=1\=1\=1\=1\= \kill
		\>\>\ \ \ \ \ \ \ \ \  
		$|$ \textit{UConsent}$_{e1,e2}$($\tilde{X}$, \textit{Purposes}, \textit{ByWhom}) \\
		\=1\=1\=1\=1\= \kill
		\>\>\ \ \ \ \ \ \ \ \ $|$  \textit{FwConsent}$_{e1,e2}$($\tilde{X}$, \textit{Purposes}, \textit{ToWhom}) \\
		\=1\=1\=1\=1\= \kill
		\>\>\ \ \ \ \ \ \ \ \  $|$ 
		\textit{Declare}$_{e1, e2}$($\tilde{X}$, \textit{DeclParams}) \\
\=1\=1\=1\=1\= \kill
\>\>\ \ \ \ \ \ \ \ \  $|$ 
		\textit{DeleteReq}$_{e1, e2}$($\tilde{X}$) \\
		\=1\=1\=1\=1\= \kill
		\>\>\ \ \ \ \ \ \ \ \ $|$ \textit{ManDelete}$_{e}$($\tilde{X}$, \textit{Places}, $dd$)  \\ 
\=1\=1\=1\=1\= \kill
\>\>\ \ \ \ \ \ \ \ \
		$|$ \textit{AutDelete}$_{e}$($\tilde{X}$, \textit{Places}, $dd$)\\ 
		\=1\=1\=1\=1\= \kill
\>\>\ \ \ \ \ \ \ \ \
		$|$ \textit{UnRegister}$_{e1,e2}$(\textit{pservices}, \textit{typeserv}).
\end{tabbing}
\end{minipage}
}
\caption{The table shows the syntax of a system architecture consisting of activities between entities. The indices $e$, $e1$, $e2$ refer to the entities $E_e$, $E_{e1}$, $E_{e2}$, respectively.}\label{fig:sysarch}
\end{figure}

Activity \textit{Own}$_{e}$($\tilde{X}$) captures that $E_e$ owns the data variable $\tilde{X}$. If we write \textit{Own}$_{e}$($X_{|\theta}$), then we can also indicate the type of the data owned by $E_e$. \textit{Register}$_{e1,e2}$(\textit{pservices}, \textit{typeserv}) captures that $E_{e1}$ registers with $E_{e2}$ for the services \textit{pservices} that requires the set of data types \textit{typeserv}. \textit{Compute}$_{e}$($\tilde{X}$ $=$ $BT$) captures that an entity $E_e$ can compute the variable $\tilde{X}$ based on the equation $\tilde{X}$ $=$ $BT$. \textit{Receive}$_{e1, e2}$($\tilde{X}$) specifies that $E_{e1}$ received $\tilde{X}$ from $E_{e2}$,    
\textit{Store}$_{e}$($\tilde{X}$, \textit{Places}) says that $E_e$ stores $\tilde{X}$ in the set of places \textit{Places}, where \textit{Places} is a set of entities $E$ $\in$ \{$E_{\textit{Main}_e}$, $E_{\textit{BckUp}_e}$\}. $E_{\textit{Main}_e}$ represents the  \textit{Storerev}$_{e}$($\tilde{X}$, \textit{Places}, \textit{TT}) captures that $E_e$ reviews the need of storing $\tilde{X}$ further in the place \textit{Places} at a predefined time \textit{TT}. 

\textit{Collect}$_{e1, e2}$($\tilde{X}$, \textit{Purposes}) captures that $E_{e1}$ collects the data $\tilde{X}$ with the purposes \textit{Purposes} from an entity $E_{e2}$. \textit{Purposes} contains the constant purpose values (\textit{Vpurp}) of the data collection. \textit{Use}$_{\textit{bywhom}}$($\tilde{X}$, \textit{Purposes}) captures the usage of $\tilde{X}$ by entities in \textit{bywhom} for the purposes \textit{Purposes}. \textit{Forward}$_{e,\textit{towhom}}$($\tilde{X}$, \textit{Purposes}) captures that $E_{e}$ forwards $\tilde{X}$ with purposes \textit{Purposes} and to the entities in the set \textit{towhom}.  The activities \textit{CConsent}$_{e1, e2}$($\tilde{X}$, \textit{Purposes}), \textit{UConsent}$_{e1, e2}$($\tilde{X}$, \textit{Purposes}) and \textit{FwConsent}$_{e1, e2}$($\tilde{X}$, \textit{Purposes}, \textit{ToWhom}) captures the consent collection activities by $E_{e1}$ from $E_{e2}$ for data collection, usage and data forwarding, respectively. \textit{Purposes} and \textit{ToWhom} are the set of purposes and entities to whom data $\tilde{X}$ will be forwarded, these are showed/revealed to the users when consents are collected from them.  

 The activity \textit{Declare}$_{e1, e2}$($\tilde{X}$, \textit{DeclParams}) captures the declaration of the parameters  \textit{DeclParams} by $E_{e1}$ to $E_{e2}$, which is a set of any parameters such as the elements from \textit{Purposes}, \textit{Places}, \textit{dd}, \textit{TT}, \textit{ByWhom}, \textit{ToWhom}. The activities \textit{DeleteReq}$_{e1, e2}$($\tilde{X}$, \textit{Places}) and \textit{ManDelete}$_{e}$($\tilde{X}$, \textit{Places},  $dd$) say that an entity $E_{e1}$ sends a deletion request to $E_{e2}$ for $\tilde{X}$  from the places \textit{Places}. \textit{ManDelete}$_{e}$($\tilde{X}$, \textit{Places},  $dd$) and \textit{AutDelete}$_{e}$($\tilde{X}$, \textit{Places},  $dd$), with a predefined deletion delay $dd$, refer to a manual and automated deletion modes, which captures that the deletion must take place within $dd$ delay after $E_{e}$ received a deletion request (in the first case), or within certain worst-case deletion delay once a certain event happens (in the latter case).  Finally, the function OWNER($\tilde{X}$) returns the owner of the data $\tilde{X}$, which is the entity that inputs this data into the system.

\subsection{Architectures Semantics}
\label{sec:semantics0}

The semantics of an architecture is based on its set of compatible traces. A trace $\Gamma$ is a sequence of high-level events 
\textit{Seq}($\epsilon$) occurring in the system as presented in the Table below.  

\begin{figure}[htbp]
\centering
\fbox{\begin{minipage}{11.87 cm}
\begin{tabbing}\\
    \=1\=1\=1\=1\= \kill
		\> $\Gamma$ ::= \textit{Seq}($\epsilon$)\\
    \>$\epsilon$ ::=
    \textit{own}$_{e}$($\tilde{X}$$:$$D$, $t$)\\ 
    \=1\=1\=1\=1\= \kill
     \=1\=1\=1\=1\= \kill
		\>\ \ \ \ \ \ \ 
	$|$ \textit{register}$_{e1,e2}$(\textit{pservices}, \textit{typeserv}, $t$)\\
		\>\ \ \ \ \ \ \ $|$ 
    \textit{compute}$_{e}$($\tilde{X}$ $=$ $BT$, $t$)\\
    \=1\=1\=1\=1\= \kill
		\>\ \ \ \ \ \ \  $|$  \textit{store}$_{e}$($\tilde{X}$$:$$D$, \textit{Places}, $t$)\\ 
	\=1\=1\=1\=1\= \kill
		\>\ \ \ \ \ \ \  $|$  \textit{receive}$_{e1,e2}$($\tilde{X}$$:$$D$, $t$)\\ 
	\=1\=1\=1\=1\= \kill
		\>\ \ \ \ \ \ \ 	
		$|$ \textit{storerev}$_{e}$($\tilde{X}$$:$$D$, \textit{Places}, $TT$, $t$) \\	
\=1\=1\=1\=1\= \kill
		\>\ \ \ \ \ \ \	$|$ \textit{collect}$_{e1,e2}$($\tilde{X}$$:$$D$, \textit{Purposes}, $t$)\\
		\=1\=1\=1\=1\=\kill
		\>\ \ \ \ \ \ \	
		$|$ \textit{use}$_{bywhom}$($\tilde{X}$$:$$D$, \textit{Purposes}, $t$)\\
		\=1\=1\=1\=1\=\kill
		\>\ \ \ \ \ \ \	
		$|$ \textit{forward}$_{e, towhom}$($\tilde{X}$$:$$D$, \textit{Purposes}, $t$)\\
\=1\=1\=1\=1\= \kill
		\>\ \ \ \ \ \ \	$|$ \textit{cconsent}$_{e1,e2}$($\tilde{X}$$:$$D$, \textit{Purposes}, $t$)\\ 
	\=1\=1\=1\=1\= \kill
		\>\ \ \ \ \ \ \	
		$|$ \textit{uconsent}$_{e1,e2}$($\tilde{X}$$:$$D$, \textit{Purposes}, \textit{ByWhom}, $t$)\\
	\=1\=1\=1\=1\= \kill
		\>\ \ \ \ \ \ \	
		$|$ \textit{fwconsent}$_{e1,e2}$($\tilde{X}$$:$$D$, \textit{Purposes}, \textit{ToWhom}, $t$)\\
		\=1\=1\=1\=1\= \kill
		\>\ \ \ \ \ \ \ $|$ 
		\textit{declare}$_{e1, e2}$($\tilde{X}$$:$$D$, \textit{DeclParams}, $t$) \\
		\=1\=1\=1\=1\=\kill
		\>\ \ \ \ \ \ \ $|$ \textit{deletereq}$_{e1,e2}$($\tilde{X}$$:$$D$, $t$)\\ 
		\=1\=1\=1\=1\=\kill
		\>\ \ \ \ \ \ \
		$|$ \textit{mandelete}$_{e}$($\tilde{X}$$:$$D$, \textit{Places}, \textit{dd}, $t$)\\
		\=1\=1\=1\=1\=\kill
		\>\ \ \ \ \ \ \
		$|$ \textit{autdelete}$_{e}$($\tilde{X}$$:$$D$, \textit{Places}, $dd$, $t$)\\
		\=1\=1\=1\=1\=\kill
		\>\ \ \ \ \ \ \ 
	$|$ \textit{unregister}$_{e1,e2}$(\textit{pservices}, \textit{typeserv}, $t$).\\
   \end{tabbing}
\end{minipage}
}
\caption{Events and event sequence for architecture.}\label{fig:eventarch}
\end{figure}

\textbf{Events:}  A trace is a operation sequence of a system, and is composed of high-level events. Events can be seen as instantiated activities defined in the Fig.~\ref{fig:sysarch} happened at time $t$, with the variables bounded to specific values (the notion instantiation here is  similar to the term instantiation used in object oriented programming languages). Events are given the same names as the corresponding activities but in lowercase letters in order to avoid confusion.  For example, the events \textit{register}$_{e1,e2}$ and \textit{unregister}$_{e1,e2}$ are the instances of the activities  \textit{Register}$_{e1,e2}$ and \textit{Unregister}$_{e1,e2}$ with the same parameters at the time $t$. The event \textit{own}$_{e}$($\tilde{X}$$:$$D$, $t$) captures  that $E_{e}$ owns $\tilde{X}$ with a value $D$ at the time $t$. $\tilde{X}$$:$$D$ means that the variable $\tilde{X}$ is bounded to the value $D$. The event \textit{store}$_{e}$($\tilde{X}$$:$$D$, \textit{Places}, $t$) saying that the service provider or an user/a client stores the data represented by $\tilde{X}$ with value $D$ in the places from the set \textit{Places} at time $t$.  \textit{receive}$_{e1,e2}$($\tilde{X}$$:$$D$, $t$) says that $E_{e1}$ received $\tilde{X}$ of a value $D$ from $E_{e2}$ at the time $t$. \textit{storerev}$_{e}$($\tilde{X}$$:$$D$, \textit{Places}, $tt$, $t$) is related to the storage review by $E_e$.  The events \textit{cconsent}$_{e1, e2}$, \textit{uconsent}$_{e1, e2}$ and \textit{fwconsent}$_{e1, e2}$ capture the events when the service provider collects the corresponding consents about data collection, usage and forwarding, respectively. The events \textit{collect} and \textit{use} can be interpreted in a similar way as the architecture, but with a certain value $D$ and time value $t$. 
Finally, the event \textit{declare}$_{e1, e2}$ is an instance of the activity \textit{Declare}$_{e1, e2}$ during the low-level system run.

An architecture is said to be consistent if each variable can be initially owned ($\textit{Own}_{e}$) by only a single entity $E_e$. In the rest of the paper,  we only consider \textit{consistent} architectures and consistent traces. 
 
\textbf{States:} The semantics of events is defined based on \textit{entity states} and the \textit{global state}. The state of an entity $E_e$, denoted by $\sigma_e$, is the variable state that assigns a value (including the undefined value $\bot$) to each variable ($\textit{State}_V$). Note that during the system operation, the variable states of different entities can change in a different way, depending on the status of the variable from the entity's perspective. For instance, let's assume that the service provider $E_{sp}$ receives a gas consumption reading $X^{xds}_{cust, gas}$ of a customer \textit{cust} that has value $V^{cust}_{cust, gas}$, then in $\sigma_e$ , $X^{xds}_{cust, gas}$ will be bound to $V^{cust}_{cust, gas}$, while $X^{xds}_{cust, gas}$ still has the undefined value for an another $E_i$ who did not receive $X^{xds}_{cust, gas}$. 

\begin{center}
\noindent\fbox{%
    \parbox{8cm}{%
    \underline{\textbf{$\sigma_e$: State of Entity $E_e$}}
\begin{align*}
    \textit{State}_V & = \, (\textit{Var} \rightarrow \textit{Val}_{\bot}).   
\end{align*}
}%
}
\end{center}

Assume that there are $m$ entities $E_{e_1}$, \dots, $E_{e_m}$ defined in the architecture. The \textit{global state} is the composition of all the entity states is defined on $\textit{State}_{V}^m$ $\times$ \textit{Time}. $\textit{State}_{V}^m$ is the composition of $m$ entity states. Global state is denoted by $\sigma$, where $\sigma$ $=$ ($\sigma_{e_1}$, \dots, $\sigma_{e_m}$, $tt$). 

\begin{center}
\noindent\fbox{%
    \parbox{8cm}{%
    \underline{\textbf{$\sigma$: Global State}}
\begin{align*}
    \textit{State} & = \, (\textit{State}_{V}^m  \times \textit{Time}).
\end{align*}
}%
}
\end{center}

The \textit{initial $($global$)$ state} for an architecture $\mathcal{P}\mathcal{A}$ is denoted by $\sigma^{init}$, and is the composition of the initial states of each defined entities. In the initial state of each entity, \textit{UnDef} denotes the undefined variable state ($\forall \tilde{X} \in \textit{Var}, \textit{UnDef}(\tilde{X}) = \bot$). This means that initially the values of all the variables defined in the architecture are undefined. Values will be assigned to variables at certain time during the system operation, which is captured by events. In addition, it contains the four initial permission groups defined by the architecture. Finally, the initial time value is also undefined.

\begin{center}
\noindent\fbox{%
    \parbox{9cm}{%
    \underline{\textbf{$\sigma^{init}$: Initial Global State}}
\begin{align*}
  \sigma^{init} & = ( \sigma^{init}_{e_1}, \ldots , \sigma^{init}_{e_m}, tt^{init})\ with\\ 
  \forall i \in [1,m],\ \sigma^{init}_{e_i} & = \textit{UnDef}\\ 
  tt^{init} & = \bot.
\end{align*}
}%
}
\end{center}

In the following, we refer to each element of $\sigma$ by $\sigma$.$\sigma_{e_1}$, \dots, $\sigma$.$\sigma_{e_m}$, and $\sigma$.$tt$.

The semantics function $S_T$ is defined in the Fig~\ref{fig:semevent}, which specifies the impact of a trace on the state of each entity $E_e$ in the architecture. It is defined as an iteration through the trace with function $S_E$ defining the impact of each type of event on the states of the users. 

\textbf{Trace and state updates:} The notation $\epsilon.\Gamma$ is used to denote a trace whose first element is $\epsilon$ and the rest of the trace is $\Gamma$. Each event modifies can either modify the global state (and entity state) or leave it unchanged. To capture the modification made by an event at time $t$ on (only) the variable state of an entity $E_e$ we write $\sigma [\sigma_e / \sigma_e[\tilde{X}/D], tt / t]$ (or $\sigma [\sigma_e / \sigma_e[\tilde{X}/\bot], tt / t]$ in the case of the undefined value, e.g., when a variable has been deleted). Intuitively, this denotation captures that the old state $\sigma_e$ is replaced with the new state $\sigma_e[\tilde{X}/D]$, in which the variable $\tilde{X}$ has been given the value $D$ (or undefined value $\bot$) as a result of the event, the time $tt$  variable is given the value $t$. Finally, $\sigma[tt / t]$ captures only a time change. 

\begin{figure}[htbp]
\centering
\fbox{\begin{minipage}{11.87 cm}
\begin{tabbing}
    \=1\=1\=1\=1\= \kill
    $S_T$ $:$ \textit{Trace} $\times$ \textit{State}  $\rightarrow$   \textit{State}\ \ \ \ \ \ \ \ \ \ \ \ \ \ \ \ \ \ \ \ \ \ \ \ \ \ \ \ \ \ $S_E$ $:$ \textit{Event} $\times$ \textit{State} $\rightarrow$  \textit{State}\\\\   
\=1\=1\=1\=1\= \kill	
$S_T$($\langle\rangle$, $\sigma$)  $=$ $\sigma$\ \ \ \ \ \ \ \ \ \ \ \ \ \ \ \ \ \ \ \ \ \ \ \ \ \ \ \ \ \ \ \ \ \ \ \ \ \ \ \ \ \ \ \ \ \ \ \ \ \ \ \ \  $S_T$($\epsilon.\theta$, $\sigma$) $=$  $S_T$($\theta$, $S_E$($\epsilon$, $\sigma$)) \\\\ 
\=1\=1\=1\=1\= \kill
$S_E$($\textit{own}_e$$\left(\tilde{X}:D, t\right)$, $\sigma$) $=$ $\sigma [\sigma_e / \sigma_e[\tilde{X}/D], tt / t]$\\\\
\=1\=1\=1\=1\= \kill	
$S_E$(\textit{register}$_{e1,e2}$(\textit{pservices}, \textit{typeserv}, $t$), $\sigma$) $=$ $\sigma[tt / t]$\\\\
\=1\=1\=1\=1\= \kill
$S_E$(\textit{compute}$_{e}$($\tilde{X}$ $=$ $BT$, $t$), $\sigma$) $=$ $\sigma$ [$\sigma_{e}$/$\sigma_{e}$[$\tilde{X}$/\textit{eval}($BT$,$\sigma_{e}$)], $tt$/$t$]\\\\ 
\=1\=1\=1\=1\= \kill
$S_E$(\textit{receive}$_{e1,e2}$($\tilde{X}:D$, $t$), $\sigma$) $=$ $\sigma [\sigma_{e1} / \sigma_{e1}[\tilde{X}/D], tt / t]$\\\\ 
\=1\=1\=1\=1\= \kill
$S_E$(\textit{store}$_{e}$($\tilde{X}:D$, \{$E_{\textit{Main}_e}$\}, $t$), $\sigma$) $=$ $\sigma [\sigma_e / \sigma_e[\tilde{X}/D], tt / t]$\\\\ 
\=1\=1\=1\=1\= \kill
$S_E$(\textit{storerev}$_{e}$($\tilde{X}:D$, \{$E_{\textit{Main}_e}$\}, $TT$, $t$),  $\sigma$) = $\sigma[tt / t]$\\\\
\=1\=1\=1\=1\= \kill
$S_E$(\textit{store}$_{e}$($\tilde{X}:D$, \{$E_{\textit{Main}_e}$, $E_{\textit{BckUp}_e}$\}, $t$), $\sigma$) $=$ $\sigma$ [$\sigma_{e}$/$\sigma_{e}$[$\tilde{X}$/$D$, $Cp$($\tilde{X}$)/$D$], $tt$/$t$]\\\\ 
\=1\=1\=1\=1\= \kill
$S_E$(\textit{storerev}$_{e}$($\tilde{X}:D$, \{$E_{\textit{Main}_e}$, $E_{\textit{BckUp}_e}$\}, $TT$, $t$),  $\sigma$) = $\sigma[tt / t]$\\\\
\=1\=1\=1\=1\= \kill
$S_E$(\textit{collect}$_{e1, e2}$($\tilde{X}:D$, \textit{Purposes},  $t$),  $\sigma$) $=$ $\sigma$ [$\sigma_{e1}$/$\sigma_{e1}$[$\tilde{X}/D$],  $tt$/$t$]\\\\ 
		\=1\=1\=1\=1\= \kill
$S_E$(\textit{use}$_{bywhom}$($\tilde{X}:D$, \textit{Purposes}, $t$),  $\sigma$) $=$ $\sigma$ [$\sigma_{e}$/($\sigma_{e}$[$\tilde{X}/D$], $tt$/$t$ \ |\ $\forall$ $E_e$ $\in$ \textit{bywhom}] \\\\ 
\=1\=1\=1\=1\= \kill
$S_E$(\textit{forward}$_{e1, towhom}$($\tilde{X}:D$, \textit{Purposes}, $t$),  $\sigma$) $=$ $\sigma$ [$\sigma_{e}$/$\sigma_{e}$[$\tilde{X}/D$], $tt$/$t$ \ |\ $\forall$ $E_e$ $\in$ \textit{towhom}] \\\\ 
\=1\=1\=1\=1\= \kill
$S_E$(\textit{cconsent}$_{e1, e2}$($\tilde{X}:D$, \textit{Purposes}, $tt$/$t$),  $\sigma$) $=$ $\sigma[tt / t]$\
\=1\=1\=1\=1\= \kill
$S_E$(\textit{uconsent}$_{e1, e2}$($\tilde{X}:D$, \textit{Purposes}, \textit{ByWhom}, $t$),  $\sigma$) $=$ $\sigma[tt / t]$\\\\ 
\=1\=1\=1\=1\= \kill
$S_E$(\textit{fwconsent}$_{e1, e2}$($\tilde{X}:D$, \textit{Purposes}, \textit{ToWhom}, $t$),  $\sigma$) $=$ $\sigma[tt / t]$\\\\ 
\=1\=1\=1\=1\= \kill
$S_E$(\textit{declare}$_{e1,e2}$($\tilde{X} : D$, \textit{DeclParam}, $t$), $\sigma$) $=$  $\sigma[tt / t]$\\\\ 
\=1\=1\=1\=1\= \kill
$S_E$(\textit{deletereq}$_{e1,e2}$($\tilde{X} : D$, $t$), $\sigma$) $=$  $\sigma[tt / t]$\\\\ 
\=1\=1\=1\=1\= \kill
$S_E$(\textit{mandelete}$_{e}$($\tilde{X}:D$, \{$E_{\textit{Main}_e}$, $E_{\textit{BckUp}_e}$\}, \textit{dd}, $t$), $\sigma$)\\
\=1\=1\=1\=1\= \kill
\>\> $=$  $\sigma$ [$\sigma_e$/$\sigma_e$[$\tilde{X}$/$\bot$, \textit{Cp}($\tilde{X}$)/$\bot$], $tt$/$t$)\ |\ $\forall$ $E_e$ $\in$ \textit{Entity}].\\\\
\=1\=1\=1\=1\= \kill
$S_E$(\textit{autdelete}$_{e}$($\tilde{X}:D$, \{$E_{\textit{Main}_e}$, $E_{\textit{BckUp}_e}$\}, \textit{dd}, $t$), $\sigma$) \\
\=1\=1\=1\=1\= \kill
\>\> $=$  $\sigma$ [$\sigma_e$ / ($\sigma^v_e$[\{$\tilde{X}$/$\bot$, \textit{Cp}($\tilde{X}$)/$\bot$\}], $tt$/$t$\ |\ $\forall$ $E_e$ $\in$ \textit{Entity}].\\\\
\=1\=1\=1\=1\= \kill
$S_E$(\textit{mandelete}$_{e}$($\tilde{X}:D$, \{\textit{Main}$_e$\}, \textit{dd}, $t$), $\sigma$) $=$  $\sigma$ [$\sigma_e$/($\sigma_e$[$\tilde{X}$/$\bot$], $tt$/$t$\ |\ $\forall$ $E_e$ $\in$ \textit{Entity}].\\\\
\=1\=1\=1\=1\= \kill
$S_E$(\textit{autdelete}$_{e}$($\tilde{X}:D$, \{\textit{Main}$_{sp}$\}, \textit{dd}, $t$), $\sigma$) $=$  $\sigma$ [$\sigma_e$/$\sigma_e$[$\tilde{X}$/$\bot$], $tt$/$t$\ |\ $\forall$ $E_e$ $\in$ \textit{Entity}]\\\\
\=1\=1\=1\=1\= \kill	
$S_E$(\textit{unregister}$_{e1,e2}$(\textit{pservices}, \textit{typeserv}, $t$), $\sigma$) $=$ $\sigma[tt / t]$.
   \end{tabbing}
\end{minipage}
}
\caption{Semantics of architectural events.}\label{fig:semevent}
\end{figure}

\begin{ttd}[Semantics of architectures]
The semantics of an architecture $\mathcal{P}\mathcal{A}$ is defined as:
$\mathcal{S}(\mathcal{P}\mathcal{A}) = \{\sigma \in \textit{State} \, | \, \exists\ \Gamma \in T(\mathcal{P}\mathcal{A}), S_T(\Gamma, \sigma^{init}) = \sigma \}$.
\end{ttd}

\textit{\textbf{Semantics of the architecture events:}} In the Figure~\ref{fig:semevent}, $S_T$ is the semantic function of event traces which takes the current trace and global state as inputs and returns an updated global state. $S_E$ is the semantic function for events, defining how each event makes changes on the global state. For the empty trace $S_T$ leaves the global state unchanged. For a non empty state with event $\epsilon$ as prefix, $S_T$ proceeds with the tail $\theta$ and the state after $\epsilon$ took effect. As a result of the event \textit{own}, the state of $E_e$ is updated by binding value $D$ to the variable $\tilde{X}$, and the variable $tt$ is bounded to $t$, while the permission state stays unchanged. The event \textit{register} only changes the time value in the global state, it has been introduced to reason about compliant traces later. The event \textit{compute} binds the value \textit{eval}($BT$, $\sigma_e$) to $\tilde{X}$ at time $t$, where \textit{eval}($BT$,$\sigma_e$) represents the evaluation of the value of $BT$ in the current variable state of $E_e$. The event \textit{receive}$_{e1,e2}$($\tilde{X}$$:$$D$, $t$) updates the state of $E_{e1}$ by given $\tilde{X}$ a value $D$. 

For the event \textit{store} we distinguish two cases, in the first case, the data is stored only in the main storage places (We let the entity $E_{\textit{Main}_e}$ represent a (group of) main storage place(s) of $E_e$). As a result of the store event, in the state of $E_e$ the value of $\tilde{X}$ is bounded to $D$.  In the second case, a copy of $\tilde{X}$ ($Cp$($\tilde{X})$) is stored at the backup place(s) of $E_e$  (denoted by $E_{\textit{BckUp}_e}$). This time, in the state of $E_e$, $Cp$($\tilde{X})$ gets the same value $D$ as $\tilde{X}$. The event \textit{collect}$_{e1, e2}$  binds the value of $\tilde{X}$ to $D$ in the variable state of $E_{e1}$, $\sigma_{e1}$. The event \textit{use}$_{\textit{bywhom}}$ binds the variable in $\sigma_{e}$ to $D$ for every $E_e$ $\in$ \textit{bywhom}. The events \textit{storerev}, \textit{cconsent}, \textit{uconsent}, \textit{fwconsent} do not affect the global state of the architecture, except for the time.       

The event \textit{deletereq}$_{e1,e2}$ does not affect the current state, while \textit{mandelete}$_{e}$ and \textit{autdelete}$_{e}$ change the value of $\tilde{X}$ to the undefined value, $\bot$, as the following: In case the deletion only takes place in the main storage place ($E_{\textit{Main}_e}$), the value of $\tilde{X}$ is bounded to $\bot$ in the state of every entity. In case the deletion is carried out from both $E_{\textit{Main}_e}$ and $E_{\textit{BckUp}_e}$, beside $\tilde{X}$, its copy $Cp$($\tilde{X}$) is also bounded to $\bot$ in the state of every entity. 
 
\begin{ttd}[Compatibility] A trace $\theta$ of length $\overline{\theta}$  is  compatible with an architecture $\mathcal{P}\mathcal{A}$ if and only if:
\begin{align*}
\forall k \in [1,\overline{\theta}], if \theta_k \neq compute_e(\tilde{X} = BT, t)\, \ then\ & \exists pa \in\, \mathcal{P}\mathcal{A}: \mathcal{C} (\theta_k, pa)
\end{align*}
where $\mathcal{C} (\epsilon, pa)$ holds if and only if $\epsilon$ can be obtained from $pa$ by giving specific values for variables and instantiating (binding) index variables to index constants. $\theta_k$ represents the $k$-th event in the trace $\theta$.  
\end{ttd}

We can now define the semantics of an architecture $\mathcal{P}\mathcal{A}$ as the set of the possible states produced by consistent traces. 

\subsection{Deduction rules for verifying privacy properties}
\label{sec:plogic0}

In this section, we outline our improved version of a privacy logic (compared to \cite{TaAntignac14}) to reason about the privacy requirements of architectures (i.e., who can have or obtain a given data based in the architecture).  

\begin{tabbing}    
    1234567\=1\=1\=1\= \kill
    \> $\phi$ ::=  \textit{HAS}$_e$($\tilde{X}$, $t$) $|$ \textit{HAS}$^{\textit{not}}_e$($\tilde{X}$, $t$) $|$ \textit{HAS}$^{\textit{never}}_e$($\tilde{X}$) $|$ $\phi_1$ $\wedge$ $\phi_2$\\ 
\end{tabbing}

\noindent Definition~\ref{ttd:logsem} provides the semantics of a property $\phi$.

\begin{ttd}
\label{ttd:logsem}  
The semantics $S(\phi)$ of a property $\phi$ is defined in Table~\ref{tab:eq:semantics-of-architecture0} as the set of architectures satisfying $\phi$.

\begin{table}[htb!]
$\mathcal{P}\mathcal{A}$ $\in$ $S(\textit{HAS}_e\left(\tilde{X}, t \right))$ $\Leftrightarrow$  $\exists$ $\sigma$ $\in$ $\mathcal{S}(\mathcal{P}\mathcal{A})$: $\sigma$.$\sigma_e$$(\tilde{X})$ $\neq$ $\bot$ $\wedge$ $\sigma$.$tt$ = $t$ $\text{and if completely defined}$ $\text{(i.e.,}$ $\tilde{X}$ $\text{does not contain any }$ $\bot$$\text{)}$\\

$\mathcal{P}\mathcal{A}$ $\in$ $S(\textit{HAS}_e^\textit{not}$ $\left(\tilde{X}, t\right))$ \, $\Leftrightarrow$ $\exists$ $\sigma$ $\in$ $\mathcal{S}(\mathcal{P}\mathcal{A}$): $\sigma$.$\sigma_e$$(\tilde{X})$ $=$ $\bot$ $\wedge$ $\sigma$.$tt$ = $t$\\

$\mathcal{P}\mathcal{A}$ $\in$ $S(\textit{HAS}_e^\textit{never}$ $\left( \tilde{X} \right))$ \, $\Leftrightarrow$  $\forall$ $\sigma$ $\in$ $\mathcal{S}(\mathcal{P}\mathcal{A}$) : $\sigma$.$\sigma_e(\tilde{X})$ = $\bot$\\

$\mathcal{P}\mathcal{A}$ $\in$ $S(\phi_1 \wedge \phi_2)$ \,  $\Leftrightarrow$  $\mathcal{P}\mathcal{A}$ $\in$ $S(\phi_1)$ $\wedge$ $\mathcal{P}\mathcal{A}$ $\in$ $S(\phi_2)$\\
\caption{Semantics of the HAS properties.}
\label{tab:eq:semantics-of-architecture0}
\end{table}
\end{ttd}

The Table~\ref{tab:eq:semantics-of-architecture0} says that an architecture satisfies the $\textit{HAS}_e(\tilde{X}, t)$ property if and only if an entity $E_e$ can obtain the value of $\tilde{X}$ in at least one compatible execution trace $\theta$ (in the global state $\sigma$ and local state $\sigma_e$ of $E_e$), whereas $ \textit{HAS}_e^{\textit{never}}(\tilde{X})$ holds if and only if there is no execution trace $\theta$ that can lead to a state in which $E_e$ obtains a 
value of $\tilde{X}$. Finally, \textit{HAS}$^{\textit{not}}_e$($\tilde{X}$, $t$) specifies the time $t$, after which $E_e$ will not be able to get a value of $\tilde{X}$. Finally, statisfying $\phi_1$ $\wedge$ $\phi_2$ means satistying both the properties $\phi_1$ and $\phi_2$ at the same time. 
     
In order to reason about the privacy property of architectures, a set of deduction rules (axioms) is provided in Table~\ref{tab:eq:axioms0}. The fact that an architecture $\mathcal{P}\mathcal{A}$ satisfies a property $\phi$ is denoted by $\mathcal{P}\mathcal{A}$ $\vdash$ $\phi$. Deduction rules ($H1$-$H15$) are defined in the Table~\ref{tab:eq:axioms0} to capture privacy related requirements, namely, investigate whether an entity can have a given data. 

\begin{figure}[htbp]
\centering
\fbox{\begin{minipage}{11.87 cm}
\begin{tabbing}
$\mathbf{H1}: \inferrule{\textit{Own}_{e} \left(\tilde{X}\right) {\in} \, \mathcal{P}\mathcal{A}}{\mathcal{P}\mathcal{A} \vdash_{\exists t} \textit{HAS}_{e}\left(\tilde{X}, t \right)}$\\\\\\
$\mathbf{H2}: \inferrule{\textit{Compute}_{e} \left(\tilde{X} = BT\right) {\in} \, \mathcal{P}\mathcal{A}}{\mathcal{P}\mathcal{A} \vdash_{\exists t} \textit{HAS}_{e}\left(\tilde{X}, t \right)}$\\\\\\
$\mathbf{H3}: \inferrule{\textit{Receive}_{e1,e2} \left(\tilde{X}\right) {\in} \, \mathcal{P}\mathcal{A}}{\mathcal{P}\mathcal{A} \vdash_{\exists t} \textit{HAS}_{e1}\left(\tilde{X}, t \right)}$\\\\\\
$\mathbf{H4}: \inferrule{\textit{Store}_{e} \left(\tilde{X}, Places\right) {\in} \, \mathcal{P}\mathcal{A}}{\mathcal{P}\mathcal{A} \vdash_{\exists t} \textit{HAS}_{e}\left(\tilde{X}, t \right)}$\\\\\\
$\mathbf{H5}: \inferrule{\textit{Collect}_{e1,e2} \left(\tilde{X}, \textit{Purposes}\right) {\in} \, \mathcal{P}\mathcal{A}}{\mathcal{P}\mathcal{A} \vdash_{\exists t} \textit{HAS}_{e1}\left(\tilde{X}, t \right)}$\\\\\\ 
$\mathbf{H6}: \inferrule{\textit{Forward}_{e1, towhom} \left(\tilde{X}, \textit{Purposes}\right) {\in} \, \mathcal{P}\mathcal{A}\ \wedge\ E_{e2} \in \textit{towhom}\ \wedge\ \mathcal{P}\mathcal{A} \vdash_{\exists t} \textit{HAS}_{e1}\left(\tilde{X}, t \right)}{\mathcal{P}\mathcal{A} \vdash_{\exists t'} \textit{HAS}_{e2}\left(\tilde{X}, t' \right)}$\\\\\\
$\mathbf{H7}: \inferrule{\exists\ G: BackgroundDeduction_e(G(\tilde{X}_1, \dots, \tilde{X}_n) \rightarrow \tilde{X})\ \wedge \\
\mathcal{P}\mathcal{A} \vdash_{\exists t} \textit{HAS}_{e}\left(\tilde{X}_1, t\right)\ \wedge \ \dots \ \wedge\ \textit{HAS}_{e}\left(\tilde{X}_n, t\right)}{\mathcal{P}\mathcal{A} \vdash_{\exists t} \textit{HAS}_{e}\left(\tilde{X}, t\right)}$\\\\\\ 
$\mathbf{H8}: \inferrule{\exists\ F: BackgroundComp_e(\tilde{X} = F(\tilde{X}_1, \dots, \tilde{X}_n))\ \wedge \\
\mathcal{P}\mathcal{A} \vdash_{\exists t} \textit{HAS}_{e}\left(\tilde{X}_1, t\right)\ \wedge \ \dots \ \wedge\ \textit{HAS}_{e}\left(\tilde{X}_n, t\right)}{\mathcal{P}\mathcal{A} \vdash_{\exists t} \textit{HAS}_{e}\left(\tilde{X}, t\right)}$\\
\end{tabbing}
\end{minipage}
}
\caption{HAS deduction rules for architectures (part 1).}
\label{tab:eq:axioms0}
\end{figure}

In Fig~\ref{tab:eq:axioms0}, 
rules $H1$, $H2$, $H3$ and $H4$ say that if the architecture contains $\textit{Own}_{e} (\tilde{X})$, $\textit{Compute}_{e}(\tilde{X} = BT)$, $\textit{Receive}_{e1,e2}(\tilde{X})$ or \textit{Store}$_{e}(\tilde{X}, \textit{Places})$, respectively, then  entity $E_e$ ($E_{e1}$) has $\tilde{X}$ at some time $t$ during the system run. $H5$ is related to the data collection by $E_{e1}$ from $E_{e2}$ and says that if there is a collection activity defined in the architecture then $E_{e1}$ has the data $\tilde{X}$ at some time. $H6$ says that if there is a forward activity in the architecture $\mathcal{P}\mathcal{A}$, then and the service provider has $\tilde{X}$ at time $t$, then the addressee $E_i$ will have $\tilde{X}$ at some time $t'$. $H7$ and $H8$ capture the background deduction and computation ability of an entity $E_e$. Specifically, $H7$ says that if there is a destructor $G(\tilde{X}_1, \dots, \tilde{X}_n)$ $\rightarrow$ $\tilde{X}$, and $E_e$ is capable to perform this operation, then when $E_e$ has the value of $\tilde{X}_1$, \dots, $\tilde{X}_n$ then it will be able to derive  the value of $\tilde{X}$. $H8$ says that if $E_e$ has the value of each $\tilde{X}_1$, \dots, $\tilde{X}_n$, then if it is capable to compute the function $F$, then it will have the value of $\tilde{X}$. 

\begin{figure}[htbp]
\centering
\fbox{\begin{minipage}{11.87 cm}
\begin{tabbing}
$\mathbf{H9}: \inferrule{\textit{Use}_{bywhom} \left(\tilde{X}, \textit{Purposes}\right) {\in} \, \mathcal{P}\mathcal{A}\ \wedge\ E_e \in \textit{bywhom}}{\mathcal{P}\mathcal{A} \vdash_{\exists t} \textit{HAS}_{e}\left(\tilde{X}, t \right)}$\\\\\\
$\mathbf{H10}: \inferrule{\text{None of the pre-conditions of H1-H9 holds for}\ \tilde{X} and\ E_e}{\mathcal{P}\mathcal{A} \vdash \textit{HAS}_e^\textit{never} \left( \tilde{X} \right)}$\\\\\\

$\mathbf{H11}: \inferrule{\{\textit{ManDelete}_{e}(\tilde{X}, \{E_{\textit{Main}_e}\}, dd), \textit{Store}_{e}(\tilde{X}, \{E_{\textit{Main}_e}, E_{\textit{BckUp}_e}\})\}\ {\subseteq} \, \mathcal{P}\mathcal{A}}{\mathcal{P}\mathcal{A} \vdash_{\exists t,\forall E_{e1} \in Entity^{SP}_{\mathcal{P}\mathcal{A}} \backslash \{E_{e}\}} \textit{HAS}^{not}_{e1}\left(\tilde{X}, t \right)}$\\\\\\
$\mathbf{H12}: \inferrule{\textit{ManDelete}_{e}(\tilde{X}, \{E_{\textit{Main}_e}, E_{\textit{BckUp}_e}\}, dd)\ {\in} \, \mathcal{P}\mathcal{A}}{\mathcal{P}\mathcal{A} \vdash_{\exists t,\forall E_{e1} \in Entity^{SP}_{\mathcal{P}\mathcal{A}}} \textit{HAS}^{not}_{e1}\left(\tilde{X}, t \right)}$\\\\\\
$\mathbf{H13}: \inferrule{\{\textit{ManDelete}_{e}(\tilde{X}, \{E_{\textit{Main}_e}\}, dd), \textit{Store}_{e}(\tilde{X}, \{E_{\textit{Main}_e}, E_{\textit{BckUp}_e}\})\}\ {\subseteq} \, \mathcal{P}\mathcal{A}}{\mathcal{P}\mathcal{A} \vdash_{\exists t, \forall E_i \in Entity^{SP}_{\mathcal{P}\mathcal{A}} \backslash \{E_{e}\}} \textit{HAS}^{not}_{e1}\left(\tilde{X}, t \right)}$\\\\\\
$\mathbf{H14}: \inferrule{\textit{AutDelete}_{e}(\tilde{X}, \{E_{\textit{Main}_e}, E_{\textit{BckUp}_e}\}, dd)\ {\in} \, \mathcal{P}\mathcal{A}}{\mathcal{P}\mathcal{A} \vdash_{\exists t,\forall E_{e1} \in Entity^{SP}_{\mathcal{P}\mathcal{A}}} \textit{HAS}^{not}_{e1}\left(\tilde{X}, t \right)}$\\\\\\
$\mathbf{H15}: \inferrule{\mathcal{P}\mathcal{A} \vdash_{\exists t, \forall E_{e1},  E_{e2} \in Entity^{SP}_{\mathcal{P}\mathcal{A}}, E_{e1} \in \textit{\textbf{PartOf}}(E_{e2})} \textit{HAS}_{e1} \left(\tilde{X}, t \right)}{\mathcal{P}\mathcal{A} \vdash_{\exists t} \textit{HAS}_{e2} \left(\tilde{X}, t \right)}$\\
\end{tabbing}
\end{minipage}
}
\caption{HAS deduction rules for architectures (part 2).}
\label{tab:eq:axioms1}
\end{figure}

In Fig~\ref{tab:eq:axioms1}, $H9$ is related to the \textit{Use}$_{bywhom}$ activity, saying that if an entity $E_e$ will use $\tilde{X}$ then it will have the value of $\tilde{X}$ at some time $t$. $H9$ captures that in case none of the $H1$-$H9$ takes place from $E_e$'s perspevtive during the system run, then $E_e$ will never have the value of $\tilde{X}$. 
    
Rules $H11$-$H14$ are related to the deletion activities. $H11$ says that if \textit{ManDelete}$_{e}(\tilde{X}, \{E_{\textit{Main}_{e}}\}, dd)$ is part of the architecture, namely, $E_e$ has the ability to delete $\tilde{X}$ from its main storage server then all entities in the system except for $E_e$ will not have $\tilde{X}$ from that point (captured by some time $t$). $E_e$ will still have $\tilde{X}$ in its backup server. On the other hand, in $H12$ if the data is deleted from both the main and the backup servers, then no entity in the system will have the value of data $\tilde{X}$. $H13$ and $H14$ are interpreted in the similar way as  $H11-H12$, but related to the automated deletion mode.    

Finally, rule $H15$ is about the \textit{PartOf} function, saying that for every $E_{e1}$ adn $E_{e2}$ from the set of all the defined entities, if $E_{e1}$ $\in$ \textit{PartOf}($E_{e2}$), then whenever $E_{e1}$ has $\tilde{X}$, $E_{e2}$ also has $\tilde{X}$.

\subsection{DPR trace compliance in architectures}
\label{tracesarch}
Similarly to the case of policies we define compliance rules for traces of events in architectures to capture a system operation that respects the data protection regulation requirements. Below, we assume that $\tilde{X}$ is a personal data.  

\begin{figure}[htbp]

\fbox{\begin{minipage}{13.87 cm}

\begin{itemize}
\item $A_1$: If $\{\textit{Collect}_{e1, e2}(\tilde{X}, \textit{Purposes}),\ \textit{CConsent}_{e3, e4}(\tilde{X}, \textit{Purposes})\}$ $\subseteq$ $\mathcal{P}\mathcal{A}$, then: 
 
 $\Gamma_i = \textit{collect}_{e1, e2}(\tilde{X}:D, \textit{Purposes}, t) \Longrightarrow
\exists \ k \ | \ \exists \ t' \ | \ \Gamma_k =  \textit{cconsent}_{e3, e4}(\tilde{X}:D, \textit{Purposes}, t')$ $\wedge$ $(t' < t)$, where $E_{e3}$ = $E_{e1}$ or $E_{e3}$ $\neq$ $E_{e1}$, $E_{e2}$ = $E_{e4}$ or $E_{e2}$ $\neq$ $E_{e4}$, and OWNER($\tilde{X}$) = $E_{e4}$.\\ 

\item $A_2$: If $\{\textit{Use}_{bywhom}(\tilde{X}, \textit{Purposes}),\ \textit{UConsent}_{e1, e2}(\tilde{X}, \textit{Purposes}, \textit{ByWhom})\}$ $\subseteq$ $\mathcal{P}\mathcal{A}$, then:

 $\Gamma_i =  \textit{use}_{bywhom}(\tilde{X}:D, \textit{Purposes}, t) 
 \Longrightarrow
\exists \ k \ | \ \exists \ t' \ | \ \Gamma_k =  \textit{uconsent}_{e1, e2}(\tilde{X}:D, \textit{Purposes}, \textit{ByWhom}, t') \wedge (t' < t)$, where \textit{bywhom} $\subseteq$ \textit{ByWhom}, and OWNER($\tilde{X}$) = $E_{e2}$.\\ 

\item $A_3$: If $\{\textit{Forward}_{e1, towhom}(\tilde{X}, \textit{Purposes}),\ \textit{FWconsent}_{e2, e3}(\tilde{X}, \textit{Purposes}, \textit{ToWhom})\}$ $\subseteq$ $\mathcal{P}\mathcal{A}$, then:

 $\Gamma_i =  \textit{forward}_{e1,towhom}(\tilde{X}:D, \textit{Purposes}, t) 
 \Longrightarrow
\exists \ k \ | \ \exists \ t' \ | \ \Gamma_k =  \textit{fwconsent}_{e2, e3}(\tilde{X}:D, \textit{Purposes}, \textit{ToWhom}, t') \wedge (t' < t)$, where \textit{towhom} $\subseteq$ \textit{ToWhom}, and OWNER($\tilde{X}$) = $E_{e3}$, and either $E_{e1}$ = $E_{e2}$ or $E_{e1}$ $\neq$ $E_{e2}$.\\ 

\item $A_4$: If $\{\textit{ManDelete}_{e1}(\tilde{X}, \textit{Places}, dd),\ \textit{DeleteReq}_{e2, e1}(\tilde{X})\}$ $\subseteq$ $\mathcal{P}\mathcal{A}$, then:

$\Gamma_i =  \textit{mandelete}_{e1}(\tilde{X}:D, \textit{Places}, \textit{dd}, t) 
 \Longrightarrow
\exists \ k \  | \ \exists \ t' \ | \exists\ \textit{DEL} = \{\textit{deletereq}_{e_i, e_j}(\tilde{X} : D, t_n)\} \ |\ \Gamma_k =  \textit{deletereq}_{e2, e1}(\tilde{X} : D, t')$ $\wedge$  (\textit{deletereq}$_{e2, e1}(\tilde{X} : D, t')$ $\in$ \textit{DEL}) $\wedge$ ($t'$ < $t$) $\wedge$  ($t$ - $t'$ $\leq$ $dd$).\\ 

\item $A_5$: If $\{$$\textit{AutDelete}_{e1}(\tilde{X}, \textit{Places}, dd)$, $\textit{UnRegister}_{e2,e3}(\textit{pservices}, \textit{typeserv})$, $\textit{Register}_{e2,e3}(\textit{pservices}, \textit{typeserv})$$\}$ $\subseteq$ $\mathcal{P}\mathcal{A}$, then:

$\Gamma_i =  \textit{autdelete}_{e1}(\tilde{X}:D, \textit{Places}, \textit{dd}, t)
 \Longrightarrow
\exists \ k \ | \ \exists \ t'\ | \ \exists \ t''  \ | \ \Gamma_k =  \textit{unregister}_{e2,e3}(\textit{pservices}, \textit{typeserv}, t') \wedge \textit{register}_{e2,e3}(\textit{pservices}, \textit{typeserv}, t'')\  \wedge\ (t'' < t' < t)\ \wedge\ (t-t' < dd)$.\\ 

\item $A_6$: If $\{\textit{Collect}_{e1, e2}(\tilde{X}, \textit{Purposes}),\ \textit{Declare}_{e3, e4}(\tilde{X}, \textit{DeclParam})\}$ $\subseteq$ $\mathcal{P}\mathcal{A}$, then: 

$\Gamma_i = \textit{collect}_{e1, e2}(\tilde{X}:D, \textit{Purposes}, t) \Longrightarrow
\exists \ k \ | \ \exists \ t' \ | \ \Gamma_k =  \textit{declare}_{e3, e4}(\tilde{X}:D, \textit{DeclParam}, t')$ $\wedge$ $(t' < t)$, where $E_{e3}$ = $E_{e1}$ or $E_{e3}$ $\neq$ $E_{e1}$, $E_{e2}$ = $E_{e4}$ or $E_{e2}$ $\neq$ $E_{e4}$, and OWNER($\tilde{X}$) = $E_{e4}$.\\  

\item $A_7$: If $\{\textit{Register}_{e1,e2}(\textit{pservices}, \textit{typeserv}, t), \textit{Collect}_{e3, e4}(\tilde{X}, \textit{Purposes})\}$ $\subseteq$ $\mathcal{P}\mathcal{A}$, then: 

$\Gamma_i =  \textit{register}_{e1,e2}(\textit{pservices}, \textit{typeserv}, t)
 \Longrightarrow
\exists \ k \ | \ \exists \ t' \ | \ \Gamma_k =  \textit{collect}_{e3, e1}(\tilde{X}:D, \textit{Purposes}, t')\ \wedge\ \textit{Purposes} \subseteq \textit{pservices}\ \wedge\ \textit{TYPE}(\tilde{X})\ \in\ \textit{typeserv}\ \wedge\ (t' < t)$, and either $E_{e2}$ = $E_{e3}$ or $E_{e2}$ $\neq$ $E_{e3}$.\\ 

\end{itemize}

\end{minipage}
}
\caption{DPR trace compliance rules in architectures.}\label{fig:comparch0}
\end{figure}
  
\begin{itemize}

\item $\textbf{A}_1$: This rule basically says that any data collection made by $E_{e1}$ should be preceded by a corresponding consent collection from the owner of the data. We recall that the owner of the data is the entity who inputs the data into the system. The rational behind the option $E_{e3}$ $\neq$ $E_{e1}$ is that the service provider can consist of several smaller components, and the component that collects the data can be different from the component that collects the consent. The case is similar for $E_{e2}$ and $E_{e4}$, as the data can be collected either from the owner or from some component of the owner.  

\item $\textbf{A}_2$: This rule says that when a data of value $D$ is used by the entities in \textit{bywhom}, the corresponding usage consent should have been collected from the owner of the data such that \textit{bywhom} is the subset of the set defined in the given consent. 

\item $\textbf{A}_3$: This rule is similar to $A_2$ but related to data forwarding.  

\item $\textbf{A}_4$: This rule says that whenever there is a deletion request from an entity $E_{e2}$ to $E_{e1}$, then this should be fulfilled within the pre-specified delay $dd$.  \textit{DEL} is a set of \textit{deletereq} events in the trace $\Gamma$. An architecture can contain several \textit{DeleteReq} activities, and the deletion request can be forwarded by several entities until it reaches $E_{e1}$. Hence, \textit{deletereq}$_{e2,e1}$($\tilde{X} : D$, $t'$) $\in$ \textit{DEL} means that $E_{e1}$ receives the request directly from $E_{e2}$, but $E_{e2}$ can also receive the deletion request from  another entity, and so on.

\item $\textbf{A}_5$:  Whenever an automated deletion takes place by an entity $E_{e1}$, there should be a corresponding event \textit{unregister}$_{e2,e3}$ that triggers the deletion delay, and  
also a corresponding event \textit{register}$_{e2,e3}$ before that in $\Gamma$.

\item $\textbf{A}_6$:  This rule says that whenever there is a collection of a personal data $\tilde{X}$, the parameters related to the \textit{Purposes}, \textit{Places}, \textit{dd}, \textit{TT}, \textit{ByWhom}, \textit{ToWhom}, must be declared before the collection to achieve transparency.       

\item $\textbf{A}_7$:  This rule captures the data minimisation requirement of data collection. Namely, the purposes of the data collection should conform with the chosen services during registration. 
\end{itemize}  

The corresponding rule $C_1$ in the policy level cannot be defined in the architecture level as it requires a comparison with the element \textit{revdata} in the policy level, which is not part of the architecture. For similar reason, the rules $C_5$, $C_8$ cannot be defined in the architecture level as they need comparison with elements in a policy.


Again, the $\mathcal{P}\mathcal{A}$ compliancy property is relevant to verify if a specific system implementation (e.g., protocol, system log procedure) complies with a given architecture. We let 

\begin{center}
\noindent\fbox{%
    \parbox{6.2cm}{%
  \ \ \ \ $\mathcal{A}$ = \{$A_1$, $A_2$, $A_3$, $A_4$, $A_5$, $A_6$, $A_7$\}
     }
}
\end{center} 

\noindent \textbf{Trace rule-compliance relation}: Let us denote by $\textit{seq}_{\theta}$ $\prec^{rule}_{arch}$ $A_i$ the fact that a given operation trace $\textit{seq}_{\theta}$ $\in$ \textit{SysImpOp}($\theta$) of a system implementation \textit{SysImp}, on a data of type $\theta$, is in compliance with a rule $A_i$ $\in$ $\mathcal{A}$.

\begin{ttd}$($\textbf{trace $\pi_{\theta}$-compliance}, $\textit{seq}_{\theta}$ $\prec_{arch}$ $\mathcal{A}$$)$

\noindent An operation trace $\textit{seq}_{\theta}$ $\in$ \textit{SysImpOp}($\theta$) of a system implementation \textit{SysImp}, on a data of type $\theta$, is in compliance with a DPR sub-policy $\pi_{\theta}$ if it satisfies all the properties in $\mathcal{A}$, $\forall$$A_i$ $\in$ $\mathcal{A}$: $\textit{seq}_{\theta}$ $\prec^{rule}_{arch}$ $A_i$.  This is denoted by $($$\textit{seq}_{\theta}$ $\prec_{arch}$ $\mathcal{A}$$)$.  
\end{ttd}

\begin{ttd}$($\textbf{system $\pi_{\theta}$-compliance}, \textit{SysImp} $\prec^{sys}_{arch}$ $\mathcal{A}$$)$

\noindent A system implementation \textit{SysImp} is compliant with a DPR sub-policy $\pi_{\theta}$ if $\forall$$\textit{seq}_{\theta}$ $\in$ \textit{SysImpOp}($\theta$): $\textit{seq}_{\theta}$ $\prec_{arch}$ $\mathcal{A}$.  We denote this by $($\textit{SysImp} $\prec^{sys}_{arch}$ $\mathcal{A}$$)$.   
\end{ttd}

\begin{ttd}$($\textbf{system $\mathcal{P}\mathcal{A}$-compliance, \textit{SysImp} $\models_{arch}$ $\mathcal{P}\mathcal{A}$}$)$

\noindent A system implementation \textit{SysImp} is in compliance with an architecture $\mathcal{P}$$\mathcal{A}$ if 

$\forall$ \textit{seq}$_{\theta}$ $\in$ \textit{SysImpOp($\theta$)}, and  $\forall$$\theta$ $\in$ \textit{TYPE}$^{SP}_{\mathcal{P}\mathcal{A}}$$:$  we have \textit{seq}$_{\theta}$ $\prec_{arch}$ $\mathcal{A}$. This is denoted by $($\textit{SysImp} $\models_{arch}$ $\mathcal{P}\mathcal{A}$$)$.   
\end{ttd}

\section{The Conformance Between Policies and Architectures}
\label{conformance} 
We distinguish three types of conformance: (i) privacy conformance, (ii) conformance with regards to the data protection regulation, and (iii) functional conformance. Privacy conformance compares the policies and the architectures based on privacy properties, namely, if at the policy level (based on a defined policy) an entity \textit{never have} a given data, then this is also true in the corresponding architecture, and vice versa. The conformance with regards to the data protection regulation deals with data protection requirements, such as consent collection, deletion/retention delay, as well as data storage for a given data. Finally, functional conformance compares the policies and the architectures based on which entities at the policy level and their corresponding entities at the architecture level have the same data type. 

In the following definition, we recall the notion of the low-level system implementations (\textit{SysImp} and \textit{SysImpOp}) outlined in the Definition~\ref{def:sysimp}.

\begin{ttd}$($\textbf{DPR conformance between $\mathcal{P}\mathcal{L}$ and $\mathcal{P}\mathcal{A}$}$)$
\label{paplconform}
\noindent We say that an architecture $\mathcal{P}\mathcal{A}$ is in DPR conformance with a given policy $\mathcal{P}\mathcal{L}$, denoted by $\mathcal{P}\mathcal{A}$ $\triangleright_{dpr}$ $\mathcal{P}\mathcal{L}$, if for all system implementations \textit{SysImp} and their \textit{SysImpOp},  and $\forall$$\theta$ $\in$ \textit{TYPE}$^{SP}_{\mathcal{P}\mathcal{L}}$ and the same  $\theta$ $\in$ \textit{TYPE}$^{SP}_{\mathcal{P}\mathcal{A}}$$:$
\begin{enumerate}
\item  whenever $\exists$\textit{seq}$_\theta$ $\in$ \textit{SysImpOp}($\theta$) such that \textit{seq}$_\theta$ $\prec^{rule}_{pol}$ $C_i$,  where 
$C_i$ $\in$ $\mathcal{C}$ $\backslash$ $\{$$C_1$, $C_5$, $C_8$$\}$, then 
$\exists$$A_j$ $\in$  $\mathcal{A}$, such that \textit{seq}$_\theta$ $\prec^{rule}_{arch}$ $A_j$, and 

\begin{enumerate}
\item \textit{Storerev}$_{e}$($X_{|\theta}$, \textit{Places}, \textit{TT}) $\in$ $\mathcal{P}\mathcal{A}$ $\wedge$  \textit{TT} $\in$ $\pi_{\theta}.\pi_{str}.\textit{revdate}$; and  
\item \textit{Use}$_{\textit{bywhom}}$($X_{|\theta}$, \textit{Purposes}) $\in$ $\mathcal{P}\mathcal{A}$ $\wedge$ \textit{Purposes} $\subseteq$ $\pi_{\theta}.\pi_{use}.upurp.purpset_u$
and \textit{bywhom} $\subseteq \pi_{\theta}.\pi_{use}.\textit{whouse}.\textit{who}$; and

\item \textit{Forward}$_{e,\textit{towhom}}$($X_{|\theta}$, \textit{Purposes}) $\in$ $\mathcal{P}\mathcal{A}$ $\wedge$ \textit{Purposes} $\subseteq$ $\pi_{\theta}.\pi_{fw}.\textit{fwpurp}$ 
$\wedge$  \textit{towhom} $\subseteq$ $\pi_{\theta}.\pi_{fw}.\textit{3rdparty}$,
\item If \textit{seq}$_\theta$ $\prec^{rule}_{pol}$ $C_2$, then \textit{seq}$^{par}_\theta$ $\prec^{rule}_{arch}$ $A_6$, where \textit{seq}$^{par}_\theta$ is the same as \textit{seq}$_\theta$, but the action  in \textit{seq}$^{par}_\theta$ corresponding to the event \textit{declare}$_{e3,e4}$($\tilde{X}$$:$$D$, \textit{DeclParam}, $t'$) in $A_6$ does not have the following parameters in the set \textit{DeclParam}: $\pi_{str}$.\textit{ho}.\textit{declparam1}, $\pi_{str}$.\textit{wh}.\textit{declparam2}, and $\pi_{del}$.\textit{how}.\textit{declparam3} $($where \textit{declparam1} $\in$ $\{$\textit{hidden}, \textit{nohidden}$\}$,\textit{declparam2} $\in$ $\{$\textit{sploc}, \textit{clientloc}$\}$, \textit{declparam3} $=$ $\{$$p_1$, $p_2$$\}$ as defined in the Section~\ref{sec:syntaxpp}$)$.      
\end{enumerate} 
\end{enumerate}   
\end{ttd}

The Definition~\ref{paplconform} says that for a low-level implementation whenever an operation trace $seq_{\theta}$ on the data type $\theta$ satisfies a policy compliance rule, then it also satisfies a corresponding architecture compliance rule. The reason behind not taking into account the rules $C_1A$, $C_5$, and $C_8$ is because there is no corresponding architecture rule in $\mathcal{A}$. Instead, to fullfil $C_1$ we have the point 1./(a), for      $C_5$ we have 1./(b), and for $C_8$ we have 1./(c).  Finally, 1./(d) is related to the fulfillment of the rules $C_2$ and $A_6$, however, since the architecture does not contains information about the values $\pi_{str}$.\textit{ho}.\textit{declparam1}, $\pi_{str}$.\textit{wh}.\textit{declparam2}, and $\pi_{del}$.\textit{how}.\textit{declparam3}, we remove them in the set \textit{DeclParam} in case of $A_6$ in the actions in $seq_{\theta}$.  Basically, here we allow a ``weak'' conformance that tolerates the difference between the declared parameters for transparency.  

\subsection{Mapping Between the Policy Level and the Architecture Level}
\label{mapping}

We provide a mapping between the policy level to the architecture level, and some formal properties of the mapping. 

\textit{Syntax Mapping}: 
The set of entities in the policy level can be mapped to the set of entities in the architecture as follows: 

\begin{ttd} $($Entity mapping$)$

\noindent For a given service provider system $SP$, let \textit{Entity}$^{SP}_{\mathcal{P}\mathcal{L}}$  and  \textit{Entity}$^{SP}_{\mathcal{A}\mathcal{P}}$ be the set of defined entities in the policy and architecture, respectively.  
We define a mapping function $\mathbb{M}$ that maps an $e$, $e$ $\in$ \textit{Entity}$^{SP}_{\mathcal{P}\mathcal{L}}$, to a corresponding $E_e$, $E_e$ $\in$ \textit{Entity}$^{SP}_{\mathcal{P}\mathcal{A}}$.       

\begin{center}
$\mathbb{M}$$:$ $e$  $\leftrightarrow_{\mathbb{M}}$ $E_e$, $($i.e., $\mathbb{M}(e)$ $=$ $E_e$, and  $\mathbb{M}(E_e)$ $=$ $e$$)$.

$\mathbb{M}\mathbb{S}$: $\{$$e_1$,\dots, $e_m$$\}$  $\leftrightarrow_{\mathbb{M}\mathbb{S}}$  $\{$$E_{e_1}$,\dots, $E_{e_m}$$\}$, where $\{$$e_1$,\dots, $e_m$$\}$ $\subseteq$ \textit{Entity}$^{SP}_{\mathcal{P}\mathcal{L}}$, and $\{$$E_{e_1}$,\dots, $E_{e_m}$$\}$ $\subseteq$ \textit{Entity}$^{SP}_{\mathcal{P}\mathcal{A}}$. 
\end{center}

\end{ttd}

Then, we define the data mapping relation between the policy level and the architecture level as follows: 

\begin{ttd} $($Data mapping$)$
    
\noindent Let Data$_{\mathcal{P}\mathcal{L}}$ and Data$_{\mathcal{A}\mathcal{P}}$ be the set of the defined data in the policy level and architecture level, respectively. 
We define a mapping function $\mathbb{D}$ that maps a data $dt_{|_{\theta}}$, $dt_{|_{\theta}}$ $\in$ \textit{Data}$_{\mathcal{P}\mathcal{L}}$, to a corresponding $X_{|_{\theta}}$, $X_{|_{\theta}}$ $\in$ \textit{Data}$_{\mathcal{P}\mathcal{A}}$.       

\begin{center}
$\mathbb{D}$$:$ $dt_{|_{\theta}}$  $\leftrightarrow_{\mathbb{D}}$ $X_{|_{\theta}}$.

\end{center}    
\end{ttd} 

Finally, the following two definitions provide a strict and a loose mapping relation between the policy and the corresponding architecture.

\begin{ttd} $($\textbf{\underline{Strict-mapping} between policies and architectures}$)$
\label{map}    
\noindent Let us consider the policy $\mathcal{P}\mathcal{L}$ and the architecture $\mathcal{P}\mathcal{A}$. We say that there is a strict-mapping between $\mathcal{P}\mathcal{L}$ and $\mathcal{P}\mathcal{A}$, denoted by $\mathcal{P}\mathcal{A}$ $\rightarrow_{\mathbb{C}_{strict}}$ $\mathcal{P}\mathcal{L}$, if \textit{Services}$^{SP}_{\mathcal{P}\mathcal{L}}$ $\subseteq$ \textit{Services}$^{SP}_{\mathcal{P}\mathcal{A}}$, and for each $\pi_{\theta}$ $\in$ $\mathcal{P}\mathcal{L}$, we have:

\begin{enumerate} 

\item If $\exists$ $\pi_{\theta}$.$\pi_{col}$, where $\pi_{col}$.\textit{cons} = Y, and $\pi_{col}$.\textit{cpurp}.\textit{decl} = Y then
\begin{itemize} 
\item $\exists$ $X_{|_{\theta}}$ $\in$ \textit{Data}$_{\mathcal{P}\mathcal{A}}$, $\exists$ $E_{e1}$,  $E_{e3}$, $E_{e3'}$ $\in$ \textit{PartOf}($E_{sp}$), where  $E_{e1}$ = $E_{e3}$ or $E_{e1}$ $\neq$ $E_{e3}$, OWNER($X_{|_{\theta}}$) = $E_{e4}$, $E_{e3}$ = $E_{e3'}$ or $E_{e3'}$ $\neq$ $E_{e3}$ and 

\item $\{$\textit{Collect}$_{e1,e2}$$(X_{|_{\theta}}$, \textit{CPurposes}$)$, \textit{CConsent}$_{e3,e4}$$(X_{|_{\theta}}$, \textit{CPurposes}), 

\textit{Declare}$_{e3',e4}$$(X_{|_{\theta}}$, \textit{DeclParam}$)\}$ $\subseteq$ $\mathcal{P}\mathcal{A}$ such that \textit{CPurposes} $\subseteq$ \textit{DeclParam}, and \textit{CPurposes} $\in$  $\pi_{col}$.\textit{cpurp}.\textit{purpset}$_c$.   
\end{itemize}

\item If $\exists$ $\pi_{\theta}$.$\pi_{use}$, where $\pi_{use}$.\textit{cons} = Y, 
 and $\pi_{use}$.\textit{upurp}.\textit{decl} = Y,  and 
 
 $\pi_{use}$.\textit{whouse}.\textit{decl} = Y then 
 
\begin{itemize} 
\item $\exists$$X_{|_{\theta}}$, $\exists$$E_{e1}$, $E_{e1'}$ $\in$ \textit{PartOf}($E_{sp}$), $E_{e1}$ = $E_{e1'}$ or $E_{e1'}$ $\neq$ $E_{e1}$, OWNER($X_{|_{\theta}}$) = $E_{e2}$ and 

\item $\{$\textit{Use}$_{\textit{bywhom}}$$($$X_{|_{\theta}}$, \textit{UPurposes}), \textit{UConsent}$_{e1,e2}$$($$X_{|_{\theta}}$, \textit{UPurposes}, \textit{ByWhom}), \textit{Declare}$_{e1',e2}$$(X_{|_{\theta}}$, \textit{DeclParam}$)\}$ $\subseteq$ $\mathcal{P}\mathcal{A}$ such that  \textit{UPurposes} $\cup$ \textit{ByWhom} $\subseteq$ \textit{DeclParam}, and \textit{UPurposes} $\in$  $\pi_{col}$.\textit{upurp}.\textit{purpset}$_u$, and \textit{ByWhom}  $\subseteq$ $\pi_{use}$.\textit{whouse}. 

\end{itemize} 

\item If $\exists$ $\pi_{\theta}$.$\pi_{str}$, and 

\begin{enumerate} 
\item $\pi_{str}$.\textit{wh} = $($\textit{clientloc}, \textit{places}, Y$)$ then $\exists$ $X_{|_{\theta}}$,  $\exists$$E_e$ $\notin$ \textit{PartOf}($E_{sp}$), $\exists$$E_e'$ $\in$ \textit{PartOf}($E_{sp}$) and  $\{$\textit{Store}$_{e}$$(X_{|_{\theta}}$, \textit{Places}), \textit{Declare}$_{e',e}$$(X_{|_{\theta}}$, \textit{DeclParam}$)\}$ $\subseteq$ $\mathcal{P}\mathcal{A}$, where \textit{Places} $\subseteq$ \textit{DeclParam}. 

\item $\pi_{str}$.\textit{wh} = $($\textit{sploc}, \textit{places}, Y$)$ and $\pi_{str}$.\textit{ho} = $($\textit{nohidden}, Y$)$ then $\exists$$X_{|_{\theta}}$, $\exists$$E_{e1}$, $E_{e2}$ $\in$ \textit{PartOf}($E_{sp}$), $\exists$$E_{e3}$ $\notin$ \textit{PartOf}($E_{sp}$), OWNER($X_{|_{\theta}}$) = $E_{e3}$, and $\{$\textit{Store}$_{e1}$$(X_{|_{\theta}}$, \textit{Places}$)$, \textit{Declare}$_{e2,e3}$$(X_{|_{\theta}}$, \textit{DeclParam})$\}$ $\subseteq$ $\mathcal{P}\mathcal{A}$, where \textit{Places} $\subseteq$ \textit{DeclParam}, and \textit{Places} $\subseteq$ \textit{places}. 

\item $\pi_{str}$.\textit{wh} = $($\textit{sploc}, \textit{places}, Y$)$ and $\pi_{str}$.\textit{ho} = $($\textit{nohidden}, Y$)$ then 

\begin{itemize} 
\item $\exists$$X_{|_{\theta}}$, $\exists$$E_{e1}$, $E_{e2}$ $\in$ \textit{PartOf}($E_{sp}$), where $E_{e1}$ = $E_{e2}$ or $E_{e1}$ $\neq$ $E_{e2}$ and 

\item $\{$\textit{Store}$_{e1}$$(X_{|_{\theta}}$, \textit{Places}$)$, \textit{Compute}$_{e2}$$($$\tilde{X}$ $=$ F$(X_{|_{\theta}})$$)$$\}$ $\subseteq$ $\mathcal{P}\mathcal{A}$, where \textit{Places} $\subseteq$ \textit{places} and 

\item $\exists\ \tilde{X}_1, \dots, \tilde{X}_n, G: BackgroundDeduction_{e3}(G(F(X_{|_{\theta}}), \tilde{X}_1, \dots, \tilde{X}_n) \rightarrow X_{|_{\theta}})\ \wedge 
\mathcal{P}\mathcal{A} \vdash_{\exists t} \textit{HAS}_{e3}\left(\tilde{X}_1, t\right)\ \wedge \ \dots \ \wedge\ \textit{HAS}_{e3}\left(\tilde{X}_n, t\right)$, where $E_{e3}$ $\in$ \textit{PartOf}($E_{sp}$), and $E_{e3}$ and $E_{e1}$ or $E_{e2}$ can be the same.

\end{itemize} 

\item $\pi_{str}$.\textit{wh} = $($\textit{sploc}, \textit{places}, Y$)$ and $\pi_{str}$.\textit{ho} = $($\textit{hidden}, Y$)$ then \begin{itemize} 
\item $\exists$$X_{|_{\theta}}$, $\exists$$E_{e1}$, $E_{e2}$ $\in$ \textit{PartOf}($E_{sp}$), where $E_{e1}$ = $E_{e2}$ or $E_{e1}$ $\neq$ $E_{e2}$ and 

\item $\{$\textit{Store}$_{e1}$$(X_{|_{\theta}}$, \textit{Places}$)$,  \textit{Declare}$_{e2,e3}$$(X_{|_{\theta}}$, \textit{DeclParam})$\}$ $\subseteq$ $\mathcal{P}\mathcal{A}$,  where where \textit{Places} $\subseteq$ \textit{DeclParam}, and \textit{Places} $\subseteq$ \textit{places}, OWNER($X_{|_{\theta}}$) = $E_{e3}$
\end{itemize}

\item $\pi_{str}$.\textit{wh} = $($\textit{sploc}, \textit{places}, Y$)$ and $\pi_{str}$.\textit{ho} = $($\textit{hidden}, Y$)$ then \begin{itemize} 
\item $\exists$$X_{|_{\theta}}$, $\exists$$E_{e1}$, $E_{e2}$ $\in$ \textit{PartOf}($E_{sp}$), where $E_{e1}$ = $E_{e2}$ or $E_{e1}$ $\neq$ $E_{e2}$ and 

\item $\{$\textit{Store}$_{e1}$$(X_{|_{\theta}}$, \textit{Places}$)$, \textit{Compute}$_{e2}$$($$\tilde{X}$ $=$ F$(X_{|_{\theta}})$$)$$\}$ $\subseteq$ $\mathcal{P}\mathcal{A}$,  where \textit{Places} $\subseteq$ \textit{places}, and 

\item $\nexists\ \tilde{X}_1, \dots, \tilde{X}_n, G: BackgroundDeduction_{e3}(G(F(X_{|_{\theta}}), \tilde{X}_1, \dots, \tilde{X}_n) \rightarrow X_{|_{\theta}})\ \wedge 
\mathcal{P}\mathcal{A} \vdash_{\exists t} \textit{HAS}_{e3}\left(\tilde{X}_1, t\right)\ \wedge \ \dots \ \wedge\ \textit{HAS}_{e3}\left(\tilde{X}_n, t\right)$, where $E_{e3}$ $\in$ \textit{PartOf}($E_{sp}$), and $E_{e3}$ and $E_{e1}$ or $E_{e2}$ can be the same.

\end{itemize}

\item $\pi_{str}$.\textit{storerev} = $($$[$t1, t2 $]$, \textit{places}, Y$)$ then $\exists$$X_{|_{\theta}}$, $\exists$$E_{e1}$, $E_{e2}$ $\in$ \textit{PartOf}($E_{sp}$), OWNER($X_{|_{\theta}}$) = $E_{e3}$ and $\{$\textit{Storerev}$_{e1}$($X_{|_{\theta}}$, $PL$, \textit{TT}), \textit{Declare}$_{e2,e3}$$(X_{|_{\theta}}$, \textit{DeclParam})$\}$ $\subseteq$ $\mathcal{P}\mathcal{A}$, where $PL$ $\subseteq$ \textit{DeclParam}, and $($$PL$ = \textit{places}$)$ and $($t1 $\leq$ \textit{TT} $\leq$ t2$)$.  
\end{enumerate}

\item If $\exists$ $\pi_{\theta}$.$\pi_{del}$, and 
\begin{enumerate}
\item $\pi_{del}$.\textit{how} = $($\textit{man}, \textit{full}, Y$)$ and $\pi_{del}$.\textit{deld} = $($$dd$, \textit{decl}$)$ then 
$\exists$ $X_{|_{\theta}}$: $\{$\textit{DeleteReq}$_{e1,e2}$$($$X_{|_{\theta}}$$)$, \textit{ManDelete}$_{e2}$$($$X_{|_{\theta}}$, $\{$E$_{Main_{sp}}$, E$_{BckUp_{sp}}$$\}$, $dd$$)$,  \textit{Declare}$_{e3,e4}$$(X_{|_{\theta}}$, \textit{DeclParam})$\}$ $\subseteq$ $\mathcal{P}\mathcal{A}$, where $\{$E$_{Main_{sp}}$, E$_{BckUp_{sp}}$, $dd$$\}$ $\subseteq$ \textit{DeclParam}. 

\item $\pi_{del}$.\textit{how} = $($\textit{man}, \textit{partly}, Y$)$ and $\pi_{del}$.\textit{deld} = $($$dd$, \textit{decl}$)$ then 
$\exists$ $X_{|_{\theta}}$: $\{$\textit{DeleteReq}$_{e1,e2}$$($$X_{|_{\theta}}$$)$, \textit{ManDelete}$_{e2}$$($$X_{|_{\theta}}$, $\{$E$_{Main_{sp}}$$\}$, $dd$$)$, \textit{Declare}$_{e3,e4}$$(X_{|_{\theta}}$, \textit{DeclParam})$\}$ $\subseteq$ $\mathcal{P}\mathcal{A}$, where  $\{$E$_{Main_{sp}}$, $dd$$\}$ $\subseteq$ \textit{DeclParam}.

\item $\pi_{del}$.\textit{how} = $($\textit{aut}, \textit{full}, Y$)$ and $\pi_{del}$.\textit{gdeld} = $($$($\textit{activity}, $gd$$)$, Y$)$ then $\exists$ $X_{|_{\theta}}$:
$\{$\textit{Unregister}$_{e2,e3}$(\textit{pservices}, \textit{typeserv}), \textit{AutDelete}$_{e1}$$($$X_{|_{\theta}}$, $\{$E$_{Main_{sp}}$, E$_{BckUp_{sp}}$$\}$, $gd$$)$, \textit{Declare}$_{e3,e4}$$(X_{|_{\theta}}$, \textit{DeclParam})$\}$ $\subseteq$ $\mathcal{P}\mathcal{A}$, where $\{$E$_{Main_{sp}}$, E$_{BckUp_{sp}}$, $gd$$\}$ $\subseteq$ \textit{DeclParam}.

\item $\pi_{del}$.\textit{how} = $($\textit{aut}, \textit{partly}, Y$)$ and $\pi_{del}$.\textit{gdeld} = $($$gd$, Y$)$ then $\exists$ $X_{|_{\theta}}$: 

$\{$\textit{Unregister}$_{e2,e3}$(\textit{pservices}, \textit{typeserv}), \textit{AutDelete}$_{e1}$$($$X_{|_{\theta}}$, $\{$E$_{Main_{sp}}$$\}$, $gd$$)$, \textit{Declare}$_{e3,e4}$$(X_{|_{\theta}}$, \textit{DeclParam})$\}$ $\subseteq$ $\mathcal{P}\mathcal{A}$, where $\{$E$_{Main_{sp}}$, $gd$$\}$ $\subseteq$ \textit{DeclParam}.
\end{enumerate}

\item If $\exists$ $\pi_{\theta}$.$\pi_{fw}$, and  
$\pi_{fw}$ = $($Y, \textit{fwpurp}, \textit{3rdparty}$)$ then $\exists$$X_{|_{\theta}}$: 

$\{$\textit{Forward}$_{e1,\textit{towhom}}$$($$X_{|_{\theta}}$, \textit{Purposes}$)$,  \textit{FwConsent}$_{e2,e3}$$($$X_{|_{\theta}}$, \textit{Purposes}, \textit{ToWhom}), \textit{Declare}$_{e3,e4}$$(X_{|_{\theta}}$,  \textit{DeclParam})$\}$ $\subseteq$ $\mathcal{P}\mathcal{A}$, where $\{$\textit{Purposes}, \textit{ToWhom}$\}$ $\subseteq$ \textit{DeclParam},  \textit{Purposes} $\subseteq$  $\pi_{fw}$.\textit{fwpurp}, and \textit{ToWhom} $\subseteq$ $\pi_{fw}$.\textit{3rdparty}.     
\end{enumerate}

\noindent where  for the types and entities in the points above, we have the following assumptions:

\begin{itemize}
\item each $\theta$ $\in$ \textit{TYPE}$^{SP}_{\mathcal{P}\mathcal{L}}$ is the same as each $\theta$ $\in$ \textit{TYPE}$^{SP}_{\mathcal{P}\mathcal{A}}$. 

\item  $\exists$ $\mathbb{M}$: $sp$ $\leftrightarrow_{\mathbb{M}}$ $E_{sp}$, $e$ $\leftrightarrow_{\mathbb{M}}$ $E_e$, $e'$ $\leftrightarrow_{\mathbb{M}}$ $E_{e'}$, $e_1$ $\leftrightarrow_{\mathbb{M}}$ $E_{e1}$, $e_{1'}$ $\leftrightarrow_{\mathbb{M}}$ $E_{e1'}$, $e_2$ $\leftrightarrow_{\mathbb{M}}$ $E_{e2}$,  $e_3$ $\leftrightarrow_{\mathbb{M}}$ $E_{e3}$, $e_{3'}$ $\leftrightarrow_{\mathbb{M}}$ $E_{e3'}$,  $e_4$ $\leftrightarrow_{\mathbb{M}}$ $E_{e4}$.
\end{itemize}

\end{ttd}

\begin{ttd} $($\textbf{\underline{Loose-mapping} between policies and architectures}$)$
\label{maploose} 
Let us consider the policy $\mathcal{P}\mathcal{L}$ and the architecture $\mathcal{P}\mathcal{A}$. We say that there is a loose-mapping between $\mathcal{P}\mathcal{L}$ and $\mathcal{P}\mathcal{A}$, denoted by $\mathcal{P}\mathcal{A}$ $\rightarrow_{\mathbb{C}_{loose}}$ $\mathcal{P}\mathcal{L}$, if it fulfills all the requirements in the Definition~\ref{map}, except for the points 3/(c), 3/(d), and 3/(e). 

\end{ttd}  

The loose-mapping is lenient with regards to the form a data is stored. The intuition behind this is that although making the content of the collected data avaiable to a service provider is not perfect from privacy perspectives,  it does not always violate the data protection regulations as sometimes a service provider needs to know the data content to ba able to provide the service or for accountability purposes. To conform with the data protection regulations, a service provider can still protect the data by encrypting it with its own key.  

\subsection{Conformances between policies and architectures}
\label{conformances}

As mentioned before, we distinguish three types of conformance: (i) privacy conformance, (ii) conformance with regards to the data protection regulation (DPR), and (iii) functional conformance. In the rest of the paper, we refer to the second case as \textit{DPR conformance}. Further, we also distinguish between strong and weak DPR conformances. 

\begin{ttd} $($\textbf{Privacy conformance}$)$
Let a policy $\mathcal{P}\mathcal{L}$ = $\{\pi_{\theta_1}, \dots, \pi_{\theta_m}\}$ defined in the policy language $\mathcal{P}_{DC}$, and an   
architecture $\mathcal{P}\mathcal{A}$ in the architecture language $\mathcal{A}_{DC}$. Further, let \textit{PartOf} be a function defined on \textit{Entity}$^{SP}_{\mathcal{P}\mathcal{A}}$.  We say that $\mathcal{P}\mathcal{A}$ conforms with $\mathcal{P}\mathcal{L}$ with regards to the privacy property, denoted by  $\mathcal{P}\mathcal{A}$ $\triangleright_{priv}$ $\mathcal{P}\mathcal{L}$:

If for a given $e$ and $dt_{|_{\theta}}$, where $\theta$ $\in$ $\{$$\theta_1$, \dots, $\theta_m$$\}$, $e$ $\in$ \textit{Entity}$^{SP}_{\mathcal{P}\mathcal{L}}$ and $dt_{|_{\theta}}$ $\in$ \textit{Data}$^{SP}_{\mathcal{P}\mathcal{L}}$,  we have that if $[e$ $\neg$has $dt_{|_{\theta}}]$$_{\mathcal{P}\mathcal{L}}$, then 
\begin{enumerate}
\item $\exists$ $\mathbb{M}$, $\mathbb{D}$: $e$ $\leftrightarrow_\mathbb{M}$ $E_e$,  and  $dt_{|_{\theta}}$ $\leftrightarrow_\mathbb{D}$ $X_{|_{\theta}}$, such that $\mathcal{P}\mathcal{A}$ $\vdash$ $\textit{HAS}_e^\textit{never} \left(X_{|_{\theta}} \right)$; and 

\item $\forall$ $E_i$ $\in$ \textit{Entity}$_{\mathcal{P}\mathcal{A}}$, $E_i$ $\neq$ $E_e$$:$ \textit{PartOf}($E_e$) = $E_i$,     then $\mathcal{P}\mathcal{A}$ $\vdash$ $\textit{HAS}_i^\textit{never} \left(X_{|_{\theta}} \right)$.
\end{enumerate}

\end{ttd}





\begin{ttpp}\label{prop:1} $($\textbf{DPR conformance - Strict mapping}$)$

\noindent Given a policy $\mathcal{P}\mathcal{L}$ = $\{\pi_{\theta_1}, \dots, \pi_{\theta_m}\}$ defined in the policy language $\mathcal{P}_{DC}$, and an   
architecture $\mathcal{P}\mathcal{A}$ in the architecture language $\mathcal{A}_{DC}$.  $\mathcal{P}\mathcal{A}$ strongly DPR conforms with  $\mathcal{P}\mathcal{L}$, $\mathcal{P}\mathcal{A}$ $\triangleright_{dpr}$ $\mathcal{P}\mathcal{L}$, if for each $\pi_{\theta}$ $\in$ $\mathcal{P}\mathcal{L}$, $\exists$ $\mathbb{C}_{strict}$: $\mathcal{P}\mathcal{A}$ $\rightarrow_{\mathbb{C}_{strict}}$ $\mathcal{P}\mathcal{L}$.
\end{ttpp}
\begin{proof}
Following  the Definition~\ref{paplconform}, let \textit{SysImp} be a system implementation, and let  \textit{SysImpOp}($\theta$) be the system operation with a data $dt$ of type $\theta$ ($dt_{|\theta}$).  
We will show that if $\exists$ $\mathbb{C}_{strict}$: $\mathcal{P}\mathcal{A}$ $\rightarrow_{\mathbb{C}_{strict}}$ $\mathcal{P}\mathcal{L}$ holds, then the Definition~\ref{paplconform} is fulfilled.

Due to the strict mapping properties in the Definition~\ref{map}, whenever \textit{SysImpOp}($\theta$) fulfils the rule $A_1$ in the Fig.~\ref{fig:comparch0} it also fulfils the rule $C_3$ in the Fig.~\ref{fig:compsn0}. Similarly, whenever \textit{SysImpOp}($\theta$) satisfies $A_2$, $A_3$, $A_4$,  $A_5$, $A_6$, $A_7$, it also satisfies the corresponding rules $C_4$, $C_{9}$,  $C_{6}$,  $C_{7}$, $C_{2}$, and $C_{10}$, respectively.    These come from the assumptions made in the Definition~\ref{map}, on the parameter sets such as \textit{purposes}, \textit{towhom}, \textit{bywhom}, etc. Further, point 2 of the Definition~\ref{paplconform} is satisfied because of the point 3/(e) of the Definition~\ref{map}. Point 3 of the Definition~\ref{paplconform} is satisfied because of the point 2 of the Definition~\ref{map}. Finally,  point 4 of the Definition~\ref{paplconform} is satisfied because of the point 5 of the Definition~\ref{map}. 
\end{proof}

\begin{ttpp}\label{prop:2} $($\textbf{DPR conformance - Loose mapping}$)$

\noindent Given a policy $\mathcal{P}\mathcal{L}$ = $\{\pi_{\theta_1}, \dots, \pi_{\theta_m}\}$ defined in the policy language $\mathcal{P}_{DC}$, and an   
architecture $\mathcal{P}\mathcal{A}$ in the architecture language $\mathcal{A}_{DC}$.  $\mathcal{P}\mathcal{A}$ DPR conforms with  $\mathcal{P}\mathcal{L}$, $\mathcal{P}\mathcal{A}$ $\triangleright_{dpr}$ $\mathcal{P}\mathcal{L}$, if for each $\pi_{\theta}$ $\in$ $\mathcal{P}\mathcal{L}$, $\exists$ $\mathbb{C}_{loose}$: $\mathcal{P}\mathcal{A}$ $\rightarrow_{\mathbb{C}_{loose}}$ $\mathcal{P}\mathcal{L}$.
\end{ttpp}
\begin{proof}
The proof is similar to the previous proposition. This proposition says that to achieve DPR conformance, it is enough to satisfy the loose mapping requirements without requiring personal data always being stored hidden from the service provider. 
\end{proof}

\begin{ttd}\label{func} $($\textbf{Functional conformance}$)$
Given a policy $\mathcal{P}\mathcal{L}$ = $\{\pi_{\theta_1}, \dots, \pi_{\theta_m}\}$ defined in the policy language $\mathcal{P}_{DC}$, and an   
architecture $\mathcal{P}\mathcal{A}$ in the architecture language $\mathcal{A}_{DC}$. We say that $\mathcal{P}\mathcal{A}$ conforms with $\mathcal{P}\mathcal{L}$ with regards to functionality, denoted by  $\mathcal{P}\mathcal{A}$ $\triangleright_{func}$ $\mathcal{P}\mathcal{L}$,  we have:

If for a given $e$ and $dt_{|_{\theta}}$ such that $e$ $\in$ \textit{Entity}$^{SP}_{\mathcal{P}\mathcal{L}}$ and $dt_{|_{\theta}}$ $\in$ \textit{Data}$^{SP}_{\mathcal{P}\mathcal{L}}$, we have $[e$ has $dt_{|_{\theta}}]$$_{\mathcal{P}\mathcal{L}}$, then for the corresponding $E_e$ and  $X_{|_{\theta}}$ in the architecture level $($$\exists$ $\mathbb{M}$, $\mathbb{D}$: $e$ $\leftrightarrow_\mathbb{M}$ $E_e$,  and  $dt_{|_{\theta}}$ $\leftrightarrow_\mathbb{D}$ $X_{|_{\theta}}$$)$, we have $\mathcal{P}\mathcal{A} \vdash_{\exists t} \textit{HAS}_{e}\left(X_{|_{\theta}}, t \right)$.  

\end{ttd}

\section{Case Study: A Privacy Policy and Architecture for a Smart Metering Service} 
\label{pol:smartmeter}

The following is an excerpt from the ICO guideline~\cite{icoper} on personal data: 
\begin{ttd}
\label{pdt}
A data can be seen as a personal data if it satisfies some of the following points: 
\begin{enumerate}
\item ``A living individual \underline{\textit{can be identified}} from the data, or, from the data and other information in your possession, or likely to come into your possession'', and 

\item ``the data \textit{`\underline{relates to}'} the identifiable living
individual, whether in personal or family life,
business or profession'', or 

\item ``the data is \underline{\textit{used}}, or is it \underline{\textit{to be used}}, to \underline{\textit{inform or influence}} actions or decisions affecting an identifiable individual''
\end{enumerate}
\end{ttd}

\subsection{Policy Level}
Let denote the service provider by \textit{SP} referring to an energy service provider, and let the set of services it provides be: 

\begin{center}
\noindent\fbox{%
    \parbox{9cm}{%
\textit{PServices}$^{SP}_{\mathcal{P}\mathcal{L}}$ = \{\textit{reg} (\textit{registration}),\\
\tab[2.6cm]\textit{ecr} (\textit{energy consumption reading}),\\ 
\tab[2.6cm]\textit{cds} (\textit{consumption data storage}),\\ 
\tab[2.6cm]\textit{ecn} (\textit{energy consumption notification}),\\  
\tab[2.6cm]\textit{bc} (\textit{balance calculation}),\\
\tab[2.6cm]\textit{bn} (\textit{balance notification}),\\
\tab[2.6cm]\textit{bil} (\textit{billing})\}\\
\tab[2.6cm]\textit{ref} (\textit{refund})\}.    
}
}
\end{center} 
 
The corresponding set of supported data types are: 

\begin{center}
\noindent\fbox{%
    \parbox{12cm}{%
    \begin{center} 
\textit{TYPE}$^{SP}_{\mathcal{P}\mathcal{L}}$ = \{$\theta_{pi}$, $\theta_{ec}$, $\theta_{bal}$, $\theta_{bill}$, $\theta^{dshide}_{ec}$, $\theta^{dshide}_{bal}$\}    
  \end{center}   
}
}
\end{center} 

where $\theta_{pi}$ $=$ \textit{personal info} (which covers  the information required at the registration phase e.g., names, phone number, address, birth date, etc.), $\theta_{ec}$ $=$ \textit{energy consumption reading}, and $\theta_{bal}$ = \textit{balance}, $\theta_{bill}$ = type fill the bills. 

\textit{Type for data hidden from the service provider}: For each type $\theta$, we define its corresponding data type $\theta^{dshide}$. The service provider will not be able to link a data of this type with any involved data subject in it, namely,  the ID of a \textit{data subject} is hidden from the service provider.  For instance, $\theta^{dshide}_{ec}$ is the type of an energy consumption reading in which the information about the related customer is hidden from the service provider (the content of the reading is available, but the (real) identity of the data subject  not). Of course, for some data type such as personal information giving at the registration phase, the data subject should not be hidden. Finally, bills should also be available to the service provider for accountability purposes, e.g., in case of dispute.   

Based on the set of supported data types, the privacy policy defined for $SP$ 
is as follows: 

\begin{center}
\noindent\fbox{%
    \parbox{11cm}{%
    \begin{center}
$\mathcal{P}\mathcal{L}$ = \{$\pi_{\theta_{pi}}$, $\pi_{\theta_{ec}}$, $\pi_{\theta_{bal}}$, $\pi_{\theta_{bill}}$,  $\pi_{\theta^{dshide}_{ec}}$, $\pi_{\theta^{dshide}_{bal}}$\}   
\end{center} 
     }
}
\end{center} 

The system is composed of three categories of entities: the \textit{service provider}, \textit{customers}, and the\textit{ authority} (as a third-party organisation) to whom the service provider forward the data in case of any criminal offence or dispute. Hence   

\begin{center}
\noindent\fbox{%
    \parbox{4.8cm}{%
\textit{Entity}$^{SP}_{\mathcal{P}\mathcal{L}}$ = \{\textit{sp}, \textit{cust},  \textit{auth}\}    
     }
}
\end{center}  

For the policies  $\pi_{\theta_{ec}}$, $\pi_{\theta_{bal}}$, data collection, storage, usage, deletion and forwarding are not allowed (\textit{NAD}), namely:

\begin{center}
\noindent\fbox{%
    \parbox{8cm}{%
  \ \ \ $\pi_{\theta_{ec}}$ $=$ $\pi_{\theta_{bal}}$ $=$ (\textit{NAD}, \textit{NAD}, \textit{NAD}, \textit{NAD}, \textit{NAD})     
     }
}
\end{center} 

\subsubsection{Personal information} 

For personal information, $\pi_{\theta_{pi}}$, we have  

\begin{center}
\noindent\fbox{%
    \parbox{5cm}{%
$\pi_{\theta_{pi}}$ $=$ ($\pi_{col}$, $\pi_{use}$, $\pi_{str}$, $\pi_{del}$, $\pi_{fw}$)     
     }
}
\end{center}  

with the following sub-policies: 

\begin{enumerate}
\item For data collection policy, $\pi_{\theta_{pi}}$.$\pi_{col}$, we have 
\begin{itemize}
\item \textit{cons} = $Y$, based on the GDPR, a consent is required when collecting personal data.  
\item  \textit{cpurp}.\textit{purpset}$_c$ = \{\textit{reg}\}, meaning that the personal info is collected for registration purposes.  
\item \textit{cpurp}.\textit{decl} = $Y$, meaning that the collection purposes are declared towards the customers, and available to them (transparency).  
\end{itemize}

\item For data usage policy, $\pi_{\theta_{pi}}$.$\pi_{use}$, we have 
\begin{itemize}
\item \textit{cons} = $Y$ 
\item  \textit{cpurp}.\textit{purpset}$_u$ = \{\textit{ecr}, \textit{gcr}, \textit{bil}\}, meaning that the personal info is used for electricity, gas consumption reading, notification and billing purposes. 
\item \textit{cpurp}.\textit{decl} = $Y$.  
\end{itemize}

\item For data storage policy, $\pi_{\theta_{pi}}$.$\pi_{str}$, we have 
\begin{itemize} 
\item \textit{wh} = (\textit{sploc}, \{\textit{Main}, \textit{BckUp}\}, $Y$), meaning that the personal info is stored at the service provider's (\textit{EN}) site, both in the main and backup storage places. This has been also declared to the customers (data subject).     

\item \textit{ho} $=$ (\textit{hidden}, $Y$), meaning that the personal information is stored hidden from the service provider (unknown to service provider), and this information should be declared to the customers. 

\item \textit{revdate} = ([t1, t2], \{\textit{Main}, \textit{BckUp}\}, $Y$), where t1 = $tx_{col}$ $+$ $n$ $\times$ $2$ years, where n={1, 2,\dots,} and t2 = t1 $+$ $1$ month. Namely, the review period for the need of storing personal info, in both the main and backup servers, is  after each 2 years from the first storing time, and should be carried out within 1 month. $tx_{col}$ is a time variable capturing the time that a personal info is collected for the first time.
\end{itemize}   

\item For data deletion policy, $\pi_{\theta_{pi}}$.$\pi_{del}$, we have 
\begin{itemize} 
\item \textit{how} $=$ ((\textit{aut}, \textit{full}), (\textit{man}, (\textit{partly}, \{\textit{bil}, \textit{ref}\})), $Y$). Namely, the policy defines both automated and fully as well as manual and partly deletion modes for personal info carried out by the service provider, and this fact is declared for the customer. In the manual mode, the customers can request deletion of their personal info when they initiate the request (e.g., by clicking on delete button) from the services, however, the policy says that the deletion request by the customer only takes place from the main servers, but not from the backup places for billing and refunding reasons. The personal info will only be deleted fully from the backup places when it is no longer necessary for any service purposes.   

\item \textit{deld} = (\textit{1m}, $Y$) represents the delay for the deletion after the customer initiates the request. The personal info must be deleted from the main servers within 1 minute from the request.        

\item \textit{gdeld} = (\{\textit{storerev}, \textit{DF}\}, $Y$). After a review of the necessity of storage, the personal info will be fully deleted from the backup places after a defined but not numerical delay.    
\end{itemize}   

\item For data forwarding, $\pi_{\theta_{pi}}$.$\pi_{fw}$, we have  

\begin{itemize}
\item ($Y$, \{\textit{bil, ref}\}, \{\textit{auth}\}), in case of dispute/offence on billing, the personal info of the particular customer is forwarded to the authority for billing (clarification) purposes. Again, consent should be collected.     
\end{itemize}   
\end{enumerate}

\subsubsection{Energy Bills}  For the bills, $\pi_{\theta_{bill}}$, we have 

\begin{center}
\noindent\fbox{%
    \parbox{6.7cm}{%
\ \ \ \ \ \ \ \ \ $\pi_{\theta_{bill}}$ $=$ ($\pi_{use}$, $\pi_{str}$, $\pi_{del}$, $\pi_{fw}$)     
     }
}
\end{center} 

There is no collection policy for $\pi_{\theta_{bill}}$ because a bill is calculated instead of collected. However, it can be seen as a personal data since it shows the amount a related customer should pay, hence let defines its sub-policies as follows: 

\begin{enumerate}
\item For data usage, $\pi_{\theta_{bill}}$.$\pi_{use}$ we have 
\begin{itemize}
\item \textit{cons} = $Y$, consent is required for bills as they are personal data. Bills usually contain the customer's name, address and consumption info.   
\item  \textit{cpurp}.\textit{purpset}$_u$ = \{\textit{bil}\}
\item \textit{cpurp}.\textit{decl} = $Y$.  
\end{itemize}

\item For data storage, $\pi_{\theta_{bill}}$.$\pi_{str}$ we have 
\begin{itemize} 
\item \textit{wh} = (\textit{sploc}, \{\textit{Main}, \textit{BckUp}\}, $Y$).     

\item \textit{ho} $=$ (\textit{unhidden}, $Y$). The bills are stored at the service provider's site, unhidden from the service provider (e.g., encrypted with a key available to the service provider). 

\item \textit{revdate} = ([t1, t2], \{\textit{Main}, \textit{BckUp}\}, $Y$), where t1 = $tx_{str}$ $+$ $n$ $\times$ $2$ years, where n={1, 2,\dots,} and t2 = t1 $+$ $1$ month. Namely, the review period for the need of storing bills is   after each 2 years from the first storing time, and should be carried out within 1 month. $tx_{str}$ is a time variable capturing the time that a bill is stored for the first time.
\end{itemize}   

\item For data deletion, $\pi_{\theta_{bill}}$.$\pi_{del}$, we have 
\begin{itemize} 
\item \textit{how} $=$ ((\textit{aut}, \textit{full}), (\textit{man}, (\textit{partly}, \{\textit{bil}, \textit{ref}\})), $Y$).
 
\item \textit{deld} = (\textit{1 min}, $Y$).  When a customer unregisters from the system the bills are deleted from the main servers but not from the backup servers. 

\item \textit{gdeld} = (\{\textit{unregister}, \textit{DF}\}, $Y$).  The  bills are kept by the service provider in its backup servers until they are required for the services (i.e., there is defined delay (DF) but not a numerical value).
\end{itemize}   

\item For data forwarding, $\pi_{\theta_{bill}}$.$\pi_{fw}$ we have  

\begin{itemize}
\item ($Y$, \{\textit{bil, ref}\}, \{\textit{auth}\}). Bills are forwarded to the authority in case of dispute for billing or refunding purposes.
\end{itemize}   

\end{enumerate}

\subsubsection{Hidden Energy Consumption Readings} 
The hidden energy consumption reading data have the same policy, hence for $\pi^{dshide}_{\theta_{ec}}$, we have 

\begin{center}
\noindent\fbox{%
    \parbox{6.7cm}{%
\ \ \ \ \ \ $\pi_{\theta^{dshide}_{ec}}$ $=$ ($\pi_{col}$, $\pi_{use}$, $\pi_{str}$, $\pi_{del}$, $\pi_{fw}$)     
     }
}
\end{center}  

where, because the data subject is hidden/unlinkable with the readings, the sub-policies are:  

\begin{enumerate}
\item For data collection, $\pi_{\theta^{dshide}_{ec}}$.$\pi_{col}$, we have 
\begin{itemize}
\item \textit{cons} = $N$, because it's not a personal data.  
\item  \textit{cpurp}.\textit{purpset}$_c$ = \{\textit{ecn},  \textit{cds}, \textit{bc}, \textit{bil}\}, meaning that the consumption data is collected for notification, storage, balance calculation and billing purposes.  
\item \textit{cpurp}.\textit{decl} = $Y$.  
\end{itemize}

\item For data usage, $\pi_{\theta^{dshide}_{ec}}$.$\pi_{use}$, we have 
\begin{itemize}
\item \textit{cons} = $N$ 
\item  \textit{cpurp}.\textit{purpset}$_u$ = \{\textit{ecn},   \textit{bc}, \textit{bil}\}. 
\item \textit{cpurp}.\textit{decl} = $Y$.  
\end{itemize}

\item For data storage, $\pi_{\theta^{dshide}_{ec}}$.$\pi_{str}$, we have 
\begin{itemize} 
\item \textit{wh} = (\textit{sploc}, \{\textit{Main}, \textit{BckUp}\}, $Y$).     

\item \textit{ho} $=$ (\textit{unhidden},  $Y$). The content of the reading is available to the service provider (but not the data subject).

\item \textit{revdate} = ([t1, t2], \{\textit{Main}, \textit{BckUp}\}, $Y$), where t1 and t2 are similar as in $\pi_{\theta_{pi}}$.   
\end{itemize}   

\item For data deletion, $\pi_{\theta^{dshide}_{ec}}$.$\pi_{del}$, we have 
\begin{itemize} 
\item \textit{how} $=$ ((\textit{aut}, \textit{full}), (\textit{man}, (\textit{partly}, \{\textit{bil}, \textit{ref}\})), $Y$). 

\item \textit{deld} = (\textit{1 min}, $Y$).  After a customer initiates a deletion request, the reading will be deleted from the main server of the service provider within 1 minute, but not from the backup server.          

\item \textit{gdeld} = (\{\textit{unregister}, \textit{DF}\}, $Y$).     The deletion from the back up server is defined (\textit{DF}), but again, it is not a numerical value.  
\end{itemize}   

\item For data forwarding, $\pi_{\theta^{dshide}_{ec}}$.$\pi_{fw}$, we have  

\begin{itemize}
\item ($Y$, \{\textit{bil, ref}\}, \{\textit{auth}\}). Consent is still required for this when forwarded to the authority as the authority will be able to link the (real) customer with the reading, in case of dispute.    
\end{itemize}   

\end{enumerate}

\subsubsection{Hidden Temporary Balance} 
For a temporary balance, $\pi_{\theta^{dshide}_{bal}}$, we have 

\begin{center}
\noindent\fbox{%
    \parbox{6.7cm}{%
\ \ \ \ \ \ \ \ \ $\pi_{\theta^{dshide}_{bal}}$ $=$ ($\pi_{use}$, $\pi_{str}$, $\pi_{del}$, $\pi_{fw}$)     
     }
}
\end{center} 

Again, there is no collection policy for $\pi_{\theta^{dshide}_{bal}}$ because a temporary balance is calculated. Further, it is not a personal data since the link between data subject and balance is hidden from the service provider, hence: 

\begin{enumerate}
\item For data usage, $\pi_{\theta^{dshide}_{bal}}$.$\pi_{use}$, we have 
\begin{itemize}
\item \textit{cons} = $N$. 
\item  \textit{cpurp}.\textit{purpset}$_u$ = \{\textit{bn}\}, meaning that the usage purpose is notifying the customer about the current balance. 
\item \textit{cpurp}.\textit{decl} = $Y$.  
\end{itemize}

\item For data storage, $\pi_{\theta^{dshide}_{bal}}$.$\pi_{str}$, we have 
\begin{itemize} 
\item \textit{wh} = (\textit{sploc}, \{\textit{Main}, \textit{BckUp}\}, $Y$).     

\item \textit{ho} $=$ (\textit{unhidden}, $Y$). The temporary balances are stored at the service provider's site, encrypted with the service provider's key. 

\item \textit{revdate} = ([t1, t2],  \{\textit{Main}, \textit{BckUp}\}, $Y$), where t1 and t2 are similar as in $\pi_{\theta_{pi}}$.   
\end{itemize}   

\item For data deletion, $\pi_{\theta^{dshide}_{bal}}$.$\pi_{del}$, we have 
\begin{itemize} 
\item \textit{how} $=$ (\{(\textit{aut}, \textit{full})\})\}, $Y$). The service provider keeps these temporary  balances for notification purpose whenever the customer download them. 

\item \textit{gdeld} = (\{\textit{unregister}, \textit{1 day}\}, $Y$).  Once the customers unregister themselves, they do not need to download these pieces of data anymore, hence the service provider deletes them within a short time period, e.g., 1 day.  
\end{itemize}   

\item For data forwarding, $\pi_{\theta^{dshide}_{bal}}$.$\pi_{fw}$, we have  $\pi_{\theta^{dshide}_{bal}}$.$\pi_{fw}$ = \textit{NAD}. 

\begin{itemize}
\item A temporary consumption balance is not allowed to be forwarded to any third-party organisation. This is only for notifying a user about their curent consumption to help them save money. In case of dispute, the authority will use the forwarded bills and energy consumption readings for verification purposes.    
\end{itemize}

\end{enumerate}

\subsection{The Architecture Level and Conformance Checks} 
\label{arch:smartmeter}

Let us consider an example architectures on which we will demonstrate  the specification based on the proposed language, as well as carry out a conformance checks against the policy above.

\subsubsection{Architecture} 
\label{smartarch1} 

\begin{figure}[htb!]
    \begin{center}
        \includegraphics[width=1\textwidth]{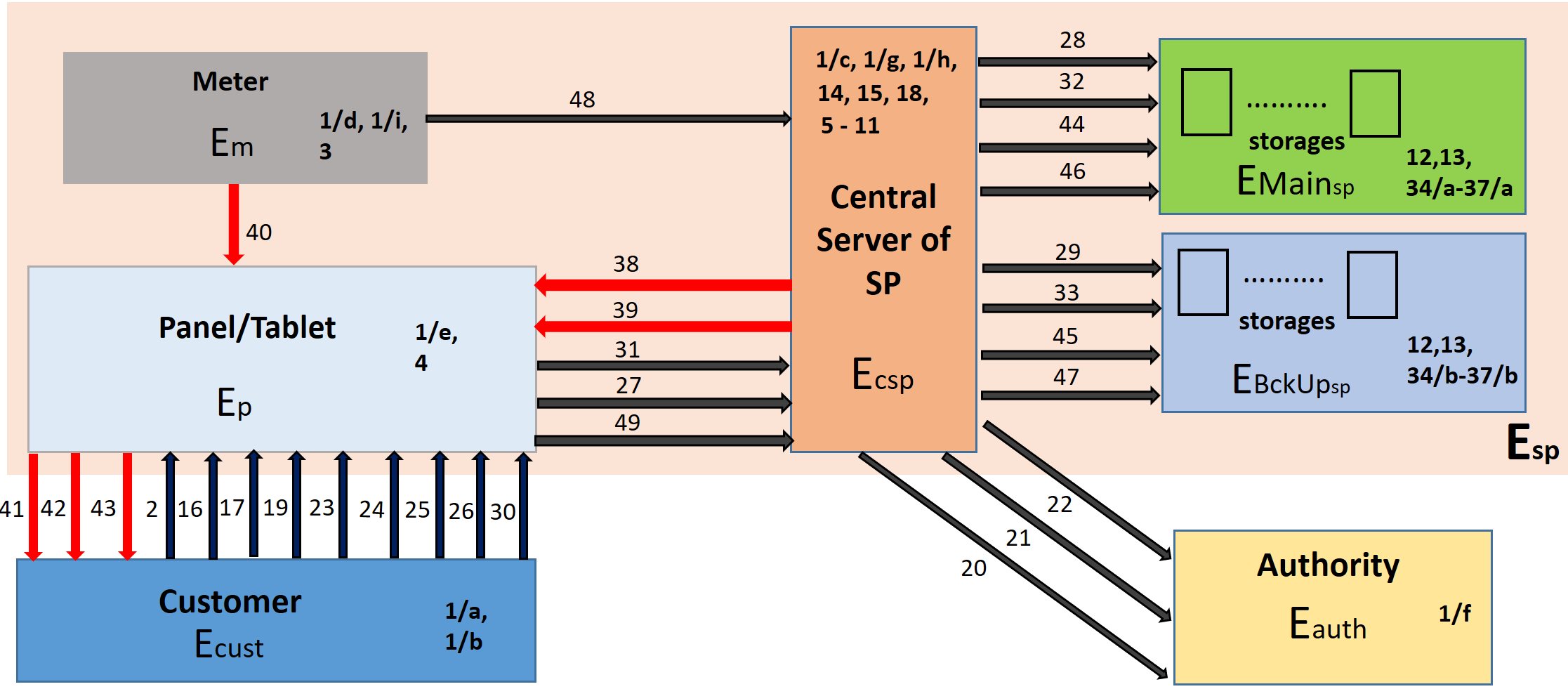}
    \end{center}
    \caption{The architecture of the first example smart meter system. We assume a long-term  customer ID (cust), which is valid throughout the service. The arrows capture the communication and direction of message exchange between entities, while the numbers on the arrows and the boxes represent the corresponding activities in the Fig.~\ref{fig:smartarch1}.}
    \label{fig:smartmeter1}
\end{figure}
 
The architecture ($\mathcal{P}\mathcal{A}_{1}$) is formally defined as follows:  

\begin{enumerate}
\item Let \textit{Entity}$^{SP}_{\mathcal{P}\mathcal{A}_1}$ = \{$E_{csp}$, $E_{m}$, $E_{p}$, $E_{\textit{Main}_{sp}}$, $E_{\textit{BckUp}_{sp}}$, $E_{cust}$, $E_{auth}$, $E_{sp}$\}, where 
 \textit{PartOf}($E_{sp}$) = \{$E_{csp}$, $E_{m}$, $E_{p}$, $E_{\textit{Main}_{sp}}$, $E_{\textit{BckUp}_{sp}}$\}. This says that  $E_{csp}$, $E_{m}$, $E_{p}$, $E_{\textit{Main}_{sp}}$, and $E_{\textit{BckUp}_{sp}}$ are parts of the service provider $E_{sp}$. $E_{csp}$ represents central server of the service provider, $E_{m}$ represents a smart meter installed at the customers premises, and $E_{p}$ the tablet/panel used by the customer to keep track of the consumption, and get notifications. $E_{\textit{Main}_{sp}}$ and  $E_{\textit{BckUp}_{sp}}$ represent the main and backup servers of the service provider, while $E_{cust}$ and $E_{auth}$ the customer and the authority, respectively. The architecture is depicted in the Fig~\ref{fig:smartmeter1}.
 
\item Let \textit{Pservices}$^{SP}_{\mathcal{P}\mathcal{A}_1}$ = \textit{Pservices}$^{SP}_{\mathcal{P}\mathcal{L}_1}$.

\item Let  \textit{TYPE}$^{SP}_{\mathcal{P}\mathcal{A}_1}$ = \{$\theta_{pi}$, $\theta_{ec}$, $\theta_{bal}$, $\theta_{bill}$, $\theta_{Ckey}$, $\theta_{Skey}$, $\theta_{SAkey}$, $\theta_{\textit{tariff}}$, $\theta_{ecSkey}$, $\theta_{piSkey}$,  $\theta_{piSAkey}$, $\theta_{balSAkey}$, $\theta_{billSAkey}$\}, where the first four types are the same as the first four types in \textit{ TYPE}$^{SP}_{\mathcal{P}\mathcal{L}}$. The following  $\theta_{Ckey}$, $\theta_{Skey}$ and $\theta_{SAkey}$ are the types of the customer keys, the server keys and the shared keys between the CSP server and the authority's server. $\theta_{\textit{tariff}}$ is the type of the current tariff for calculation the current balance. Finally, $\theta_{ecSkey}$, $\theta_{piSkey}$,  $\theta_{piSAkey}$, $\theta_{balSAkey}$ and $\theta_{billSAkey}$ are the types of the energy reading, personal information, balance and bill encrypted with the data of types $\theta_{Skey}$, $\theta_{Skey}$, $\theta_{SAkey}$, $\theta_{SAkey}$ and $\theta_{SAkey}$, respectively.
\end{enumerate}

A customer is assigned a long-term ID (cust), and this is shared with the service provider. The smart meter ($E_{m}$) regularly sends the energy consumption readings to the central server ($E_{csp}$). Based on this information, the central server calculates the temporary balance and then sends it to the panel for notifying the customer.  A monthly bill will be issued for the customer at the end.      

Let the set of services provided by the architecture, \textit{PServices}$^{SP}_{\mathcal{P}\mathcal{A}}$, is the same as the services defined in the policy, \textit{PServices}$^{SP}_{\mathcal{P}\mathcal{L}}$: \textit{PServices}$^{SP}_{\mathcal{P}\mathcal{A}}$ $=$ \textit{PServices}$^{SP}_{\mathcal{P}\mathcal{L}}$. Formally, $\mathcal{P}\mathcal{A}_{1}$ is defined by the set of \textit{activities} in the Fig.~\ref{fig:smartarch1}.  

\begin{figure}[htbp]
\centering
\fbox{\begin{minipage}{15.87 cm}
\begin{tabbing}  
    \=1\=1\=1\= \kill\\
    \>\> $\mathcal{P}\mathcal{A}_{SP}$ ::=  \{\textbf{1/a.} \textit{Own}$_{cust}$($X_{|\theta_{pi}}$), \textbf{1/b.} 
    \textit{Own}$_{cust}$($X_{|\theta_{Ckey}}$), \textbf{1/c.} \textit{Own}$_{csp}$($X_{|\theta_{Skey}}$), 
    \\ 
\ \ \ \ \textbf{1/d.} \textit{Own}$_{m}$($X_{|\theta_{Skey}}$), \textbf{1/e.} \textit{Own}$_{p}$($X_{|\theta_{Skey}}$), \textbf{1/f.}
    \textit{Own}$_{auth}$($X_{|\theta_{SAkey}}$),\\ 
\ \ \ \ \textbf{1/g.} \textit{Own}$_{csp}$($X_{|\theta_{SAkey}}$), \textbf{1/h.} \textit{Own}$_{csp}$($X_{|\theta_{\textit{tariff}}}$), \textbf{1/i.} \textit{Own}$_{m}$($X_{|\theta_{ec}}$),\\   
\ \ \ \ 2. \textit{Register}$_{cust, p}$(\textit{pservices}, \textit{typeserv}), where   \textit{pservices} $\subseteq$   \textit{PServices}$^{SP}_{\mathcal{P}\mathcal{A}}$,  \\
\ \ \ \ 3. \textit{Compute}$_{m}$($X_{|\theta_{ecSkey}}$ $=$ \textit{Enc}$(X_{|\theta_{ec}}, X_{|\theta_{Skey}})$),\\ 
\ \ \ \ 4. \textit{Compute}$_{p}$($X_{|\theta_{piSkey}}$ $=$ \textit{Enc}$(X_{|\theta_{pi}}, X_{|\theta_{Skey}})$),\\ 
\ \ \ \ 5. \textit{Compute}$_{csp}$($X_{|\theta_{ecSAkey}}$ $=$ \textit{Enc}$(X_{|\theta_{ec}}, X_{|\theta_{SAkey}})$),\\ 
\ \ \ \ 6. \textit{Compute}$_{csp}$($X_{|\theta_{piSAkey}}$ $=$ \textit{Enc}$(X_{|\theta_{pi}}, X_{|\theta_{SAkey}})$),\\ 
\ \ \ \ 7. \textit{Compute}$_{csp}$($X_{|\theta_{bal}}$ $=$ \textit{Bal}$(X_{|\theta_{ec}}, X_{|\theta_{\textit{tariff}}})$),\\ 
\ \ \ \ 8. \textit{Compute}$_{csp}$($X_{|\theta_{balSAkey}}$ $=$ \textit{Enc}$(X_{|\theta_{bal}}, X_{|\theta_{SAkey}})$),\\ 
\ \ \ \ 9. \textit{Compute}$_{csp}$($X_{|\theta_{bill}}$ $=$ \textit{Bill}$(X_{|\theta_{ec}}, X_{|\theta_{\textit{tariff}}})$),\\ 
\ \ \ \ 10. \textit{Compute}$_{csp}$($X_{|\theta_{billSAkey}}$ $=$ \textit{Enc}$(X_{|\theta_{bill}}, X_{|\theta_{SAkey}})$),\\
\ \ \ \ 11. \textit{Compute}$_{csp}$($X_{|\theta_{billSkey}}$ $=$ \textit{Enc}$(X_{|\theta_{bill}}, X_{|\theta_{Skey}})$),\\
\ \ \ \ 12. \textit{Store}$_{\{Main_{sp}, BckUp_{sp}\}}$($X_{|\theta_{ecSkey}}$),\\ 
\ \ \ \ 13. \textit{Store}$_{\{Main_{sp}, BckUp_{sp}\}}$($X_{|\theta_{piSkey}}$),\\ 
\ \ \ \ 14. \textit{Storerev}$_{csp}$($X_{|\theta_{ecSkey}}$, \{$E_{\textit{Main}_{sp}}$,  $E_{\textit{BckUp}_{sp}}$\}, \textit{tt}),\\ 
\ \ \ \ 15. \textit{Storerev}$_{csp}$($X_{|\theta_{piSkey}}$, \{$E_{\textit{Main}_{sp}}$,  $E_{\textit{BckUp}_{sp}}$\}, \textit{tt}),
\\ \ \ \ \ 16. \textit{Collect}$_{p,cust}$($X_{|\theta_{pi}}$, \textit{Purposes}), \textit{Purposes} = \{\textit{reg}\},\\
\ \ \ \  17. \textit{CConsent}$_{p,cust}$($X_{|\theta_{pi}}$, \textit{Purposes}), \textit{Purposes} = \{\textit{reg}\},\\ 
\ \ \ \ 18/a. \textit{Use}$_{\textit{csp}}$($X_{|\theta_{pi}}$, \textit{Purposes}), \textit{Purposes} = \{\textit{ecr},   \textit{gcr}, \textit{bil}\},\\ 
\ \ \ \ 18/b. \textit{Use}$_{\textit{csp}}$($X_{|\theta_{bill}}$, \textit{Purposes}), \textit{Purposes} = \{\textit{ecr},   \textit{gcr}, \textit{bil}\},\\ 
\ \ \ \ 19. \textit{UConsent}$_{p,cust}$($X_{|\theta_{pi}}$, \textit{Purposes}, \{$E_{sp}$\}), \textit{Purposes} = \{\textit{ecr},   \textit{gcr}, \textit{bil}\},\\ 
\ \ \ \ 20. \textit{Forward}$_{csp,\{\textit{auth}\}}$($X_{|\theta_{piSAkey}}$, \textit{Purposes}), \textit{Purposes} = \{\textit{bil},  \textit{ref}\},\\ 
\ \ \ \ 21. \textit{Forward}$_{csp,\{\textit{auth}\}}$($X_{|\theta_{ecSAkey}}$, \textit{Purposes}), \textit{Purposes} = \{\textit{bil},  \textit{ref}\},\\ 
\ \ \ \ 22. \textit{Forward}$_{csp,\{\textit{auth}\}}$($X_{|\theta_{billSAkey}}$, \textit{Purposes}), \textit{Purposes} = \{\textit{bil},  \textit{ref}\},\\ 
\ \ \ \ 23. \textit{FwConsent}$_{p,cust}$($X_{|\theta_{piSAkey}}$, \textit{Purposes}, \{$E_{auth}$\}), \textit{Purposes} = \{\textit{bil},  \textit{ref}\},\\ 
\ \ \ \ 24. \textit{FwConsent}$_{p,cust}$($X_{|\theta_{ecSAkey}}$,  \textit{Purposes} = \{\textit{bil},  \textit{ref}\},\\ 
\ \ \ \ 25. \textit{FwConsent}$_{p,cust}$($X_{|\theta_{billSAkey}}$, \textit{Purposes}, \{$E_{auth}$\}),  \textit{Purposes} = \{\textit{bil},  \textit{ref}\},\\ 
\ \ \ \ 26. \textit{DeleteReq}$_{cust, p}$($X_{|\theta_{piSkey}}$),\\
\ \ \ \ 27. \textit{DeleteReq}$_{p, csp}$($X_{|\theta_{piSkey}}$),\\
\ \ \ \ 28. \textit{DeleteReq}$_{csp, \textit{Main}_{sp}}$($X_{|\theta_{piSkey}}$),\\
\ \ \ \ 29. \textit{DeleteReq}$_{csp, \textit{BckUp}_{sp}}$($X_{|\theta_{piSkey}}$),\\
\ \ \ \ 30. \textit{DeleteReq}$_{cust, p}$($X_{|\theta_{ecSkey}}$),\\
\ \ \ \ 31. \textit{DeleteReq}$_{p, csp}$($X_{|\theta_{ecSkey}}$),\\
\ \ \ \ 32. \textit{DeleteReq}$_{csp, \textit{Main}_{sp}}$($X_{|\theta_{ecSkey}}$),\\
\ \ \ \ 33. \textit{DeleteReq}$_{csp, \textit{BckUp}_{sp}}$($X_{|\theta_{ecSkey}}$),\\
\ \ \ \ \textbf{34/a.} \textit{ManDelete}$_{\textit{Main}_{sp}}$($X_{|\theta_{piSkey}}$, $dd$), \textbf{34/b.} \textit{ManDelete}$_{\textit{BckUp}_{sp}}$($X_{|\theta_{piSkey}}$, $dd$),\\
\ \ \ \ \textbf{35/a.} \textit{ManDelete}$_{\textit{Main}_{sp}}$($X_{|\theta_{ecSkey}}$, $dd$), \textbf{35/b.} \textit{ManDelete}$_{\textit{BckUp}_{sp}}$($X_{|\theta_{ecSkey}}$, $dd$),\\
\ \ \ \ \textbf{36/a.} \textit{AutDelete}$_{\textit{Main}_{sp}}$($X_{|\theta_{piSkey}}$, $dd$), \textbf{36/b.} \textit{AutDelete}$_{\textit{BckUp}_{sp}}$($X_{|\theta_{piSkey}}$, $dd$),\\
\ \ \ \ \textbf{37/a}. \textit{AutDelete}$_{\textit{Main}_{sp}}$($X_{|\theta_{ecSkey}}$, $dd$), \textbf{37/b}. \textit{AutDelete}$_{\textit{BckUp}_{sp}}$($X_{|\theta_{ecSkey}}$, $dd$),\\
\ \ \ \ 38. \textit{Receive}$_{p,csp}$($X_{|\theta_{balSkey}}$),\\ 
\ \ \ \ 39. \textit{Receive}$_{p,csp}$($X_{|\theta_{billSkey}}$),\\  
\ \ \ \ 40. \textit{Receive}$_{p,m}$($X_{|\theta_{ecSkey}}$),\\
\ \ \ \ 41. \textit{Receive}$_{cust,p}$($X_{|\theta_{bill}}$),\\
\ \ \ \ 42. \textit{Receive}$_{cust,p}$($X_{|\theta_{ec}}$),\\  
\ \ \ \ 43. \textit{Receive}$_{cust,p}$($X_{|\theta_{bal}}$).\\  
\ \ \ \ 44. \textit{Receive}$_{\textit{Main}_{sp},csp}$($X_{|\theta_{ecSkey}}$),\\  
\ \ \ \ 45. \textit{Receive}$_{\textit{BckUp}_{sp},csp}$($X_{|\theta_{ecSkey}}$).\\ 
\ \ \ \ 46. \textit{Receive}$_{\textit{Main}_{sp},csp}$($X_{|\theta_{piSkey}}$),\\  
\ \ \ \ 47. \textit{Receive}$_{\textit{BckUp}_{sp},csp}$($X_{|\theta_{piSkey}}$).\\  
\ \ \ \ 48. \textit{Receive}$_{m,csp}$($X_{|\theta_{ecSkey}}$),\\  
\ \ \ \ 49. \textit{Receive}$_{p,csp}$($X_{|\theta_{piSkey}}$).\}  
\end{tabbing}
\end{minipage}
}
\caption{The architecture defined in the proposed language syntax.}\label{fig:smartarch1}
\end{figure}

\begin{figure}[htbp]
\centering
\fbox{\begin{minipage}{11.87 cm}
\begin{tabbing}    
    \=123456\=1\=1\=1\= \kill
		\>\> \underline{\textit{Destructors}} for $E_{sp}$: \\\\
		\=123456\=1\=1\=1\= \kill
		\>\>  \textbf{Dest.} \textit{Dec}(\textit{Enc}$(X_{|\theta}, X_{|\theta_{key}}), X_{|\theta_{key}}$) $\rightarrow$ $X_{|\theta}$ 
\end{tabbing}
\end{minipage}
}
\caption{Destructor defined for tservice provider. It says that the service provider has the ability decrypt the ciphertext using the right key.}\label{fig:Destructor}
\end{figure}
 
The following conformances properties hold between the architecture the policy: 

\begin{ttp} \textbf{(Privacy conformance)}
\label{propprivconf}
The architecture $\mathcal{P}$$\mathcal{A}_1$ does \textit{not privacy conform} with the policy $\mathcal{P}$$\mathcal{L}$, namely, $\mathcal{P}$$\mathcal{A}_1$ $\ntriangleright_{priv}$ $\mathcal{P}$$\mathcal{L}$.   
\end{ttp}

\begin{proof}
Let the mapping $\mathbb{M}$ be as follows: \textit{sp} $\leftrightarrow_\mathbb{M}$ $E_{sp}$, \textit{cust} $\leftrightarrow_\mathbb{M}$ $E_{cust}$, \textit{auth} $\leftrightarrow_\mathbb{M}$ $E_{auth}$.  In the policy, we have $\pi_{\theta_{ec}}$ $=$ $\pi_{\theta_{bal}}$ $=$ (\textit{NAD}, \textit{NAD}, \textit{NAD}, \textit{NAD}, \textit{NAD}), which means that the service provider is not allowed to collect, use, store, delete or forward any energy comsumption  reading, and any temporary balance. Hence, we have $[e$ $\neg$has $dt_{|_{\theta_{ec}}}]$$_{\mathcal{P}\mathcal{L}}$.     

However, in the architecture, based on the rule H5 in the Fig.
~\ref{tab:eq:axioms0} and the row 5. in the Fig.~\ref{fig:smartarch1} as well as the destructor in the Fig.~\ref{fig:Destructor}, we have $\mathcal{P}\mathcal{A} \vdash_{\exists t} \textit{HAS}_{csp}\left(X_{|_{_{\theta_{ecSAkey}}}}, t \right)$. Further, since $E_{sp}$ $\in$ \textit{PartOf}($E_{sp}$), based on the rule H15 in the Fig.
~\ref{tab:eq:axioms0}, we also have $\mathcal{P}\mathcal{A} \vdash_{\exists t} \textit{HAS}_{sp}\left(X_{|_{_{\theta_{ecSAkey}}}}, t \right)$. Then, following this and the rule H7 (Fig.
~\ref{tab:eq:axioms0}), we have $\mathcal{P}\mathcal{A} \vdash_{\exists t} \textit{HAS}_{sp}\left(X_{|_{_{\theta_{ec}}}}, t \right)$. Similarly, for  $\pi_{\theta_{bal}}$, we have $\mathcal{P}\mathcal{A} \vdash_{\exists t} \textit{HAS}_{sp}\left(X_{|_{_{\theta_{bal}}}}, t \right)$ based on the rule H5 in the Fig.
~\ref{tab:eq:axioms0} and the row 7. in the Fig.~\ref{fig:smartarch1}. 
\end{proof} 

\begin{ttp} \textbf{(DPR conformance)}
The architecture $\mathcal{P}$$\mathcal{A}_1$ is \textit{not in DPR conformance} with the policy $\mathcal{P}$$\mathcal{L}$, namely, $\mathcal{P}$$\mathcal{A}_1$ $\ntriangleright_{dpr}$ $\mathcal{P}$$\mathcal{L}$.   
\end{ttp}
\begin{proof} 
It is easy to see that the point 1./(d) of the Definition~\ref{paplconform} does not hold, as there was no activity Declare  defined in $\mathcal{P}\mathcal{A}_1$. 
Interestingly, there is not any mapping $\mathbb{C}_{strict}$ such that $\mathcal{P}\mathcal{A}$ $\rightarrow_{\mathbb{C}_{strict}}$ $\mathcal{P}\mathcal{L}$. This is because of the violation of the point 3/d of the Definition~\ref{map}, since for a data of type $\theta_{ec}$ the policy requires that it is stored hidden from (not available to) the service provider, however, this is not the case in the architecture, as the service provider can decrypt the encrypted energy readings using its own key. 

\end{proof}

\begin{ttp} \textbf{(Functional conformance)}
The architecture $\mathcal{P}$$\mathcal{A}_1$ does \textit{not functionally conform} with the policy $\mathcal{P}$$\mathcal{L}$, namely, $\mathcal{P}$$\mathcal{A}_1$ $\ntriangleright_{func}$ $\mathcal{P}$$\mathcal{L}$.   
\end{ttp}
\begin{proof} 
The proof of this property is similar to the Property~\ref{propprivconf}. It can be shown that $[e$ has $dt_{|_{\theta^{dshide}_{ec}}}]$$_{\mathcal{P}\mathcal{L}}$, but $\mathcal{P}\mathcal{A}$ $\vdash$ $\textit{HAS}_e^\textit{never} \left(X_{|_{\theta^{dshide}_{ec}}} \right)$, hence, not satisfying the requirements set in the Definition~\ref{func}.    
\end{proof} 

\section{Conclusion and Future Works}
In this paper, we proposed a policy language and an architecture languages for specifying and reasoning about data protection properties in the policy and architecture levels. We demonstrated their expressive syntax and semantics by specifying a DPR policy and architecture for a smart metering service as case study. Different variants of conformance relations between the defined architecture and policy have been proved and refuted using our definitions,  propositions and inference rules. 

This paper is one of the first works of its kind and we believe that in the near future we will propose a software tool based on the initial  theoretical results given in this paper, also extended with automated reasoning algorithms. A further possibility is to extend and improve our previous work on conformance check between architectures and system implementations \cite{TaAntignac14} with DPR properties. As a final result, we will be able to verify if a concrete system implementation conform with the current data protection regulations. Finally, we also intend to extend our languages with the dynamic data control sub-policies given in \cite{TaArxiv15}.

\bibliography{datatest}

\end{document}